\documentclass[10pt]{article}
\usepackage{graphicx}
\usepackage{amsmath}
\usepackage{amssymb}
\usepackage{caption2}
\begin{document}
\bibliographystyle{prsty}
\begin{center}
{\large {\bf \sc{  Analysis of  the ${\frac{3}{2}}^{\pm}$  pentaquark states in the diquark-diquark-antiquark model with QCD sum rules }}} \\[2mm]
Zhi-Gang Wang \footnote{E-mail: zgwang@aliyun.com.  }     \\
 Department of Physics, North China Electric Power University, Baoding 071003, P. R. China
\end{center}

\begin{abstract}
In this article, we   construct    the  axialvector-diquark-scalar-diquark-antiquark type and the axialvector-diquark-axialvector-diquark-antiquark type  interpolating currents, and  study  the $J^P={\frac{3}{2}}^\pm$   hidden-charm pentaquark states with the strangeness  $S=0,\,-1,\,-2,\,-3$ systematically using the QCD sum rules. The predicted masses of the pentaquark states $P_{uud}^{10\frac{1}{2}}\left({\frac{3}{2}^-}\right)$, $P_{uud}^{11\frac{1}{2}}\left({\frac{3}{2}^-}\right)$ and $P_{uud}^{11\frac{3}{2}}\left({\frac{3}{2}^-}\right)$ are compatible with the experimental value  $M_{P_c(4380)}=4380\pm 8\pm 29\,\rm{MeV}$
from the LHCb collaboration, more experimental data  are still needed to identify the $P_c(4380)$ unambiguously.
\end{abstract}



\section{Introduction}

In 2015,  the  LHCb collaboration studied the $\Lambda_b^0\to J/\psi K^- p$ decays with the data sample corresponds to an integrated luminosity of $3\, \rm{fb}^{-1}$ acquired with the LHCb detector from 7 and 8 TeV $pp$ collisions, and observed  two pentaquark candidates $P_c(4380)$ and $P_c(4450)$ in the $J/\psi p$ invariant mass distributions    with the significances of more than 9 standard deviations \cite{LHCb-4380}. They performed the amplitude analysis   on all relevant
masses and decay angles of the six-dimensional   data using the helicity formalism
and Breit-Wigner amplitudes to describe all resonances.
The  Breit-Wigner   masses and widths are  $M_{P_c(4380)}=4380\pm 8\pm 29\,\rm{MeV}$, $M_{P_c(4450)}=4449.8\pm 1.7\pm 2.5\,\rm{MeV}$, $\Gamma_{P_c(4380)}=205\pm 18\pm 86\,\rm{MeV}$, and  $\Gamma_{P_c(4450)}=39\pm 5\pm 19\,\rm{MeV}$, respectively.
The  preferred quantum numbers of the  $P_c(4380)$ and $P_c(4450)$ are   $J^P={\frac{3}{2}}^-$ and ${\frac{5}{2}}^+$, respectively.

Recently, the LHCb collaboration  inspected the $\Lambda_b^0\to J/\psi K^- p$ decays for the presence of $J/\psi p$  or $J/\psi K^-$  contributions with minimal assumptions about $K^- p$  contributions (such as their number, their resonant or nonresonant nature,
or their lineshapes), and observed  that at more than 9 standard deviations the $\Lambda_b^0\to J/\psi K^- p$   decays cannot be described with the $K^- p$  contributions alone, and the $J/\psi p$  contributions play a dominant role in this incompatibility,  and obtained model-independent support for the evidences of the $P_c^+(4380/4500)\to J/\psi p$ decays  \cite{LHCb-1604}.

Furthermore, the LHCb collaboration performed a full amplitude analysis of the $\Lambda_b^0\to J/\psi \pi^- p$ decays allowing
for previously observed conventional ($p\pi^-$) and exotic ($J/\psi p$ and $J/\psi \pi^-$) resonances,  and observed that a
significantly better description of the data can be achieved by either including the two $P_c^+(4380/4450)$
states    or the $Z_c^-(4200)$ state \cite{LHCb-1606}.

 There have been  several  possible  assignments since the observations of the $P_c(4380)$ and $P_c(4450)$, such as the  molecule-like  pentaquark  states \cite{mole-penta} (or not the molecular pentaquark states \cite{mole-penta-No}), the diquark-triquark  type  pentaquark states \cite{di-tri-penta},  the diquark-diquark-antiquark type pentaquark states \cite{Maiani1507,di-di-anti-penta,Wang1508,WangHuang1508,Wang1509}, re-scattering effects \cite{rescattering-penta}, etc. We can diagnose their resonant or non-resonant nature  using photoproduction (or pionproduction) off a proton target \cite{Test-Penta}. The photoproduction process
does not satisfy the so-called anomalous triangle singularity condition, their presence in the $J/\psi$  photoproduction is crucial to conforming the resonant nature of such states as opposed to their being kinematical effects.

The QCD sum rules is a powerful theoretical tool in studying the
ground state heavy baryon states \cite{SVZ79,PRT85,ColangeloReview,NarisonBook}.  We usually take the diquarks as the basic constituents in constructing the baryon currents and pentaquark currents.  The
attractive interactions induced by  one-gluon exchange  favor  formation of
the diquark states in  color antitriplet, flavor
antitriplet and spin singlet $\varepsilon^{ijk}q^{T}_j C\gamma_5 q^{\prime}_k$ or in color antitriplet, flavor
sextet   and spin triplet $\varepsilon^{ijk}q^{T}_j C\gamma_\mu q^{\prime}_k$, where the $i$, $j$ and $k$ are color indexes  \cite{One-gluon},
  the QCD sum rules also favor formation of the diquark states  $\varepsilon^{ijk}q^{T}_j C\gamma_5 q^{\prime}_k$ and  $\varepsilon^{ijk}q^{T}_j C\gamma_\mu q^{\prime}_k$    \cite{WangDiquark,WangLDiquark}, so we can take the scalar or axialvector diquark states as the basic constituents in constructing currents to interpolate the lowest pentaquark states.   In our previous works, we observed that  the  heavy-light scalar and axialvector  diquark states have almost  degenerate masses \cite{WangDiquark}, the masses of the  light  axialvector  diquark states lie   $(150-200)\,\rm{MeV}$ above that of the corresponding  light scalar diquark states \cite{WangLDiquark} from the QCD sum rules.
We expect that the ground state diquark-diquark-antiquark type hidden-charm pentaquark states consist of a light scalar (or axialvector) diquark, a charm scalar (or axialvector) diquark and an anti-charm-quark.

 In Refs.\cite{Wang1508,WangHuang1508}, we take the light scalar diquark states, heavy axialvector diquark states and heavy scalar diquark states as the basic constituents,  construct    the scalar-diquark-axialvector-diquark-antiquark type and scalar-diquark-scalar-diquark-antiquark type   currents to study the  $J^P={\frac{3}{2}}^-$, ${\frac{5}{2}}^+$ and ${\frac{1}{2}}^\pm$ hidden-charm pentaquark states    using  the QCD sum rules.  The predicted masses  $M_{P_c({\frac{3}{2}}^-)}=4.38\pm0.13\,\rm{GeV}$
and $M_{P_c({\frac{5}{2}}^+)}=4.44\pm0.14 \,\rm{GeV}$   support   assigning  the $P_c(4380)$ and $P_c(4450)$ to be the ${\frac{3}{2}}^-$ and ${\frac{5}{2}}^+$ hidden-charm pentaquark states, respectively \cite{Wang1508}. In Ref.\cite{Wang1508}, we observe that the empirical energy scale formula,
 $\mu=\sqrt{M_{X/Y/Z}^2-(2{\mathbb{M}}_Q)^2}$   with  the effective heavy quark mass ${\mathbb{M}}_Q$, in determining the ideal  energy scales of the QCD spectral densities in the QCD sum rules for the tetraquark states $X$, $Y$ and $Z$ \cite{Wang-tetraquark}, can be successfully applied to study the hidden-charm pentaquark states with a slight modification  $\mu=\sqrt{M_{P}^2-(2{\mathbb{M}}_c)^2}$.
  In Ref.\cite{Wang1509}, we take the light axialvector  diquark states, heavy axialvector diquark states and heavy scalar diquark states as the basic constituents, construct   the   axialvector-diquark-axialvector-diquark-antiquark type and axialvector-diquark-scalar-diquark-antiquark type interpolating currents to study the   $J^P={\frac{1}{2}}^\pm$ hidden-charm pentaquark states    with the QCD sum rules in a systematic way.

  In this article, we resort to  the same routine as that in Refs.\cite{Wang1508,WangHuang1508,Wang1509}  to study the masses and pole residues of the $J^P={\frac{3}{2}}^\pm$ hidden-charm pentaquark states  with the QCD sum rules by constructing  the axialvector-diquark-scalar-diquark-antiquark type and axialvector-diquark-axialvector-diquark-antiquark type interpolating currents, and revisit the assignment of the $P_c(4380)$.

 The article is arranged as follows:
  we derive the QCD sum rules for the masses and pole residues of  the
$ {\frac{3}{2}}^{\pm}$ hidden-charm pentaquark states in Sect.2;  in Sect.3, we present the numerical results and discussions; and Sect.4 is reserved for our
conclusion.

\section{QCD sum rules for  the $ {\frac{3}{2}}^{\pm}$ hidden charm pentaquark states}

Firstly, let us write down the two-point correlation functions $\Pi_{q_1q_2q_3,\mu\nu}^{j_Lj_H j}(p)$ in the QCD sum rules,
\begin{eqnarray}
\Pi_{q_1q_2q_3,\mu\nu}^{j_Lj_H j}(p)&=&i\int d^4x e^{ip \cdot x} \langle0|T\left\{J_{q_1q_2q_3,\mu}^{j_Lj_Hj}(x)\bar{J}_{q_1q_2q_3,\nu}^{j_Lj_Hj}(0)\right\}|0\rangle \, ,
\end{eqnarray}
where  the $J_{q_1q_2q_3,\mu}^{j_Lj_Hj}(x)$ are currents to interpolate the $ J^P={\frac{3}{2}}^{\pm}$ hidden-charm tetraquark states, the superscripts $j_L$ and $j_H$  are  the spins of the light diquark state and the charm (or heavy) diquark state,  respectively, $\vec{j}=\vec{j}_H +\vec{j}_{\bar{c}}$, the $j_{\bar{c}}$ is the spin of the charm (or heavy) antiquark, the subscripts $q_1$, $q_2$, $q_3$ are the light quark constituents $u$, $d$ or $s$. Now we write down the interpolating currents $J_{q_1q_2q_3,\mu}^{j_Lj_Hj}(x)$ explicitly,
\begin{eqnarray}
J_{uuu,\mu}^{10\frac{1}{2}}(x)&=&\frac{\varepsilon^{ila} \varepsilon^{ijk}\varepsilon^{lmn}}{\sqrt{3}}   u^T_j(x) C\gamma_\mu u_k(x) u^T_m(x) C\gamma_5 c_n(x) C\bar{c}^{T}_{a}(x) \, , \nonumber \\
J^{10\frac{1}{2}}_{uud,\mu}(x)&=&\frac{\varepsilon^{ila} \varepsilon^{ijk}\varepsilon^{lmn}}{\sqrt{3}} \left[ u^T_j(x) C\gamma_\mu u_k(x) d^T_m(x) C\gamma_5 c_n(x)+2u^T_j(x) C\gamma_\mu d_k(x) u^T_m(x) C\gamma_5 c_n(x)\right]    C\bar{c}^{T}_{a}(x) \, ,  \nonumber\\
J^{10\frac{1}{2}}_{udd,\mu}(x)&=&\frac{\varepsilon^{ila} \varepsilon^{ijk}\varepsilon^{lmn}}{\sqrt{3}} \left[ d^T_j(x) C\gamma_\mu d_k(x) u^T_m(x) C\gamma_5 c_n(x)+2d^T_j(x) C\gamma_\mu u_k(x) d^T_m(x) C\gamma_5 c_n(x)\right]    C\bar{c}^{T}_{a}(x) \, , \nonumber \\
J_{ddd,\mu}^{10\frac{1}{2}}(x)&=&\frac{\varepsilon^{ila} \varepsilon^{ijk}\varepsilon^{lmn}}{\sqrt{3}}   d^T_j(x) C\gamma_\mu d_k(x) d^T_m(x) C\gamma_5 c_n(x) C\bar{c}^{T}_{a}(x) \, ,
\end{eqnarray}

\begin{eqnarray}
J^{10\frac{1}{2}}_{uus,\mu}(x)&=&\frac{\varepsilon^{ila} \varepsilon^{ijk}\varepsilon^{lmn}}{\sqrt{3}} \left[ u^T_j(x) C\gamma_\mu u_k(x) s^T_m(x) C\gamma_5 c_n(x)+2u^T_j(x) C\gamma_\mu s_k(x) u^T_m(x) C\gamma_5 c_n(x)\right]    C\bar{c}^{T}_{a}(x) \, ,  \nonumber\\
J^{10\frac{1}{2}}_{uds,\mu}(x)&=&\frac{\varepsilon^{ila} \varepsilon^{ijk}\varepsilon^{lmn}}{\sqrt{3}} \left[ u^T_j(x) C\gamma_\mu d_k(x) s^T_m(x) C\gamma_5 c_n(x)+u^T_j(x) C\gamma_\mu s_k(x) d^T_m(x) C\gamma_5 c_n(x)\right.  \nonumber\\
&&\left.+d^T_j(x) C\gamma_\mu s_k(x) u^T_m(x) C\gamma_5 c_n(x)\right]    C\bar{c}^{T}_{a}(x) \, ,  \nonumber\\
J^{10\frac{1}{2}}_{dds,\mu}(x)&=&\frac{\varepsilon^{ila} \varepsilon^{ijk}\varepsilon^{lmn}}{\sqrt{3}} \left[ d^T_j(x) C\gamma_\mu d_k(x) s^T_m(x) C\gamma_5 c_n(x)+2d^T_j(x) C\gamma_\mu s_k(x) d^T_m(x) C\gamma_5 c_n(x)\right]    C\bar{c}^{T}_{a}(x) \, ,  \nonumber\\
\end{eqnarray}

\begin{eqnarray}
J^{10\frac{1}{2}}_{uss,\mu}(x)&=&\frac{\varepsilon^{ila} \varepsilon^{ijk}\varepsilon^{lmn}}{\sqrt{3}} \left[ s^T_j(x) C\gamma_\mu s_k(x) u^T_m(x) C\gamma_5 c_n(x)+2s^T_j(x) C\gamma_\mu u_k(x) s^T_m(x) C\gamma_5 c_n(x)\right]    C\bar{c}^{T}_{a}(x) \, , \nonumber \\
J^{10\frac{1}{2}}_{dss,\mu}(x)&=&\frac{\varepsilon^{ila} \varepsilon^{ijk}\varepsilon^{lmn}}{\sqrt{3}} \left[ s^T_j(x) C\gamma_\mu s_k(x) d^T_m(x) C\gamma_5 c_n(x)+2s^T_j(x) C\gamma_\mu d_k(x) s^T_m(x) C\gamma_5 c_n(x)\right]    C\bar{c}^{T}_{a}(x) \, , \nonumber \\
\end{eqnarray}

\begin{eqnarray}
J^{10\frac{1}{2}}_{sss,\mu}(x)&=&\frac{\varepsilon^{ila} \varepsilon^{ijk}\varepsilon^{lmn}}{\sqrt{3}}   s^T_j(x) C\gamma_\mu s_k(x) s^T_m(x) C\gamma_5 c_n(x) C\bar{c}^{T}_{a}(x) \, ,
\end{eqnarray}

\begin{eqnarray}
J^{11\frac{1}{2}}_{uuu,\mu}(x)&=&\varepsilon^{ila} \varepsilon^{ijk}\varepsilon^{lmn}  u^T_j(x) C\gamma_\mu u_k(x)u^T_m(x) C\gamma_\alpha c_n(x)  \gamma_5\gamma^\alpha C\bar{c}^{T}_{a}(x) \, , \nonumber\\
J^{11\frac{1}{2}}_{uud,\mu}(x)&=&\frac{\varepsilon^{ila} \varepsilon^{ijk}\varepsilon^{lmn}}{\sqrt{3}} \left[ u^T_j(x) C\gamma_\mu u_k(x)d^T_m(x) C\gamma_\alpha c_n(x)+2u^T_j(x) C\gamma_\mu d_k(x)u^T_m(x) C\gamma_\alpha c_n(x) \right] \gamma_5\gamma^\alpha C\bar{c}^{T}_{a}(x) \, , \nonumber\\
J^{11\frac{1}{2}}_{udd,\mu}(x)&=&\frac{\varepsilon^{ila} \varepsilon^{ijk}\varepsilon^{lmn}}{\sqrt{3}} \left[ d^T_j(x) C\gamma_\mu d_k(x)u^T_m(x) C\gamma_\alpha c_n(x)+2d^T_j(x) C\gamma_\mu u_k(x)d^T_m(x) C\gamma_\alpha c_n(x) \right] \gamma_5\gamma^\alpha C\bar{c}^{T}_{a}(x) \, , \nonumber\\
J^{11\frac{1}{2}}_{ddd,\mu}(x)&=&\varepsilon^{ila} \varepsilon^{ijk}\varepsilon^{lmn}  d^T_j(x) C\gamma_\mu d_k(x)d^T_m(x) C\gamma_\alpha c_n(x)  \gamma_5\gamma^\alpha C\bar{c}^{T}_{a}(x) \, ,
\end{eqnarray}

\begin{eqnarray}
 J^{11\frac{1}{2}}_{uus,\mu}(x)&=&\frac{\varepsilon^{ila} \varepsilon^{ijk}\varepsilon^{lmn}}{\sqrt{3}} \left[ u^T_j(x) C\gamma_\mu u_k(x)s^T_m(x) C\gamma_\alpha c_n(x)+2u^T_j(x) C\gamma_\mu s_k(x)u^T_m(x) C\gamma_\alpha c_n(x) \right] \gamma_5\gamma^\alpha C\bar{c}^{T}_{a}(x) \, , \nonumber\\
 J^{11\frac{1}{2}}_{uds,\mu}(x)&=&\frac{\varepsilon^{ila} \varepsilon^{ijk}\varepsilon^{lmn}}{\sqrt{3}} \left[ u^T_j(x) C\gamma_\mu d_k(x)s^T_m(x) C\gamma_\alpha c_n(x)+u^T_j(x) C\gamma_\mu s_k(x)d^T_m(x) C\gamma_\alpha c_n(x) \right.\nonumber\\
 &&\left.+d^T_j(x) C\gamma_\mu s_k(x)u^T_m(x) C\gamma_\alpha c_n(x) \right] \gamma_5\gamma^\alpha C\bar{c}^{T}_{a}(x) \, ,\nonumber\\
 J^{11\frac{1}{2}}_{dds,\mu}(x)&=&\frac{\varepsilon^{ila} \varepsilon^{ijk}\varepsilon^{lmn}}{\sqrt{3}} \left[ d^T_j(x) C\gamma_\mu d_k(x)s^T_m(x) C\gamma_\alpha c_n(x)+2d^T_j(x) C\gamma_\mu s_k(x)d^T_m(x) C\gamma_\alpha c_n(x) \right] \gamma_5\gamma^\alpha C\bar{c}^{T}_{a}(x) \, , \nonumber\\
\end{eqnarray}

\begin{eqnarray}
  J^{11\frac{1}{2}}_{uss,\mu}(x)&=&\frac{\varepsilon^{ila} \varepsilon^{ijk}\varepsilon^{lmn}}{\sqrt{3}} \left[ s^T_j(x) C\gamma_\mu s_k(x)u^T_m(x) C\gamma_\alpha c_n(x)+2s^T_j(x) C\gamma_\mu u_k(x)s^T_m(x) C\gamma_\alpha c_n(x) \right] \gamma_5\gamma^\alpha C\bar{c}^{T}_{a}(x) \, , \nonumber\\
    J^{11\frac{1}{2}}_{dss,\mu}(x)&=&\frac{\varepsilon^{ila} \varepsilon^{ijk}\varepsilon^{lmn}}{\sqrt{3}} \left[ s^T_j(x) C\gamma_\mu s_k(x)d^T_m(x) C\gamma_\alpha c_n(x)+2s^T_j(x) C\gamma_\mu d_k(x)s^T_m(x) C\gamma_\alpha c_n(x) \right] \gamma_5\gamma^\alpha C\bar{c}^{T}_{a}(x) \, , \nonumber\\
 \end{eqnarray}

\begin{eqnarray}
  J^{11\frac{1}{2}}_{sss,\mu}(x)&=&\varepsilon^{ila} \varepsilon^{ijk}\varepsilon^{lmn}  s^T_j(x) C\gamma_\mu s_k(x)s^T_m(x) C\gamma_\alpha c_n(x)  \gamma_5\gamma^\alpha C\bar{c}^{T}_{a}(x) \, ,
 \end{eqnarray}

\begin{eqnarray}
  J^{11\frac{3}{2}}_{uuu,\mu}(x)&=&\varepsilon^{ila} \varepsilon^{ijk}\varepsilon^{lmn}  u^T_j(x) C\gamma_\alpha u_k(x)u^T_m(x) C\gamma_\mu c_n(x)  \gamma_5\gamma^\alpha C\bar{c}^{T}_{a}(x) \, , \nonumber\\
  J^{11\frac{3}{2}}_{uud,\mu}(x)&=&\frac{\varepsilon^{ila} \varepsilon^{ijk}\varepsilon^{lmn}}{\sqrt{3}} \left[ u^T_j(x) C\gamma_\alpha u_k(x)d^T_m(x) C\gamma_\mu c_n(x)+2u^T_j(x) C\gamma_\alpha d_k(x)u^T_m(x) C\gamma_\mu c_n(x) \right] \gamma_5\gamma^\alpha C\bar{c}^{T}_{a}(x) \, , \nonumber\\
  J^{11\frac{3}{2}}_{udd,\mu}(x)&=&\frac{\varepsilon^{ila} \varepsilon^{ijk}\varepsilon^{lmn}}{\sqrt{3}} \left[ d^T_j(x) C\gamma_\alpha d_k(x)u^T_m(x) C\gamma_\mu c_n(x)+2d^T_j(x) C\gamma_\alpha u_k(x)d^T_m(x) C\gamma_\mu c_n(x) \right] \gamma_5\gamma^\alpha C\bar{c}^{T}_{a}(x) \, , \nonumber\\
  J^{11\frac{3}{2}}_{ddd,\mu}(x)&=&\varepsilon^{ila} \varepsilon^{ijk}\varepsilon^{lmn}  d^T_j(x) C\gamma_\alpha d_k(x)d^T_m(x) C\gamma_\mu c_n(x)  \gamma_5\gamma^\alpha C\bar{c}^{T}_{a}(x) \, ,
\end{eqnarray}

\begin{eqnarray}
 J^{11\frac{3}{2}}_{uus,\mu}(x)&=&\frac{\varepsilon^{ila} \varepsilon^{ijk}\varepsilon^{lmn}}{\sqrt{3}} \left[ u^T_j(x) C\gamma_\alpha u_k(x)s^T_m(x) C\gamma_\mu c_n(x)+2u^T_j(x) C\gamma_\alpha s_k(x)u^T_m(x) C\gamma_\mu c_n(x) \right] \gamma_5\gamma^\alpha C\bar{c}^{T}_{a}(x) \, , \nonumber\\
  J^{11\frac{3}{2}}_{uds,\mu}(x)&=&\frac{\varepsilon^{ila} \varepsilon^{ijk}\varepsilon^{lmn}}{\sqrt{3}} \left[ u^T_j(x) C\gamma_\alpha d_k(x)s^T_m(x) C\gamma_\mu c_n(x)+u^T_j(x) C\gamma_\alpha s_k(x)d^T_m(x) C\gamma_\mu c_n(x) \right. \nonumber\\
  &&\left.+d^T_j(x) C\gamma_\alpha s_k(x)u^T_m(x) C\gamma_\mu c_n(x) \right] \gamma_5\gamma^\alpha C\bar{c}^{T}_{a}(x) \, , \nonumber\\
   J^{11\frac{3}{2}}_{dds,\mu}(x)&=&\frac{\varepsilon^{ila} \varepsilon^{ijk}\varepsilon^{lmn}}{\sqrt{3}} \left[ d^T_j(x) C\gamma_\alpha d_k(x)s^T_m(x) C\gamma_\mu c_n(x)+2d^T_j(x) C\gamma_\alpha s_k(x)d^T_m(x) C\gamma_\mu c_n(x) \right] \gamma_5\gamma^\alpha C\bar{c}^{T}_{a}(x) \, , \nonumber\\
 \end{eqnarray}

\begin{eqnarray}
  J^{11\frac{3}{2}}_{uss,\mu}(x)&=&\frac{\varepsilon^{ila} \varepsilon^{ijk}\varepsilon^{lmn}}{\sqrt{3}} \left[ s^T_j(x) C\gamma_\alpha s_k(x)u^T_m(x) C\gamma_\mu c_n(x)+2s^T_j(x) C\gamma_\alpha u_k(x)s^T_m(x) C\gamma_\mu c_n(x) \right] \gamma_5\gamma^\alpha C\bar{c}^{T}_{a}(x) \, , \nonumber\\
   J^{11\frac{3}{2}}_{dss,\mu}(x)&=&\frac{\varepsilon^{ila} \varepsilon^{ijk}\varepsilon^{lmn}}{\sqrt{3}} \left[ s^T_j(x) C\gamma_\alpha s_k(x)d^T_m(x) C\gamma_\mu c_n(x)+2s^T_j(x) C\gamma_\alpha d_k(x)s^T_m(x) C\gamma_\mu c_n(x) \right] \gamma_5\gamma^\alpha C\bar{c}^{T}_{a}(x) \, , \nonumber\\
 \end{eqnarray}

\begin{eqnarray}
  J^{11\frac{3}{2}}_{sss,\mu}(x)&=&\varepsilon^{ila} \varepsilon^{ijk}\varepsilon^{lmn}  s^T_j(x) C\gamma_\alpha s_k(x)s^T_m(x) C\gamma_\mu c_n(x)  \gamma_5\gamma^\alpha C\bar{c}^{T}_{a}(x) \, ,
 \end{eqnarray}
where the $i$, $j$, $k$, $l$, $m$, $n$ and $a$ are color indices, the $C$ is the charge conjugation matrix.

We take the isospin limit by assuming the $u$ and $d$ quarks have the degenerate masses,
and classify the above thirty currents  couple to the hidden-charm pentaquark states with degenerate masses  into the following  twelve types,
\begin{eqnarray}
1. && J_{uuu,\mu}^{10\frac{1}{2}}(x)\, , \, \, \, J^{10\frac{1}{2}}_{uud,\mu}(x)\, , \, \, \, J^{10\frac{1}{2}}_{udd,\mu}(x)\, , \, \, \,
J_{ddd,\mu}^{10\frac{1}{2}}(x)\, ; \nonumber\\
2.&&J^{10\frac{1}{2}}_{uus,\mu}(x)\, , \, \, \, J^{10\frac{1}{2}}_{uds,\mu}(x)\, , \, \, \, J^{10\frac{1}{2}}_{dds,\mu}(x) \, ; \nonumber\\
3.&&J^{10\frac{1}{2}}_{uss,\mu}(x)\, , \, \, \, J^{10\frac{1}{2}}_{dss,\mu}(x) \, ; \nonumber \\
4.&&J^{10\frac{1}{2}}_{sss,\mu}(x) \, ; \nonumber\\
5.&&J^{11\frac{1}{2}}_{uuu,\mu}(x)\, , \, \, \, J^{11\frac{1}{2}}_{uud,\mu}(x)\, , \, \, \, J^{11\frac{1}{2}}_{udd,\mu}(x)\, , \, \, \,
J^{11\frac{1}{2}}_{ddd,\mu}(x) \, ; \nonumber\\
6.&&J^{11\frac{1}{2}}_{uus,\mu}(x)\, , \, \, \,  J^{11\frac{1}{2}}_{uds,\mu}(x)\, , \, \, \,  J^{11\frac{1}{2}}_{dds,\mu}(x) \, ; \nonumber\\
7.&&  J^{11\frac{1}{2}}_{uss,\mu}(x)\, , \, \, \,   J^{11\frac{1}{2}}_{dss,\mu}(x)\, ; \nonumber\\
8.&&  J^{11\frac{1}{2}}_{sss,\mu}(x) \, ; \nonumber\\
9.&&  J^{11\frac{3}{2}}_{uuu,\mu}(x)\, , \, \, \, J^{11\frac{3}{2}}_{uud,\mu}(x)\, , \, \, \,  J^{11\frac{3}{2}}_{udd,\mu}(x)\, , \, \, \, J^{11\frac{3}{2}}_{ddd,\mu}(x) \, ; \nonumber\\
10.&& J^{11\frac{3}{2}}_{uus,\mu}(x)\, , \, \, \,  J^{11\frac{3}{2}}_{uds,\mu}(x)\, , \, \, \,  J^{11\frac{3}{2}}_{dds,\mu}(x) \, ; \nonumber\\
11.&&  J^{11\frac{3}{2}}_{uss,\mu}(x)\, , \, \, \,  J^{11\frac{3}{2}}_{dss,\mu}(x) \, ; \nonumber\\
 12.&&  J^{11\frac{3}{2}}_{sss,\mu}(x) \, .
 \end{eqnarray}
In calculations, we choose the first current in each type for simplicity.

The currents $J_{q_1q_2q_3,\mu}^{j_Lj_Hj}(0)$     couple potentially to the $J^P={\frac{3}{2}}^\pm$ and $J^P={\frac{1}{2}}^\pm$  hidden-charm  pentaquark  states $P_{q_1q_2q_3}^{j_Lj_H j}( {\frac{3}{2}}^\pm)$ and $P_{q_1q_2q_3}^{j_Lj_H j}( {\frac{1}{2}}^\pm)$, respectively,
\begin{eqnarray}
\langle 0| J_{q_1q_2q_3,\mu}^{j_Lj_Hj}(0)|P_{q_1q_2q_3}^{j_Lj_H j}( {\frac{3}{2}}^-)(p)\rangle &=&\lambda^{-}_{P}\, U^{-}_\mu(p,s) \, , \nonumber\\
\langle 0| J_{q_1q_2q_3,\mu}^{j_Lj_H j}(0)|P_{q_1q_2q_3}^{j_Lj_H j}( {\frac{3}{2}}^+)(p)\rangle &=&\lambda^{+}_{P}i\gamma_5 \,U^{+}_\mu(p,s) \, ,\nonumber\\
\langle 0| J_{q_1q_2q_3,\mu}^{j_Lj_Hj}(0)|P_{q_1q_2q_3}^{j_Lj_H j}( {\frac{1}{2}}^+)(p)\rangle &=&\widetilde{\lambda}^{+}_{P} \,U^{+}(p,s)\, p_\mu \, , \nonumber\\
\langle 0| J_{q_1q_2q_3,\mu}^{j_Lj_H j}(0)|P_{q_1q_2q_3}^{j_Lj_H j}( {\frac{1}{2}}^-)(p)\rangle &=&\widetilde{\lambda}^{-}_{P}i\gamma_5 \,U^{-}(p,s)\,p_\mu \, ,
\end{eqnarray}
where the $\lambda^{\pm}_P$ and $\widetilde{\lambda}^{\pm}_P$ are the pole residues or current-hadron coupling constants, the $U_\mu^{\pm}(p,s)$ and $U^{\pm}(p,s)$ are the Dirac spinors  \cite{Chung82,Bagan93,Oka96,WangHbaryon}.  The Dirac  spinors $U_\mu^{\pm}(p,s)$   obey  the Rarita-Schwinger equation $(\not\!\!p-M_{P,\pm})U^{\pm}_\mu(p)=0$,  and the relations $\gamma^\mu U^{\pm}_\mu(p,s)=0$,
$p^\mu U^{\pm}_\mu(p,s)=0$; while the Dirac  spinors $U^{\pm}(p,s)$   obey  the Dirac equation $(\not\!\!p-\widetilde{M}_{P,\pm})U^{\pm}(p)=0$.

 At the phenomenological side, we  can insert  a complete set  of intermediate hidden-charm pentaquark states with the
same quantum numbers as the current operators $J_{q_1q_2q_3,\mu}^{j_Lj_Hj}(x)$   and $i\gamma_5 J_{q_1q_2q_3,\mu}^{j_Lj_Hj}(x)$ into the correlation functions
$\Pi_{q_1q_2q_3,\mu\nu}^{j_Lj_H j}(p)$   to obtain the hadronic representation
\cite{SVZ79,PRT85}. After isolating the pole terms of the lowest
states of the hidden-charm  pentaquark states with $J^P={\frac{3}{2}}^\pm$ and $J^P={\frac{1}{2}}^\pm$ respectively, we obtain the
following results:
\begin{eqnarray}
  \Pi_{q_1q_2q_3,\mu\nu}^{j_Lj_H j}(p) & = & \left( \lambda^{-}_{P}{}^2  {\!\not\!{p}+ M_{P,-} \over M_{P,-}^{2}-p^{2}  } +\lambda^{+}_{P}{}^2  {\!\not\!{p}- M_{P,+} \over M_{P,+}^{2}-p^{2}  }\right) \left( -g_{\mu\nu}+\frac{p_\mu p_\nu}{p^2}\right)       \nonumber\\
  &&+\left(  \widetilde{\lambda}^{+}_{P}{}^2    {\!\not\!{p}+ \widetilde{M}_{P,+} \over \widetilde{M}_{P,+}^{2}-p^{2}  } +\widetilde{\lambda}^{-}_{P}{}^2  {\!\not\!{p}- \widetilde{M}_{P,-} \over \widetilde{M}_{P,-}^{2}-p^{2}  }\right) p_\mu p_\nu+\cdots\nonumber\\
  &=&\Pi_{q_1q_2q_3}^{j_Lj_H j}(p^2)\left( -g_{\mu\nu}+\frac{p_\mu p_\nu}{p^2}\right)     +\widetilde{\Pi}_{q_1q_2q_3}^{j_Lj_H j}(p^2)\,p_\mu p_\nu  \, .
    \end{eqnarray}
 In this article, we choose the component $\Pi_{q_1q_2q_3}^{j_Lj_H j}(p^2)$ associated with the tensor structure $-g_{\mu\nu}+\frac{p_\mu p_\nu}{p^2}$ to study the hidden-charm  pentaquark states with $J^P={\frac{3}{2}}^\pm$, there are no contaminations come from the hidden-charm  pentaquark states with $J^P={\frac{1}{2}}^\pm$ in the component $\widetilde{\Pi}_{q_1q_2q_3}^{j_Lj_H j}(p^2)$ associated with the tensor structure $p_\mu p_\nu$. It is easy to project out the component $\Pi_{q_1q_2q_3}^{j_Lj_H j}(p^2)$,
 \begin{eqnarray}
 \Pi_{q_1q_2q_3}^{j_Lj_H j}(p^2)&=&\frac{1}{3}\left( -g_{\mu\nu}+\frac{p_\mu p_\nu}{p^2}\right)\Pi_{q_1q_2q_3,\mu\nu}^{j_Lj_H j}(p)\, .
 \end{eqnarray}

We can obtain the hadronic spectral densities  through  dispersion relation,
\begin{eqnarray}
\frac{{\rm Im}\Pi_{q_1q_2q_3}^{j_Lj_H j}(s)}{\pi}&=&\!\not\!{p} \left[{\lambda^{-}_{P}}^2 \delta\left(s-M_{P,-}^2\right)+{\lambda^{+}_{P}}^2 \delta\left(s-M_{P,+}^2\right)\right] \nonumber\\
&& +\left[M_{P,-}{\lambda^{-}_{P}}^2 \delta\left(s-M_{P,-}^2\right)-M_{P,+}{\lambda^{+}_{P}}^2 \delta\left(s-M_{P,+}^2\right)\right]\, , \nonumber\\
&=&\!\not\!{p} \,\rho_{q_1q_2q_3,H}^{j_Lj_H j,1}(s)+\rho_{q_1q_2q_3,H}^{j_Lj_H j,0}(s) \, ,
\end{eqnarray}
where the subscript index $H$ denotes the hadron side (or phenomenological side).  Then we introduce the weight function $\exp\left(-\frac{s}{T^2}\right)$ to obtain the phenomenological  side of the QCD sum rules,
\begin{eqnarray}
\int_{4m_c^2}^{s_0}ds \left[\sqrt{s}\rho_{q_1q_2q_3,H}^{j_Lj_H j,1}(s)+\rho_{q_1q_2q_3,H}^{j_Lj_H j,0}(s)\right]\exp\left( -\frac{s}{T^2}\right)
&=&2M_{P,-}\lambda^{-}_{P}{}^2\exp\left( -\frac{M_{P,-}^2}{T^2}\right) \, , \\
\int_{4m_c^2}^{s_0}ds \left[\sqrt{s}\rho_{q_1q_2q_3,H}^{j_Lj_H j,1}(s)-\rho_{q_1q_2q_3,H}^{j_Lj_H j,0}(s)\right]\exp\left( -\frac{s}{T^2}\right)
&=&2M_{P,+}\lambda^{+}_{P}{}^2\exp\left( -\frac{M_{P,+}^2}{T^2}\right) \, ,
\end{eqnarray}
where the $T^2$ are the Borel parameters and the $s_0$ are the continuum threshold parameters. In Eqs.(19-20), we separate the contributions of   the
negative-parity  pentaquark states from that of the positive-parity  pentaquark states explicitly. They do not contaminate each other, and warrant robust predictions.

Now we briefly outline  the operator product expansion for the correlation functions $ \Pi_{q_1q_2q_3,\mu\nu}^{j_Lj_H j}(p)$ and use the example $\Pi_{sss,\mu\nu}^{10\frac{1}{2}}(p)$ to illustrate the routine. Firstly,  we contract the   $s$ and $c$ quark fields in the correlation function $\Pi_{sss,\mu\nu}^{10\frac{1}{2}}(p)$    with Wick theorem,
\begin{eqnarray}
\Pi_{sss,\mu\nu}^{10\frac{1}{2}}(p)&=&i\,\varepsilon^{ila}\varepsilon^{ijk}\varepsilon^{lmn}\varepsilon^{i^{\prime}l^{\prime}a^{\prime}}\varepsilon^{i^{\prime}j^{\prime}k^{\prime}}
\varepsilon^{l^{\prime}m^{\prime}n^{\prime}}\int d^4x e^{ip\cdot x}C C_{a^{\prime}a}^T(-x)C \nonumber\\
&&\left\{ 2  Tr\left[\gamma_\mu S_{kk^\prime}(x) \gamma_\nu C S^{T}_{jj^\prime}(x)C\right] \,Tr\left[\gamma_5 C_{nn^\prime}(x) \gamma_5 C S^{T}_{mm^\prime}(x)C\right] \right. \nonumber\\
&&\left.-4  Tr \left[\gamma_\mu S_{kk^\prime}(x) \gamma_\nu C S^{T}_{mj^\prime}(x)C \gamma_5C_{nn^\prime}(x) \gamma_5 C S^{T}_{jm^\prime}(x)C\right]  \right\} \, ,
\end{eqnarray}
where the  $S_{ij}(x)$ and $C_{ij}(x)$ are the full  $s$ and $c$ quark propagators, respectively,
 \begin{eqnarray}
S_{ij}(x)&=& \frac{i\delta_{ij}\!\not\!{x}}{ 2\pi^2x^4}
-\frac{\delta_{ij}m_s}{4\pi^2x^2}-\frac{\delta_{ij}\langle
\bar{s}s\rangle}{12} +\frac{i\delta_{ij}\!\not\!{x}m_s
\langle\bar{s}s\rangle}{48}-\frac{\delta_{ij}x^2\langle \bar{s}g_s\sigma Gs\rangle}{192}+\frac{i\delta_{ij}x^2\!\not\!{x} m_s\langle \bar{s}g_s\sigma
 Gs\rangle }{1152}\nonumber\\
&& -\frac{ig_s G^{a}_{\alpha\beta}t^a_{ij}(\!\not\!{x}
\sigma^{\alpha\beta}+\sigma^{\alpha\beta} \!\not\!{x})}{32\pi^2x^2}  -\frac{1}{8}\langle\bar{s}_j\sigma^{\mu\nu}s_i \rangle \sigma_{\mu\nu}+\cdots \, ,
\end{eqnarray}

\begin{eqnarray}
C_{ij}(x)&=&\frac{i}{(2\pi)^4}\int d^4k e^{-ik \cdot x} \left\{
\frac{\delta_{ij}}{\!\not\!{k}-m_c}
-\frac{g_sG^n_{\alpha\beta}t^n_{ij}}{4}\frac{\sigma^{\alpha\beta}(\!\not\!{k}+m_c)+(\!\not\!{k}+m_c)
\sigma^{\alpha\beta}}{(k^2-m_c^2)^2}\right.\nonumber\\
&&\left. -\frac{g_s^2 (t^at^b)_{ij} G^a_{\alpha\beta}G^b_{\mu\nu}(f^{\alpha\beta\mu\nu}+f^{\alpha\mu\beta\nu}+f^{\alpha\mu\nu\beta}) }{4(k^2-m_c^2)^5}+\cdots\right\} \, ,\nonumber\\
f^{\alpha\beta\mu\nu}&=&(\!\not\!{k}+m_c)\gamma^\alpha(\!\not\!{k}+m_c)\gamma^\beta(\!\not\!{k}+m_c)\gamma^\mu(\!\not\!{k}+m_c)\gamma^\nu(\!\not\!{k}+m_c)\, ,
\end{eqnarray}
and  $t^n=\frac{\lambda^n}{2}$, the $\lambda^n$ is the Gell-Mann matrix   \cite{PRT85}, then we compute all the integrals  in the coordinate space and in the momentum space  to obtain the correlation function $\Pi_{sss,\mu\nu}^{10\frac{1}{2}}(p)$ at the quark level. The calculations are   straightforward  but  tedious. All the correlation functions $\Pi_{q_1q_2q_3,\mu\nu}^{j_Lj_H j}(p)$ at the QCD side can be written as
 \begin{eqnarray}
  \Pi_{q_1q_2q_3,\mu\nu}^{j_Lj_H j}(p)   &=&\Pi_{q_1q_2q_3,QCD}^{j_Lj_H j}(p^2)\left( -g_{\mu\nu}+\frac{p_\mu p_\nu}{p^2}\right)     +\cdots \, ,
    \end{eqnarray}
 where we add the notation $QCD$ to denote the QCD side.  Once the analytical expressions of the correlation functions $\Pi_{q_1q_2q_3,\mu\nu}^{j_Lj_H j}(p) $ are gotten, we can obtain the QCD spectral densities $\rho_{q_1q_2q_3}^{j_Lj_H j,1}(s)$ and $\widetilde{\rho}_{q_1q_2q_3}^{j_Lj_H j,0}(s)$ through   dispersion  relation,
   \begin{eqnarray}
\frac{{\rm Im}\Pi_{q_1q_2q_3,QCD}^{j_Lj_H j}(s)}{\pi}&=&\!\not\!{p} \,\rho_{q_1q_2q_3}^{j_Lj_H j,1}(s)+m_c\widetilde{\rho}_{q_1q_2q_3}^{j_Lj_H j,0}(s) \, .
\end{eqnarray}
The lengthy  expressions of the   QCD spectral densities $\rho_{q_1q_2q_3}^{j_Lj_H j,1}(s)$ and $\widetilde{\rho}_{q_1q_2q_3}^{j_Lj_H j,0}(s)$  are given explicitly in the Appendix. We carry out the operator product expansion to the vacuum condensates  up to dimension 10, and
assume vacuum saturation for the  higher dimension vacuum condensates.
The   gluon condensates  are accompanied  with  large numerical denominators, their contributions to the total QCD spectral densities are  less (or much less) than the contributions of the dimension 10 vacuum condensates  $D_{10}$ \cite{Wang1508,WangHuang1508}, see Eq.(26), so we  neglect the vacuum condensates involving the gluon condensate $\langle \frac{\alpha_sGG}{\pi}\rangle$, $\langle \bar{q}q\rangle\langle \frac{\alpha_sGG}{\pi}\rangle$, $\langle \bar{s}s\rangle\langle \frac{\alpha_sGG}{\pi}\rangle$, $\langle \bar{q}q\rangle^2\langle \frac{\alpha_sGG}{\pi}\rangle$, $\langle \bar{q}q\rangle\langle \bar{s}s\rangle\langle \frac{\alpha_sGG}{\pi}\rangle$, $\langle \bar{s}s\rangle^2\langle \frac{\alpha_sGG}{\pi}\rangle$, the predictive ability will not be impaired.
In calculations,  we take into account the contributions of the terms $D_0$, $D_3$, $D_5$, $D_6$, $D_8$, $D_9$ and $D_{10}$,
where
\begin{eqnarray}
D_0&=& {\rm perturbative \,\,\,\, terms}\, , \nonumber\\
D_3&\propto& \langle \bar{q}q\rangle\, , \,\langle \bar{s}s\rangle\, ,  \nonumber\\
D_5&\propto& \langle \bar{q}g_s\sigma Gq\rangle\, , \,\langle \bar{s}g_s\sigma Gs\rangle \, , \nonumber\\
D_6&\propto& \langle \bar{q}q\rangle^2\, , \, \langle \bar{q}q\rangle \langle \bar{s}s\rangle\, , \,\langle \bar{s}s\rangle^2  \, , \nonumber\\
D_8&\propto& \langle\bar{q}q\rangle\langle \bar{q}g_s\sigma Gq\rangle\, , \,\langle\bar{s}s\rangle\langle \bar{q}g_s\sigma Gq\rangle\, , \,\langle\bar{q}q\rangle\langle \bar{s}g_s\sigma Gs\rangle\, , \, \langle\bar{s}s\rangle\langle \bar{s}g_s\sigma Gs\rangle\, ,   \nonumber\\
D_9&\propto& \langle \bar{q}q\rangle^3\, , \,\langle \bar{q}q\rangle\langle \bar{s}s\rangle^2\, , \, \langle \bar{q}q\rangle^2 \langle \bar{s}s\rangle\, , \, \langle \bar{s}s\rangle^3\, ,  \nonumber\\
D_{10}&\propto& \langle \bar{q}g_s\sigma Gq\rangle^2\, , \, \langle \bar{q}g_s\sigma Gq\rangle \langle \bar{s}g_s\sigma Gs\rangle\, , \,\langle \bar{s}g_s\sigma Gs\rangle^2  \, .
\end{eqnarray}

 Now  we can take the quark-hadron duality below the continuum thresholds  $s_0$ and  obtain  the following QCD sum rules:
\begin{eqnarray}
2M_{P,-}\lambda^{-}_{P}{}^2\exp\left( -\frac{M_{P,-}^2}{T^2}\right)
&=& \int_{4m_c^2}^{s_0}ds \left[\sqrt{s}\rho_{q_1q_2q_3}^{j_Lj_H j,1}(s)+m_c\widetilde{\rho}_{q_1q_2q_3}^{j_Lj_H j,0}(s)\right]\exp\left( -\frac{s}{T^2}\right)\, ,\\
2M_{P,+}\lambda^{+}_{P}{}^2\exp\left( -\frac{M_{P,+}^2}{T^2}\right)
&=& \int_{4m_c^2}^{s_0}ds \left[\sqrt{s}\rho_{q_1q_2q_3}^{j_Lj_H j,1}(s)-m_c\widetilde{\rho}_{q_1q_2q_3}^{j_Lj_H j,0}(s)\right]\exp\left( -\frac{s}{T^2}\right)\, .
\end{eqnarray}

We differentiate   Eqs.(27-28) with respect to  $\frac{1}{T^2}$, then eliminate the
 pole residues $\lambda^{\pm}_{P}$ and obtain the QCD sum rules for
 the masses of the hidden-charm  pentaquark states $M^2_{P,\mp}$,
 \begin{eqnarray}
 M^2_{P,-} &=& \frac{\int_{4m_c^2}^{s_0}ds \frac{d}{d(-1/T^2)}\left[\sqrt{s}\rho_{q_1q_2q_3}^{j_Lj_H j,1}(s)+m_c\widetilde{\rho}_{q_1q_2q_3}^{j_Lj_H j,0}(s)\right]\exp\left( -\frac{s}{T^2}\right)}{\int_{4m_c^2}^{s_0}ds \left[\sqrt{s}\rho_{q_1q_2q_3}^{j_Lj_H j,1}(s)+m_c\widetilde{\rho}_{q_1q_2q_3}^{j_Lj_H j,0}(s)\right]\exp\left( -\frac{s}{T^2}\right)}\, ,\\
 M^2_{P,+} &=& \frac{\int_{4m_c^2}^{s_0}ds \frac{d}{d(-1/T^2)}\left[\sqrt{s}\rho_{q_1q_2q_3}^{j_Lj_H j,1}(s)-m_c\widetilde{\rho}_{q_1q_2q_3}^{j_Lj_H j,0}(s)\right]\exp\left( -\frac{s}{T^2}\right)}{\int_{4m_c^2}^{s_0}ds \left[\sqrt{s}\rho_{q_1q_2q_3}^{j_Lj_H j,1}(s)-m_c\widetilde{\rho}_{q_1q_2q_3}^{j_Lj_H j,0}(s)\right]\exp\left( -\frac{s}{T^2}\right)}\, .
\end{eqnarray}
We obtain the masses through  fractions, see Eqs.(29-30), the contributions involving the gluon condensate in the  numerator and  denominator are canceled  out with each other,   the main effects of the gluon condensates can be safely absorbed into the pole residues $\lambda^{\pm}_{P}$ and the remaining tiny effects on the masses can be  safely neglected.

\section{Numerical results and discussions}
The input parameters are shown explicitly in Table 1.
The quark condensates, mixed quark condensates and $\overline{MS}$ masses  evolve with the   renormalization group equation, we take into account
the energy-scale dependence according to the following equations,
\begin{eqnarray}
\langle\bar{q}q \rangle(\mu)&=&\langle\bar{q}q \rangle(Q)\left[\frac{\alpha_{s}(Q)}{\alpha_{s}(\mu)}\right]^{\frac{4}{9}}\, , \nonumber\\
 \langle\bar{s}s \rangle(\mu)&=&\langle\bar{s}s \rangle(Q)\left[\frac{\alpha_{s}(Q)}{\alpha_{s}(\mu)}\right]^{\frac{4}{9}}\, , \nonumber\\
 \langle\bar{q}g_s \sigma Gq \rangle(\mu)&=&\langle\bar{q}g_s \sigma Gq \rangle(Q)\left[\frac{\alpha_{s}(Q)}{\alpha_{s}(\mu)}\right]^{\frac{2}{27}}\, , \nonumber\\ \langle\bar{s}g_s \sigma Gs \rangle(\mu)&=&\langle\bar{s}g_s \sigma Gs \rangle(Q)\left[\frac{\alpha_{s}(Q)}{\alpha_{s}(\mu)}\right]^{\frac{2}{27}}\, , \nonumber\\
m_c(\mu)&=&m_c(m_c)\left[\frac{\alpha_{s}(\mu)}{\alpha_{s}(m_c)}\right]^{\frac{12}{25}} \, ,\nonumber\\
m_s(\mu)&=&m_s({\rm 2GeV} )\left[\frac{\alpha_{s}(\mu)}{\alpha_{s}({\rm 2GeV})}\right]^{\frac{4}{9}} \, ,\nonumber\\
\alpha_s(\mu)&=&\frac{1}{b_0t}\left[1-\frac{b_1}{b_0^2}\frac{\log t}{t} +\frac{b_1^2(\log^2{t}-\log{t}-1)+b_0b_2}{b_0^4t^2}\right]\, ,
\end{eqnarray}
  where $t=\log \frac{\mu^2}{\Lambda^2}$, $b_0=\frac{33-2n_f}{12\pi}$, $b_1=\frac{153-19n_f}{24\pi^2}$, $b_2=\frac{2857-\frac{5033}{9}n_f+\frac{325}{27}n_f^2}{128\pi^3}$,  $\Lambda=213\,\rm{MeV}$, $296\,\rm{MeV}$  and  $339\,\rm{MeV}$ for the flavors  $n_f=5$, $4$ and $3$, respectively  \cite{PDG}.
 Furthermore, we set the $m_u=m_d=0$.

\begin{table}
\begin{center}
\begin{tabular}{|c|c|c|c|}\hline\hline
    Parameters                                          & Values\\   \hline
   $\langle\bar{q}q \rangle({\rm 1GeV})$                & $-(0.24\pm 0.01\, \rm{GeV})^3$ \,\, \cite{SVZ79,PRT85,ColangeloReview}         \\  \hline
   $\langle\bar{s}s \rangle({\rm 1GeV})$                & $(0.8\pm0.1)\langle\bar{q}q \rangle({\rm 1GeV})$ \,\, \cite{SVZ79,PRT85,ColangeloReview}     \\ \hline
$\langle\bar{q}g_s\sigma G q \rangle({\rm 1GeV})$       & $m_0^2\langle \bar{q}q \rangle({\rm 1GeV})$   \,\,  \cite{SVZ79,PRT85,ColangeloReview}       \\  \hline
$\langle\bar{s}g_s\sigma G s \rangle({\rm 1GeV})$       & $m_0^2\langle \bar{s}s \rangle({\rm 1GeV})$  \,\,  \cite{SVZ79,PRT85,ColangeloReview}        \\  \hline
$m_0^2({\rm 1GeV})$                                     & $(0.8 \pm 0.1)\,\rm{GeV}^2$      \,\,  \cite{SVZ79,PRT85,ColangeloReview}    \\   \hline
   $m_{c}(m_c)$                                         & $(1.275\pm0.025)\,\rm{GeV}$ \,\, \cite{PDG}      \\    \hline
   $m_{s}({\rm 2GeV})$                                  & $(0.095\pm0.005)\,\rm{GeV}$  \,\, \cite{PDG}      \\   \hline \hline
\end{tabular}
\end{center}
\caption{ The  input parameters in the QCD sum rules, the values in the bracket denote the energy scales $\mu=1\,\rm{GeV}$, $2\,\rm{GeV}$ and $m_c$, respectively. }
\end{table}

In Refs.\cite{WangHbaryon,WangLambda},  we separate the contributions come from the positive parity and negative parity baryon states explicitly, and study the masses and pole residues of the  $J^P={1\over 2}^{\pm}$ and ${3\over 2}^{\pm}$ heavy, doubly-heavy and triply-heavy baryon states  with the QCD sum rules systematically. In calculations, we observe that the continuum threshold parameters $\sqrt{s_0}=M_{\rm{gr}}+ (0.6-0.8)\,\rm{GeV}$  can reproduce the experimental values of the masses of the observed heavy baryon states \cite{PDG},  where the subscript $\rm{gr}$ denotes the ground state  baryon states.
The pentaquark states are another type baryon states considering  the fractional spins, such as $1\over 2$, $3\over 2$, $5\over 2$. In Ref.\cite{Wang1508},  we take the continuum threshold parameters as
$\sqrt{s_0}= M_{P_c(4380/4450)}+(0.6-0.8)\,\rm{GeV}$ in the QCD sum rules for the scalar-diquark-axialvector-diquark-antiquark type hidden-charm pentaquark states,
which can reproduce the experimental values of the masses  $M_{P_c(4380/4450)}$.
In the present QCD sum rules, we take the continuum threshold parameters as $\sqrt{s_0}= M_{P}+(0.6-0.8)\,\rm{GeV}$.

The hidden-charm or hidden-bottom five-quark systems  $q_1q_2q_3Q\bar{Q}$ can be described
by a double-well potential in the heavy quark limit. The two light quarks $q_1$ and $q_2$ combine together to form a light diquark state or correlation $\mathcal{D}^i_{\bar{3}}(q_1q_2)$ in color antitriplet, the  heavy antiquark $\bar{Q}$ serves  as one  static well potential and combines with the light diquark state $\mathcal{D}^i_{\bar{3}}(q_1q_2)$  to form a heavy antitriquark state or correlation $\varepsilon_{ijk}\mathcal{D}^i_{\bar{3}}(q_1q_2)\bar{Q}^j_{\bar{3}}$ in  color triplet, where the $i$, $j$ and $k$ are color indexes. On the other hand,  the heavy quark $Q$ serves as the other  static well potential and  combines with the light quark $q_3$  to form a heavy diquark state or correlation $\mathcal{D}^k_{\bar{3}}(q_3Q)$ in  color antitriplet. The heavy diquark state $\mathcal{D}^k_{\bar{3}}(q_3Q)$ and the heavy antitriquark state $\varepsilon_{ijk}\mathcal{D}^i_{\bar{3}}(q_1q_2)\bar{Q}_{\bar{3}}^j$ combine together to form a  double  heavy physical pentaquark state $\varepsilon_{ijk}\mathcal{D}^i_{\bar{3}}(q_1q_2)\bar{Q}_{\bar{3}}^j\mathcal{D}^k_{\bar{3}}(q_3Q)$, the two heavy quarks $Q$ and $\bar{Q}$ stabilize the five-quark systems or  pentaquark states, just like   the double heavy four-quark systems or  tetraquark states \cite{Wang-tetraquark,Brodsky-2014}. In this article, we construct the interpolating currents according to such routines, see Eqs.(2-13).

In the heavy quark limit, the masses of the hidden-charm pentaquark states can be written as
\begin{eqnarray}
M_{P}&=&2m_c+\overline{\Lambda}+{\mathcal{O}}\left(\frac{1}{m_c} \right)+\cdots \, ,
\end{eqnarray}
where the $\overline{\Lambda}$ is a parameter of the order  ${\mathcal{O}}\left(m_c^0\right)$, and independent on the heavy flavor. The $m_c$ can be taken to be the pole mass according to the  reparameterization invariance. The  fitted value  from the mass of the $D$-mesons in the heavy quark effective theory is about $m_c=1.3\,\rm{GeV}$ \cite{Neubert-PRT}, which is approximate to the $\overline{MS}$  mass $m_{c}(m_c)=(1.275\pm0.025)\,\rm{GeV}$ from the Particle Data Group \cite{PDG}.
The $\overline{MS}$ mass $m_c(m_c)$ relates with the pole mass
$m_c$ through the relation \cite{PDG},
\begin{eqnarray}
m_c&=& m_c(m_c)\left[1+\frac{4 \alpha_s(m_c)}{3\pi}+\cdots\right]\, .
\end{eqnarray}
   Up to  corrections of the  order $\mathcal{O}\left(\alpha_s^3\right)$, the $\overline{MS}$ mass $m_c(m_c)=(1.275\pm0.025) \,\rm{GeV}$ corresponds to the pole mass $m_c=(1.67\pm0.07)\,\rm{GeV}$ \cite{PDG}, which is quite different from the   fitted value   $m_c =1.3 \,\rm{GeV}$ \cite{Neubert-PRT}. The heavy quark mass $m_Q$ is just a parameter.

Now we assume that there exists  an effective heavy quark mass ${\mathbb{M}}_c$ or ${\mathbb{M}}_b$, which is a free parameter, and is larger than the fitted value of the pole mass  $m_c =1.3 \,\rm{GeV}$ or $m_b =4.7\,\rm{GeV}$ \cite{Neubert-PRT} and differs  from the $\overline{MS}$ mass $m_c(m_c)$ or $m_b(m_b)$, some effects  of the order ${\mathcal{O}}\left(m_c^0\right)$ or ${\mathcal{O}}\left(m_b^0\right)$ are embodied in the  ${\mathbb{M}}_c$ or ${\mathbb{M}}_b$.
Then the double  heavy five-quark systems  $q_1q_2q_3Q\bar{Q}$ are characterized by the effective heavy quark masses ${\mathbb{M}}_Q$ and
the virtuality $V=\sqrt{M^2_{P}-(2{\mathbb{M}}_Q)^2}$, just like   the double heavy four-quark systems $q_1\bar{q}_2Q\bar{Q}$ \cite{Wang-tetraquark}.

The  QCD sum rules for the hidden-charm or hidden-bottom pentaquark states  $q_1q_2q_3Q\bar{Q}$ have three typical energy scales $\mu^2$, $T^2$, $V^2$, we can  set the energy  scales to be  $ \mu^2=V^2={\mathcal{O}}(T^2)$, and obtain an energy scale formula,
 \begin{eqnarray}
 \mu&=&\sqrt{M_{P}^2-(2{\mathbb{M}}_Q)^2}\, ,
   \end{eqnarray}
   to determine the ideal energy scales of the QCD spectral densities \cite{Wang1508,WangHuang1508,Wang1509}. Once the pentaquark mass $M_P$ is  specialized,  the effective heavy quark mass ${\mathbb{M}}_Q$ determines the energy scale $\mu$, which determines  the $\overline{MS}$ mass $m_Q(\mu)$ therefore the pentaquark mass $M_P$ based on  the QCD sum rules in return.  In this article, we take the energy scale formula
$\mu=\sqrt{M_{P}^2-(2{\mathbb{M}}_c)^2}$ as a powerful constraint to obey.   In Ref.\cite{Wang1508}, we choose  the  value ${\mathbb{M}}_c=1.8\,\rm{GeV}$ determined in the QCD sum rules for the hidden-charm  tetraquark states \cite{Wang-tetraquark} and obtain the ideal energy scales $\mu=2.5\,\rm{GeV}$ and $\mu=2.6\,\rm{GeV}$ for the  pentaquark states $P_c(4380)$  and $P_c(4450)$, respectively. The empirical energy scale formula $\mu=\sqrt{M_{X/Y/Z/P}^2-(2{\mathbb{M}}_c)^2}$  works well for  the hidden-charm  tetraquark states and hidden-charm  pentaquark states \cite{Wang1508,WangHuang1508,Wang1509,Wang-tetraquark}.

 In the present QCD sum rules, we choose the  Borel parameters $T^2$ and continuum threshold
parameters $s_0$  to obey  the  following four criteria:

$\bf 1.$ Pole dominance at the phenomenological side;

$\bf 2.$ Convergence of the operator product expansion;

$\bf 3.$ Appearance of the Borel platforms;

$\bf 4.$ Satisfying the energy scale formula.

In the QCD sum rules for the three-quark baryon states $qq^{\prime}Q$, $qQQ^\prime$, $QQ^{\prime}Q^{\prime\prime}$  \cite{WangHbaryon,WangLambda}, the predicted masses increase slowly with the increase of the Borel parameters, the Borel platforms do not appear at  the minimum values  and are not very flat, we determine the Borel windows by the criteria $\bf{1}$ and $\bf{2}$, the two necessary constraints  warrant that the extracted masses are reliable.   The pentaquark states are another type baryon states considering  the fractional spins, so their  Borel platforms maybe not  appear at the minimum values of the predicted masses.
 In this article, we choose small Borel windows $T^2_{max}-T^2_{min}=0.4\,\rm{GeV}^2$, just like in the case of the hidden-charm tetraquark states \cite{Wang-tetraquark} and hidden-charm pentaquark states \cite{Wang1508,WangHuang1508,Wang1509},  and obtain the Borel platforms  by requiring the uncertainties $\frac{\delta M_{P}}{M_{P}} $ induced by the Borel parameters are about $1\%$, the criterion $\bf 3$ is modified slightly and satisfied automatically.

We  search for the ideal   Borel parameters $T^2$ and continuum threshold parameters $s_0$ according to  the four criteria $\bf{1}$, $\bf{2}$, $\bf{3}$ and $\bf{4}$ using try and error.  The resulting Borel parameters or Borel windows $T^2$, continuum threshold parameters $s_0$, ideal energy scales of the QCD spectral densities, pole contributions of the ground state pentaquark states, and contributions of the vacuum condensates of dimension 9 and 10 in the operator product expansion are shown   explicitly in Table 2. From the table, we can see that the first two criteria of the QCD sum rules are satisfied, so we expect to make reasonable predictions.

We take into account  all uncertainties  of the input   parameters,
and obtain  the masses and pole residues of
 the $J^P={3\over 2}^{\pm}$   hidden-charm pentaquark states, which are shown explicitly in Table 3. From Table 2 and Table 3, we can see that the  criterion  $\bf 4$ is also satisfied. Now the four criteria of the QCD sum rules are all satisfied, and we expect to make reliable predictions.   The present predictions of the hidden-charm pentaquark masses can be confronted to the experimental data in the future.

In the isospin limit,
\begin{eqnarray}
M_P&=&4.39\pm0.13 \,\, {\rm{GeV}}\,\,\,\, {\rm for} \,\,\,\,P_{uuu}^{10\frac{1}{2}}\left({\frac{3}{2}^-}\right)\, , \,P_{uud}^{10\frac{1}{2}}\left({\frac{3}{2}^-}\right)\, , \, P_{udd}^{10\frac{1}{2}}\left({\frac{3}{2}^-}\right)\, , \,P_{ddd}^{10\frac{1}{2}}\left({\frac{3}{2}^-}\right)\, ,  \nonumber\\
M_P&=&4.39\pm0.14 \,\, {\rm{GeV}}\,\,\,\, {\rm for} \,\,\,\,P_{uuu}^{11\frac{1}{2}}\left({\frac{3}{2}^-}\right)\, , \,P_{uud}^{11\frac{1}{2}}\left({\frac{3}{2}^-}\right)\, , \, P_{udd}^{11\frac{1}{2}}\left({\frac{3}{2}^-}\right)\, , \,P_{ddd}^{11\frac{1}{2}}\left({\frac{3}{2}^-}\right)\, ,  \nonumber\\
M_P&=&4.39\pm0.14 \,\, {\rm{GeV}}\,\,\,\, {\rm for} \,\,\,\,P_{uuu}^{11\frac{3}{2}}\left({\frac{3}{2}^-}\right)\, , \,P_{uud}^{11\frac{3}{2}}\left({\frac{3}{2}^-}\right)\, , \, P_{udd}^{11\frac{3}{2}}\left({\frac{3}{2}^-}\right)\, , \,P_{ddd}^{11\frac{3}{2}}\left({\frac{3}{2}^-}\right)\, ,
\end{eqnarray}
which are all compatible with the experimental value $M_{P_c(4380)}=4380\pm 8\pm 29\,\rm{MeV}$ from the  LHCb collaboration \cite{LHCb-4380}. The $P_c(4380)$ can be assigned to be the pentaquark state $P_{uud}^{10\frac{1}{2}}\left({\frac{3}{2}^-}\right)$, $P_{uud}^{11\frac{1}{2}}\left({\frac{3}{2}^-}\right)$ or $P_{uud}^{11\frac{3}{2}}\left({\frac{3}{2}^-}\right)$, or the $P_c(4380)$ has some pentaquark components $P_{uud}^{10\frac{1}{2}}\left({\frac{3}{2}^-}\right)$, $P_{uud}^{11\frac{1}{2}}\left({\frac{3}{2}^-}\right)$ and  $P_{uud}^{11\frac{3}{2}}\left({\frac{3}{2}^-}\right)$.
From Eq.(35), we can see that the pentaquark states $P_{uud}^{10\frac{1}{2}}\left({\frac{3}{2}^-}\right)$, $P_{uud}^{11\frac{1}{2}}\left({\frac{3}{2}^-}\right)$ and $P_{uud}^{11\frac{3}{2}}\left({\frac{3}{2}^-}\right)$
  have almost  degenerate masses. All the pentaquark states $P_{uud}^{10\frac{1}{2}}\left({\frac{3}{2}^-}\right)$, $P_{uud}^{11\frac{1}{2}}\left({\frac{3}{2}^-}\right)$ and $P_{uud}^{11\frac{3}{2}}\left({\frac{3}{2}^-}\right)$  have the light axialvector diquark state  $\varepsilon^{ijk}u^T_jC\gamma_\mu u_k$ or $\varepsilon^{ijk}u^T_jC\gamma_\mu d_k$, moreover, the heavy-light  scalar and axialvector  diquark states have almost  degenerate masses, so the
 axialvector-diquark-scalar-diquark-antiquark type hidden-charm  pentaquark states   and the axialvector-diquark-axialvector-diquark-antiquark type hidden-charm  pentaquark states maybe have degenerate masses.   
   The mass alone cannot identify the $P_c(4380)$ unambiguously, more experimental data on its productions and decays are still needed.    On the theoretical side, we can study the two-body strong decays  $P_c(4380) \to J/\psi p$ with the three-point QCD sum rules or the light-cone QCD sum rules, which maybe give additional support for such assignment.

\begin{table}
\begin{center}
\begin{tabular}{|c|c|c|c|c|c|c|c|}\hline\hline
                                  &$T^2 \rm{GeV}^2)$     &$\sqrt{s_0}(\rm{GeV})$       &$\mu(\rm{GeV})$  &pole          &$D_9$         &$D_{10}$ \\ \hline

$P_{uuu}^{10\frac{1}{2}}\left({\frac{3}{2}^-}\right)$   &$3.3-3.7$     &$5.1\pm0.1$   &2.5   &$(41-62)\%$   &$(11-15)\%$   &$(3-4)\%$   \\ \hline
$P_{uus}^{10\frac{1}{2}}\left({\frac{3}{2}^-}\right)$   &$3.5-3.9$     &$5.2\pm0.1$   &2.7   &$(40-60)\%$   &$(7-9)\%$     &$(2-3)\%$   \\ \hline
$P_{uss}^{10\frac{1}{2}}\left({\frac{3}{2}^-}\right)$   &$3.6-4.0$     &$5.3\pm0.1$   &2.8   &$(41-61)\%$   &$(5-6)\%$     &$(1-2)\%$   \\ \hline
$P_{sss}^{10\frac{1}{2}}\left({\frac{3}{2}^-}\right)$   &$3.8-4.2$     &$5.4\pm0.1$   &3.0   &$(40-59)\%$   &$(3-4)\%$     &$\sim1\%$   \\ \hline

$P_{uuu}^{11\frac{1}{2}}\left({\frac{3}{2}^-}\right)$   & $3.2-3.6$    &$5.1\pm0.1$   &2.5   &$(42-64)\%$   &$(12-17)\%$   &$(3-5)\%$   \\ \hline
$P_{uus}^{11\frac{1}{2}}\left({\frac{3}{2}^-}\right)$   & $3.4-3.8$    &$5.2\pm0.1$   &2.7   &$(41-61)\%$   &$(7-10)\%$    &$(2-3)\%$   \\ \hline
$P_{uss}^{11\frac{1}{2}}\left({\frac{3}{2}^-}\right)$   & $3.6-4.0$    &$5.3\pm0.1$   &2.9   &$(40-59)\%$   &$(4-6)\%$     &$(1-2)\%$   \\ \hline
$P_{sss}^{11\frac{1}{2}}\left({\frac{3}{2}^-}\right)$   & $3.7-4.1$    &$5.4\pm0.1$   &3.0   &$(41-60)\%$   &$(3-4)\%$     &$(1-2)\%$   \\ \hline

$P_{uuu}^{11\frac{3}{2}}\left({\frac{3}{2}^-}\right)$   & $3.2-3.6$    &$5.1\pm0.1$   &2.5   &$(41-63)\%$   &$(12-17)\%$   &$(2-3)\%$   \\ \hline
$P_{uus}^{11\frac{3}{2}}\left({\frac{3}{2}^-}\right)$   & $3.4-3.8$    &$5.2\pm0.1$   &2.7   &$(40-60)\%$   &$(7-10)\%$    &$(1-2)\%$   \\ \hline
$P_{uss}^{11\frac{3}{2}}\left({\frac{3}{2}^-}\right)$   & $3.5-3.9$    &$5.3\pm0.1$   &2.9   &$(41-61)\%$   &$(5-7)\%$     &$\sim 1\%$   \\ \hline
$P_{sss}^{11\frac{3}{2}}\left({\frac{3}{2}^-}\right)$   & $3.7-4.1$    &$5.4\pm0.1$   &3.0   &$(40-59)\%$   &$(3-4)\%$     &$\sim 1\%$   \\ \hline

$P_{uuu}^{10\frac{1}{2}}\left({\frac{3}{2}^+}\right)$   & $3.2-3.6$    &$5.2\pm0.1$   &2.7   &$(41-62)\%$   &$(13-18)\%$   &$(3-5)\%$   \\ \hline
$P_{uus}^{10\frac{1}{2}}\left({\frac{3}{2}^+}\right)$   & $3.3-3.7$    &$5.3\pm0.1$   &2.9   &$(42-63)\%$   &$(9-12)\%$    &$(2-4)\%$   \\ \hline
$P_{uss}^{10\frac{1}{2}}\left({\frac{3}{2}^+}\right)$   & $3.5-3.9$    &$5.4\pm0.1$   &3.0   &$(40-60)\%$   &$(6-8)\%$     &$\sim2\%$   \\ \hline
$P_{sss}^{10\frac{1}{2}}\left({\frac{3}{2}^+}\right)$   & $3.6-4.0$    &$5.5\pm0.1$   &3.2   &$(41-61)\%$   &$(4-5)\%$     &$(1-2)\%$   \\ \hline

$P_{uuu}^{11\frac{1}{2}}\left({\frac{3}{2}^+}\right)$   & $3.4-3.8$    &$5.6\pm0.1$   &3.3   &$(50-69)\%$   &$-(4-6)\%$    &$(3-4)\%$   \\ \hline
$P_{uus}^{11\frac{1}{2}}\left({\frac{3}{2}^+}\right)$   & $3.6-4.0$    &$5.7\pm0.1$   &3.5   &$(48-68)\%$   &$-(2-4)\%$    &$(2-3)\%$   \\ \hline
$P_{uss}^{11\frac{1}{2}}\left({\frac{3}{2}^+}\right)$   & $3.8-4.2$    &$5.8\pm0.1$   &3.6   &$(47-66)\%$   &$-(1-2)\%$    &$(1-2)\%$   \\ \hline
$P_{sss}^{11\frac{1}{2}}\left({\frac{3}{2}^+}\right)$   & $4.0-4.4$    &$5.9\pm0.1$   &3.7   &$(46-64)\%$   &$-(1-2)\%$    &$\sim 1\%$   \\ \hline

$P_{uuu}^{11\frac{3}{2}}\left({\frac{3}{2}^+}\right)$   & $3.3-3.7$    &$5.8\pm0.1$   &3.7   &$(60-78)\%$   &$-(3-6)\%$    &$\sim -1\%$   \\ \hline
$P_{uus}^{11\frac{3}{2}}\left({\frac{3}{2}^+}\right)$   & $3.5-3.9$    &$5.9\pm0.1$   &3.8   &$(59-77)\%$   &$-(2-3)\%$    &$\sim -1\%$   \\ \hline
$P_{uss}^{11\frac{3}{2}}\left({\frac{3}{2}^+}\right)$   & $3.7-4.1$    &$6.0\pm0.1$   &3.9   &$(58-75)\%$   &$-(1-2)\%$    &$\sim -0\%$   \\ \hline
$P_{sss}^{11\frac{3}{2}}\left({\frac{3}{2}^+}\right)$   & $3.9-4.3$    &$6.1\pm0.1$   &4.0   &$(57-74)\%$   &$-1\%$        &$\sim -0\%$   \\ \hline
 \hline
\end{tabular}
\end{center}
\caption{ The Borel parameters (Borel windows), continuum threshold parameters, ideal energy scales, pole contributions,   contributions of the vacuum condensates of dimension 9 and dimension 10 for the hidden-charm pentraquark states. }
\end{table}

\begin{table}
\begin{center}
\begin{tabular}{|c|c|c|c|c|c|c|c|}\hline\hline
                                                          &$M_P(\rm{GeV})$  &$\lambda_P(10^{-3}\rm{GeV}^6)$   \\ \hline

$P_{uuu}^{10\frac{1}{2}}\left({\frac{3}{2}^-}\right)$     &$4.39\pm0.13$    &$2.25\pm0.40$ \\ \hline
$P_{uus}^{10\frac{1}{2}}\left({\frac{3}{2}^-}\right)$     &$4.51\pm0.12$    &$2.75\pm0.45$ \\ \hline
$P_{uss}^{10\frac{1}{2}}\left({\frac{3}{2}^-}\right)$     &$4.60\pm0.11$    &$3.19\pm0.50$ \\ \hline
$P_{sss}^{10\frac{1}{2}}\left({\frac{3}{2}^-}\right)$     &$4.70\pm0.11$    &$3.78\pm0.57$ \\ \hline

$P_{uuu}^{11\frac{1}{2}}\left({\frac{3}{2}^-}\right)$     &$4.39\pm0.14$    &$3.75\pm0.68$ \\ \hline
$P_{uus}^{11\frac{1}{2}}\left({\frac{3}{2}^-}\right)$     &$4.51\pm0.12$    &$4.64\pm0.77$ \\ \hline
$P_{uss}^{11\frac{1}{2}}\left({\frac{3}{2}^-}\right)$     &$4.62\pm0.11$    &$5.60\pm0.88$ \\ \hline
$P_{sss}^{11\frac{1}{2}}\left({\frac{3}{2}^-}\right)$     &$4.71\pm0.11$    &$6.47\pm1.00$ \\ \hline

$P_{uuu}^{11\frac{3}{2}}\left({\frac{3}{2}^-}\right)$     &$4.39\pm0.14$    &$3.74\pm0.70$ \\ \hline
$P_{uus}^{11\frac{3}{2}}\left({\frac{3}{2}^-}\right)$     &$4.52\pm0.12$    &$4.64\pm0.79$ \\ \hline
$P_{uss}^{11\frac{3}{2}}\left({\frac{3}{2}^-}\right)$     &$4.61\pm0.12$    &$5.52\pm0.90$   \\ \hline
$P_{sss}^{11\frac{3}{2}}\left({\frac{3}{2}^-}\right)$     &$4.72\pm0.11$    &$6.52\pm1.02$  \\ \hline

$P_{uuu}^{10\frac{1}{2}}\left({\frac{3}{2}^+}\right)$     &$4.51\pm0.13$    &$0.99\pm0.19$  \\ \hline
$P_{uus}^{10\frac{1}{2}}\left({\frac{3}{2}^+}\right)$     &$4.60\pm0.12$    &$1.21\pm0.22$ \\ \hline
$P_{uss}^{10\frac{1}{2}}\left({\frac{3}{2}^+}\right)$     &$4.72\pm0.11$    &$1.45\pm0.25$ \\ \hline
$P_{sss}^{10\frac{1}{2}}\left({\frac{3}{2}^+}\right)$     &$4.81\pm0.12$    &$1.71\pm0.29$ \\ \hline

$P_{uuu}^{11\frac{1}{2}}\left({\frac{3}{2}^+}\right)$     &$4.91\pm0.09$    &$5.04\pm0.68$ \\ \hline
$P_{uus}^{11\frac{1}{2}}\left({\frac{3}{2}^+}\right)$     &$4.99\pm0.09$    &$5.83\pm0.82$ \\ \hline
$P_{uss}^{11\frac{1}{2}}\left({\frac{3}{2}^+}\right)$     &$5.08\pm0.09$    &$6.72\pm0.98$  \\ \hline
$P_{sss}^{11\frac{1}{2}}\left({\frac{3}{2}^+}\right)$     &$5.18\pm0.09$    &$7.73\pm1.10$ \\ \hline

$P_{uuu}^{11\frac{3}{2}}\left({\frac{3}{2}^+}\right)$     &$5.14\pm0.08$    &$7.86\pm0.91$   \\ \hline
$P_{uus}^{11\frac{3}{2}}\left({\frac{3}{2}^+}\right)$     &$5.21\pm0.08$    &$8.79\pm1.07$   \\ \hline
$P_{uss}^{11\frac{3}{2}}\left({\frac{3}{2}^+}\right)$     &$5.28\pm0.08$    &$9.91\pm1.25$   \\ \hline
$P_{sss}^{11\frac{3}{2}}\left({\frac{3}{2}^+}\right)$     &$5.36\pm0.08$    &$11.20\pm1.42$   \\ \hline
 \hline
\end{tabular}
\end{center}
\caption{ The masses and pole residues of the hidden-charm pentaquark states. }
\end{table}

The $\Lambda_b^0$ can be well interpolated by the current $J(x)=\varepsilon^{ijk}u^T_i(x) C\gamma_5 d_j(x) b_k(x)$ \cite{WangLambda}, the $u$ and $d$ quarks in the $\Lambda_b^0$ form a scalar diquark $[ud]$ (or $\varepsilon^{ijk}u^T_i C\gamma_5 d_j$) in color antitriplet, the decays $\Lambda_b^0\to J/\psi  p K^-$ take  place through the following mechanism,
 \begin{eqnarray}
 \Lambda_b^0([ud] b )&\to& [ud] c\bar{c}s\to [ud] c\bar{c}u\bar{u}s\to P_c^+([ud] [uc]\bar{c})K^-(\bar{u}s)\to J/\psi  p K^- \, ,
  \end{eqnarray}
  at the quark level.  In the decays $P_c^+([ud] [uc]\bar{c})\to J/\psi  p$, the scalar diquark $[ud]$ survives in the decay chains, the decays are greatly facilitated and can take place easily. On the other hand, if there exists a light axialvector diquark $\{ud\}$ (or $\varepsilon^{ijk}u^T_i C\gamma_\mu d_j$), which has to dissolve to form a scalar diquark $[ud]$, the    decays are not facilitated and cannot take place easily, but the decays are not forbidden. It is also possible to assign the $P_c(4380)$ to be the axialvector-diquark-scalar-diquark-antiquark type pentaquark state $P_{uud}^{10\frac{1}{2}}\left({\frac{3}{2}^-}\right)$ or the axialvector-diquark-axialvector-diquark-antiquark type pentaquark states $P_{uud}^{11\frac{1}{2}}\left({\frac{3}{2}^-}\right)$ and $P_{uud}^{11\frac{3}{2}}\left({\frac{3}{2}^-}\right)$.

 In Fig.1, we plot the   masses  of the pentaquark states $P_{uud}^{10\frac{1}{2}}\left({\frac{3}{2}^-}\right)$, $P_{uud}^{10\frac{1}{2}}\left({\frac{3}{2}^+}\right)$,
$P_{uud}^{11\frac{1}{2}}\left({\frac{3}{2}^-}\right)$, $P_{uud}^{11\frac{1}{2}}\left({\frac{3}{2}^+}\right)$, $P_{uud}^{11\frac{3}{2}}\left({\frac{3}{2}^-}\right)$ and $P_{uud}^{11\frac{3}{2}}\left({\frac{3}{2}^+}\right)$ with variations of the Borel parameters $T^2$ as an example. In this article, we take the isospin limit, the masses of the pentaquark states $P_{uud}^{j_Lj_Hj}\left({\frac{3}{2}^-}\right)$ and $P_{uuu}^{j_Lj_Hj}\left({\frac{3}{2}^-}\right)$ degenerate.  From the figure, we can see that the platforms in the Borel windows shown in Table 2 appear at   the minimum values  for the  axialvector-diquark-axialvector-diquark-antiquark type pentaquark states with positive parity, and the Borel platforms are very flat; other  platforms in the Borel windows shown Table 2 are not as flat, the uncertainties $\frac{\delta M_{P}}{M_{P}} $ induced by the Borel parameters in the Borel windows are about $1\%$. In Fig.1, we plot the masses with variations of the Borel parameters at a large interval  $T^2=(2.6-4.6)\,\rm{GeV}^2$,  which is much larger than the interval of the Borel windows shown in Table 2. Compared to the values  shown in Table 3, the masses shown in Fig.1 have larger uncertainties due to the larger interval of the Borel parameters.
Furthermore, from the values in Table 3, we can see that  the mass gaps between  the pentaquark states with opposite  parity $P_{q_1q_2q_3}^{11\frac{1}{2}}\left({\frac{3}{2}^+}\right)-P_{q_1q_2q_3}^{11\frac{1}{2}}\left({\frac{3}{2}^-}\right)$ and
$P_{q_1q_2q_3}^{11\frac{3}{2}}\left({\frac{3}{2}^+}\right)-P_{q_1q_2q_3}^{11\frac{3}{2}}\left({\frac{3}{2}^-}\right)$
are   about $0.5\,\rm{GeV}$ and $0.7\,\rm{GeV}$  respectively,
which is much larger than the mass gap $0.1\,\rm{GeV}$  between  the pentaquark states with opposite  parity $P_{q_1q_2q_3}^{10\frac{1}{2}}\left({\frac{3}{2}^+}\right)-P_{q_1q_2q_3}^{10\frac{1}{2}}\left({\frac{3}{2}^-}\right)$.
An additional P-wave costs  about $0.5\,\rm{GeV}$ for the conventional baryon states, the mass spectra of the pentaquark states have new feature.

\begin{figure}
 \centering
 \includegraphics[totalheight=5cm,width=7cm]{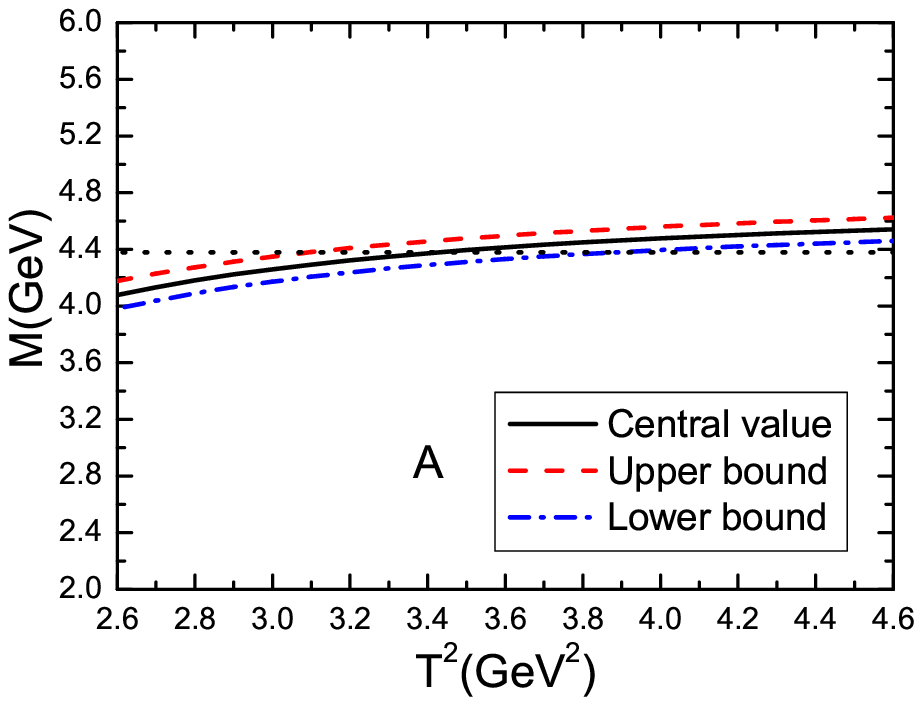}
 \includegraphics[totalheight=5cm,width=7cm]{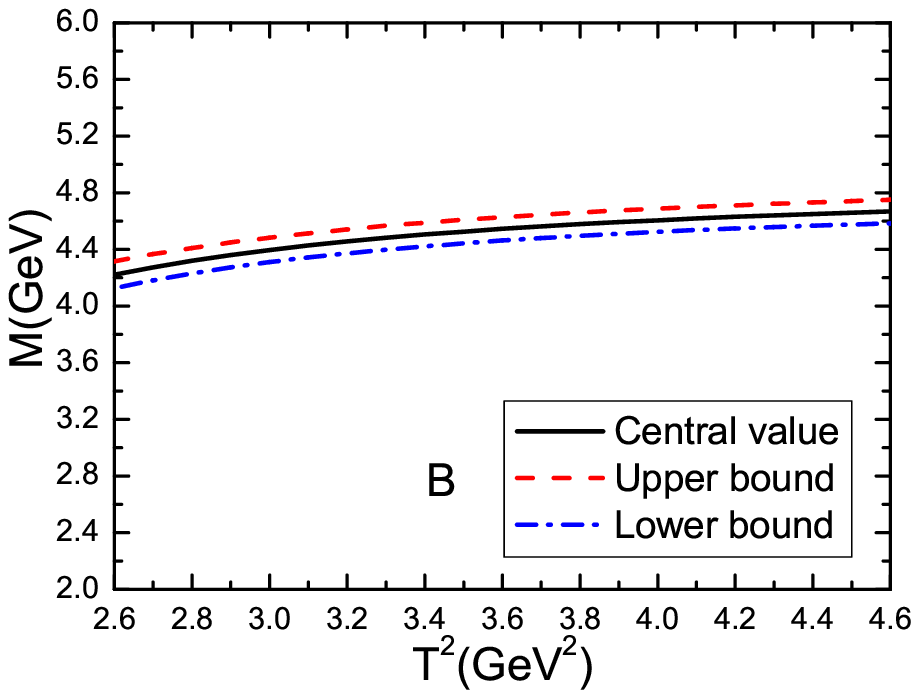}
 \includegraphics[totalheight=5cm,width=7cm]{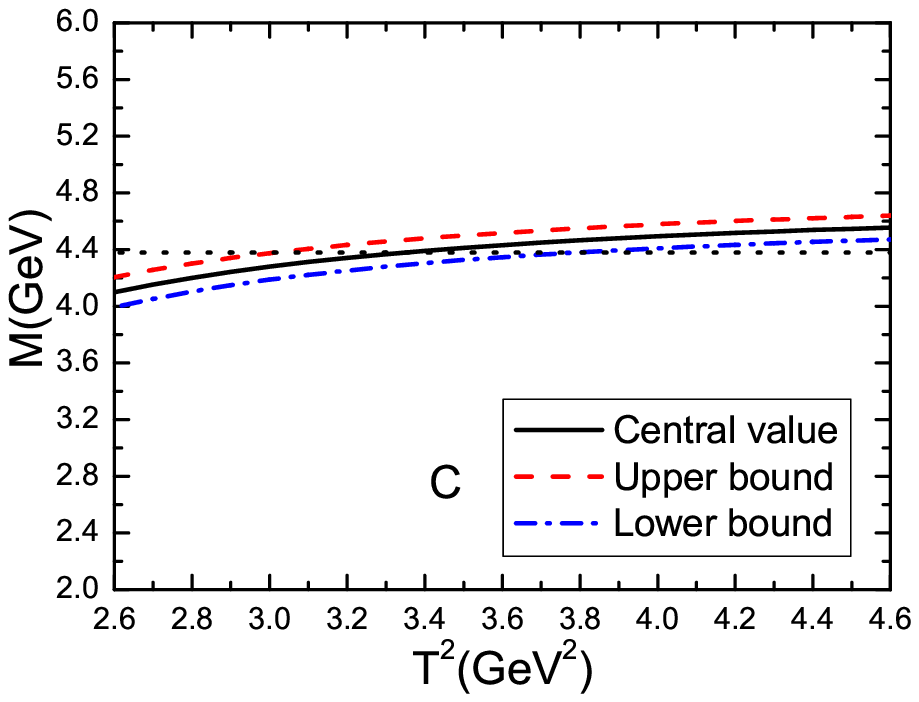}
 \includegraphics[totalheight=5cm,width=7cm]{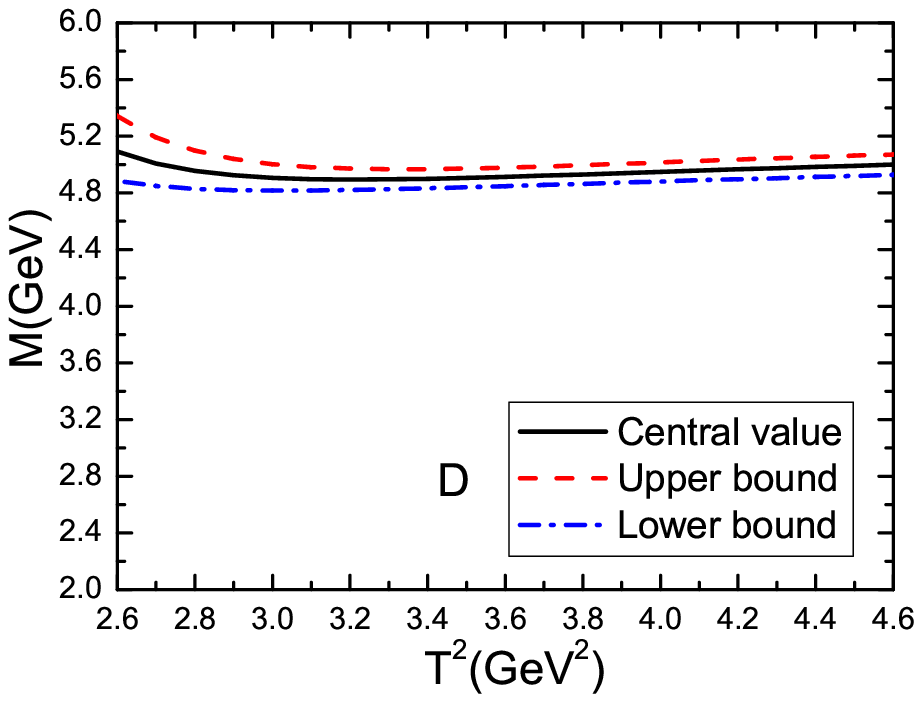}
 \includegraphics[totalheight=5cm,width=7cm]{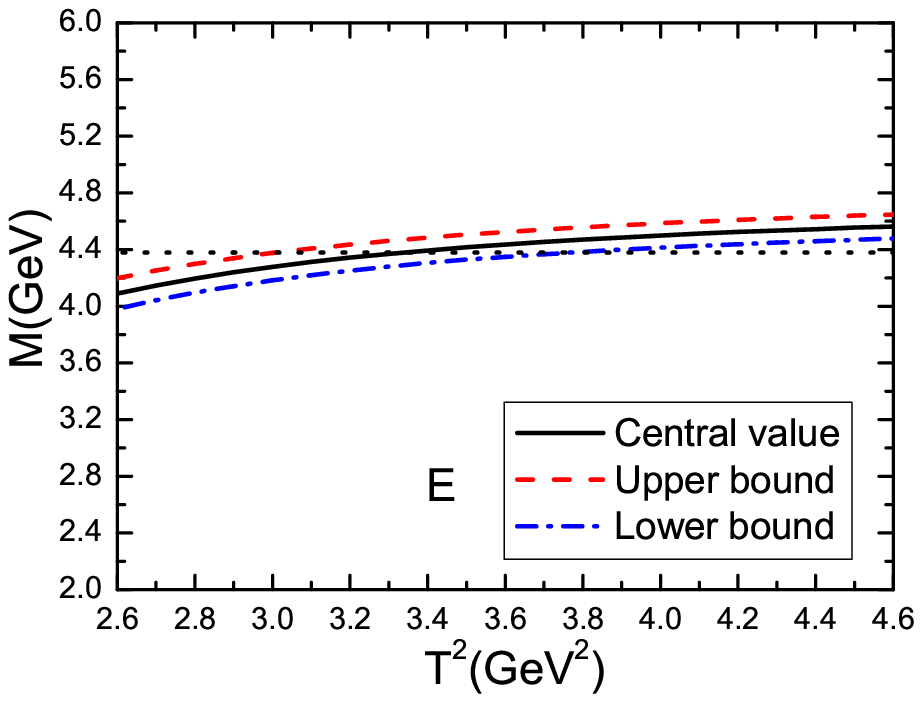}
 \includegraphics[totalheight=5cm,width=7cm]{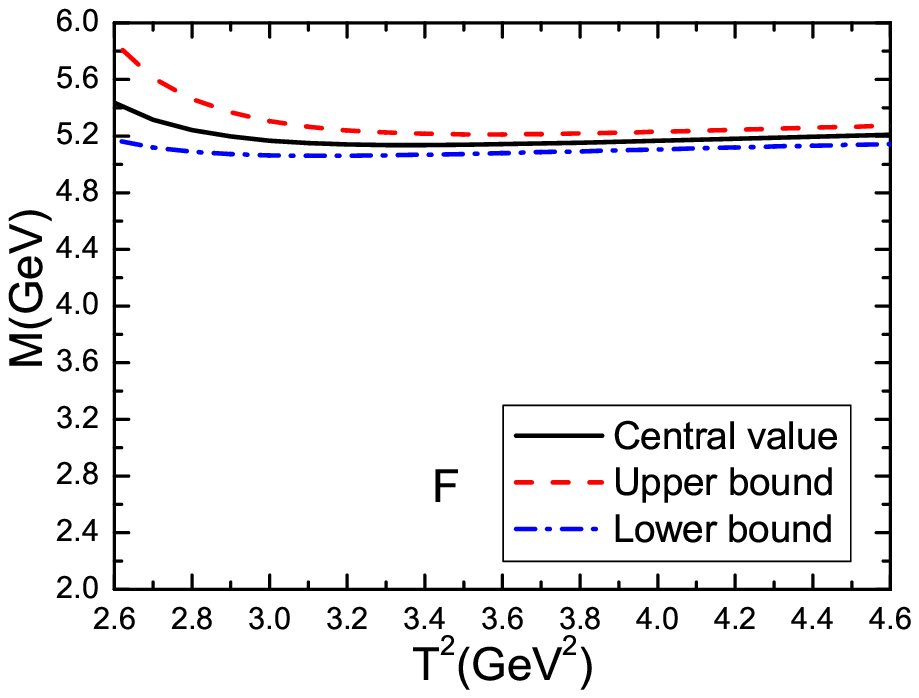}
         \caption{ The masses  of the pentaquark states  with variations of the Borel parameters $T^2$, where the $A$, $B$, $C$, $D$, $E$ and $F$  denote the pentaquark states $P_{uud}^{10\frac{1}{2}}\left({\frac{3}{2}^-}\right)$, $P_{uud}^{10\frac{1}{2}}\left({\frac{3}{2}^+}\right)$,
$P_{uud}^{11\frac{1}{2}}\left({\frac{3}{2}^-}\right)$, $P_{uud}^{11\frac{1}{2}}\left({\frac{3}{2}^+}\right)$, $P_{uud}^{11\frac{3}{2}}\left({\frac{3}{2}^-}\right)$ and $P_{uud}^{11\frac{3}{2}}\left({\frac{3}{2}^+}\right)$ respectively, the horizontal lines denote the experimental value of the mass of the $P_c(4380)$.  }
\end{figure}

\section{Conclusion}
In this article, we   construct   the axialvector-diquark-scalar-diquark-antiquark type and the axialvector-diquark-axialvector-diquark-antiquark type     currents to interpolate the $J^P={\frac{3}{2}}^\pm$   hidden-charm pentaquark states,   study the masses and pole residues  with the QCD sum rules  systematically  by calculating the contributions of the vacuum condensates up to dimension 10 in the operator product expansion. We obtain the masses of the hidden-charm pentaquark states with the strangeness  $S=0,\,-1,\,-2,\,-3$, respectively, which can be confronted to the experimental data in the future. As a byproduct, the predicted pole residues can be taken as basic input parameters in studying the strong decays of the hidden-charm pentaquark states with the three-point QCD sum rules or the light-cone QCD sum rules.  In calculations, we use the  empirical energy scale formula $\mu=\sqrt{M^2_{X/Y/Z/P}-(2{\mathbb{M}}_c)^2}$ with the effective heavy quark mass ${\mathbb{M}}_c$ to determine  the ideal energy scales of the QCD spectral densities, which works well in studying  the hidden-charm tetraquark states and hidden-charm pentaquark states.
The predicted masses of the pentaquark states $P_{uud}^{10\frac{1}{2}}\left({\frac{3}{2}^-}\right)$, $P_{uud}^{11\frac{1}{2}}\left({\frac{3}{2}^-}\right)$ and $P_{uud}^{11\frac{3}{2}}\left({\frac{3}{2}^-}\right)$ are compatible with the experimental value   $M_{P_c(4380)}=4380\pm 8\pm 29\,\rm{MeV}$
from the LHCb collaboration. We can draw the conclusion tentatively that the $P_c(4380)$ can be assigned to be the pentaquark state $P_{uud}^{10\frac{1}{2}}\left({\frac{3}{2}^-}\right)$, $P_{uud}^{11\frac{1}{2}}\left({\frac{3}{2}^-}\right)$ or $P_{uud}^{11\frac{3}{2}}\left({\frac{3}{2}^-}\right)$, or the $P_c(4380)$ has some pentaquark components $P_{uud}^{10\frac{1}{2}}\left({\frac{3}{2}^-}\right)$, $P_{uud}^{11\frac{1}{2}}\left({\frac{3}{2}^-}\right)$ and  $P_{uud}^{11\frac{3}{2}}\left({\frac{3}{2}^-}\right)$.  More experimental data on its productions and decays are still needed to identify the $P_c(4380)$ unambiguously.

\section*{Appendix}
The QCD spectral densities
$\rho^{11\frac{3}{2},1}_{sss}(s)$, $\widetilde{\rho}^{11\frac{3}{2},0}_{sss}(s)$,
$\rho^{11\frac{3}{2},1}_{uss}(s)$, $\widetilde{\rho}^{11\frac{3}{2},0}_{uss}(s)$,
$\rho^{11\frac{3}{2},1}_{uus}(s)$, $\widetilde{\rho}^{11\frac{3}{2},0}_{uus}(s)$,
$\rho^{11\frac{3}{2},1}_{uuu}(s)$, $\widetilde{\rho}^{11\frac{3}{2},0}_{uuu}(s)$,
$\rho^{11\frac{1}{2},1}_{sss}(s)$, $\widetilde{\rho}^{11\frac{1}{2},0}_{sss}(s)$,
$\rho^{11\frac{1}{2},1}_{uss}(s)$, $\widetilde{\rho}^{11\frac{1}{2},0}_{uss}(s)$,
$\rho^{11\frac{1}{2},1}_{uus}(s)$, $\widetilde{\rho}^{11\frac{1}{2},0}_{uus}(s)$,
$\rho^{11\frac{1}{2},1}_{uuu}(s)$, $\widetilde{\rho}^{11\frac{1}{2},0}_{uuu}(s)$,
$\rho^{10\frac{1}{2},1}_{sss}(s)$, $\widetilde{\rho}^{10\frac{1}{2},0}_{sss}(s)$,
$\rho^{10\frac{1}{2},1}_{uss}(s)$, $\widetilde{\rho}^{10\frac{1}{2},0}_{uss}(s)$,
$\rho^{10\frac{1}{2},1}_{uus}(s)$, $\widetilde{\rho}^{10\frac{1}{2},0}_{uus}(s)$,
$\rho^{10\frac{1}{2},1}_{uuu}(s)$ and $\widetilde{\rho}^{10\frac{1}{2},0}_{uuu}(s)$     of the hidden-charm pentaquark states,

\begin{eqnarray}
\rho^{11\frac{3}{2},1}_{sss}(s)&=&\frac{1}{122880\pi^8}\int dydz \, yz(1-y-z)^4\left(s-\overline{m}_c^2\right)^4\left(7s-2\overline{m}_c^2 \right)  \nonumber\\
&&+\frac{1}{614400\pi^8}\int dydz \, yz(1-y-z)^5\left(s-\overline{m}_c^2\right)^3\left(63s^2-51s\overline{m}_c^2+ 8\overline{m}_c^4\right)  \nonumber\\
&&- \frac{m_c\langle \bar{s}s\rangle}{768\pi^6}\int dydz \, (y+z)(1-y-z)^2\left(s-\overline{m}_c^2\right)^3  \nonumber \\
&&- \frac{m_c\langle \bar{s}s\rangle}{768\pi^6}\int dydz \, (y+z)(1-y-z)^3\left(s-\overline{m}_c^2\right)^2\left(2s-\overline{m}_c^2\right)  \nonumber \\
&&+ \frac{11m_c\langle \bar{s}g_s\sigma Gs\rangle}{4096\pi^6}\int dydz  \, (y+z)(1-y-z) \left(s-\overline{m}_c^2 \right)^2 \nonumber\\
&&+ \frac{11m_c\langle \bar{s}g_s\sigma Gs\rangle}{8192\pi^6}\int dydz  \, (y+z)(1-y-z)^2 \left(s-\overline{m}_c^2 \right)\left(5s-3\overline{m}_c^2 \right) \nonumber\\
&&+\frac{\langle\bar{s}s\rangle^2}{24\pi^4}\int dydz \,  yz(1-y-z)\left(s-\overline{m}_c^2 \right)\left(2s-\overline{m}_c^2 \right) \nonumber\\
&&-\frac{19\langle\bar{s}s\rangle\langle\bar{s}g_s\sigma Gs\rangle}{768\pi^4}\int dydz \,yz \left(3s-2\overline{m}_c^2\right)\nonumber\\
&&+\frac{\langle\bar{s}s\rangle\langle\bar{s}g_s\sigma Gs\rangle}{1536\pi^4}\int dydz \,(y+z)(1-y-z) \left(5s-4\overline{m}_c^2\right)\nonumber\\
&&-\frac{m_c\langle\bar{s}s\rangle^3}{36\pi^2}\int dy \nonumber\\
&&-\frac{13\langle\bar{s}g_s\sigma Gs\rangle^2}{13824\pi^4}\int dydz \, (y+z)\left[ 1+\frac{s}{4}\delta\left(s-\overline{m}_c^2 \right)\right] \nonumber\\
&&+\frac{11\langle\bar{s}g_s\sigma Gs\rangle^2}{1536\pi^4}\int dy \, y(1-y)\left[ 1+\frac{s}{2}\delta\left(s-\widetilde{m}_c^2 \right)\right] \nonumber\\
&&+\frac{m_s m_c}{12288\pi^8}\int dydz \, (y+z)(1-y-z)^3\left(s-\overline{m}_c^2\right)^4  \nonumber
\end{eqnarray}
\begin{eqnarray}
&&+\frac{m_s m_c}{49152\pi^8}\int dydz \, (y+z)(1-y-z)^4\left(s-\overline{m}_c^2\right)^3\left(7s-3\overline{m}_c^2\right)  \nonumber\\
&&- \frac{m_s\langle \bar{s}s\rangle}{768\pi^6}\int dydz \, yz(1-y-z)^2\left(s-\overline{m}_c^2\right)^2 \left(5s-2\overline{m}_c^2 \right) \nonumber \\
&&+ \frac{m_s\langle \bar{s}s\rangle}{768\pi^6}\int dydz \, yz(1-y-z)^3\left(s-\overline{m}_c^2\right) \left(35s^2-37s\overline{m}_c^2+8\overline{m}_c^4 \right) \nonumber \\
&&+ \frac{11m_s\langle \bar{s}g_s\sigma Gs\rangle}{1024\pi^6}\int dydz  \, yz(1-y-z) \left(s-\overline{m}_c^2 \right)\left(2s-\overline{m}_c^2 \right) \nonumber\\
&&- \frac{m_s\langle \bar{s}g_s\sigma Gs\rangle}{256\pi^6}\int dydz  \, yz(1-y-z)^2 \left(12s^2-15s\overline{m}_c^2 +4\overline{m}_c^4\right) \nonumber\\
&&- \frac{m_s\langle \bar{s}g_s\sigma Gs\rangle}{4096\pi^6}\int dydz  \, (y+z)(1-y-z)^2 \left(s-\overline{m}_c^2 \right)\left(3s-2\overline{m}_c^2 \right) \nonumber\\
&&+\frac{ m_s m_c\langle\bar{s}s\rangle^2}{24\pi^4}\int dydz \,  (y+z)\left(s-\overline{m}_c^2 \right) \nonumber\\
&&-\frac{ m_s m_c\langle\bar{s}s\rangle^2}{48\pi^4}\int dydz \,  (y+z)(1-y-z)\left(4s-3\overline{m}_c^2 \right) \nonumber\\
&&+ \frac{89m_s m_c \langle  \bar{s}s\rangle\langle \bar{s}g_s\sigma Gs\rangle}{3072\pi^4}\int dydz  \, (y+z)\left[1+\frac{s}{3}\delta \left(s-\overline{m}_c^2 \right)  \right] \nonumber\\
&&+\frac{m_s m_c\langle\bar{s}s\rangle\langle\bar{s}g_s\sigma Gs\rangle}{768\pi^4}\int dydz  \left(\frac{y}{z}+\frac{z}{y}\right)\nonumber\\
&&-\frac{199m_s m_c \langle\bar{s}s\rangle\langle\bar{s}g_s\sigma Gs\rangle}{9216\pi^4}\int dy   \nonumber\\
&&   +\frac{m_s\langle\bar{s}s\rangle^3}{18\pi^2}\int dy \,y(1-y) \left[ 1+\frac{s}{2}\delta\left(s-\widetilde{m}_c^2 \right)\right]  \nonumber\\
&&-\frac{m_s m_c\langle\bar{s}g_s\sigma Gs\rangle^2}{3072\pi^4}\int dy   \left(\frac{1-y}{y}+\frac{y}{1-y}\right)\delta \left(s-\widetilde{m}_c^2 \right)\nonumber\\
&&+\frac{17m_s m_c\langle\bar{s}g_s\sigma Gs\rangle^2}{4608\pi^4}\int dy \left(1+\frac{s}{2T^2}\right)\delta\left(s-\widetilde{m}_c^2 \right) \, ,
\end{eqnarray}

\begin{eqnarray}
\widetilde{\rho}^{11\frac{3}{2},0}_{sss}(s)&=&\frac{1}{122880\pi^8}\int dydz \, (y+z)(1-y-z)^4\left(s-\overline{m}_c^2\right)^4\left(6s-\overline{m}_c^2 \right)  \nonumber\\
&&-\frac{1}{1228800\pi^8}\int dydz \, (y+z)(1-y-z)^5\left(s-\overline{m}_c^2\right)^3\left(48s^2-31s\overline{m}_c^2+ 3\overline{m}_c^4\right)  \nonumber\\
&&- \frac{m_c\langle \bar{s}s\rangle}{192\pi^6}\int dydz \, (1-y-z)^2\left(s-\overline{m}_c^2\right)^3  \nonumber \\
&&+ \frac{m_c\langle \bar{s}s\rangle}{1152\pi^6}\int dydz \,  (1-y-z)^3\left(s-\overline{m}_c^2\right)^2\left(5s-2\overline{m}_c^2\right)  \nonumber \\
&&+ \frac{11m_c\langle \bar{s}g_s\sigma Gs\rangle}{1024\pi^6}\int dydz  \, (1-y-z) \left(s-\overline{m}_c^2 \right)^2 \nonumber\\
&&- \frac{11m_c\langle \bar{s}g_s\sigma Gs\rangle}{2048\pi^6}\int dydz  \,  (1-y-z)^2 \left(s-\overline{m}_c^2 \right)\left(2s-\overline{m}_c^2 \right) \nonumber\\
&&+\frac{\langle\bar{s}s\rangle^2}{48\pi^4}\int dydz \,  (y+z)(1-y-z)\left(s-\overline{m}_c^2 \right)\left(3s-\overline{m}_c^2 \right) \nonumber
\end{eqnarray}
\begin{eqnarray}
&&-\frac{19\langle\bar{s}s\rangle\langle\bar{s}g_s\sigma Gs\rangle}{768\pi^4}\int dydz \,(y+z) \left(2s-\overline{m}_c^2\right)\nonumber\\
&&-\frac{m_c\langle\bar{s}s\rangle^3}{9\pi^2}\int dy \nonumber\\
&&+\frac{11\langle\bar{s}g_s\sigma Gs\rangle^2}{3072\pi^4}\int dy  \left[ 1+s\,\delta\left(s-\widetilde{m}_c^2 \right)\right] \nonumber\\
&&+\frac{m_s m_c}{3072\pi^8}\int dydz \, (1-y-z)^3\left(s-\overline{m}_c^2\right)^4  \nonumber\\
&&-\frac{m_s m_c}{12288\pi^8}\int dydz \, (1-y-z)^4\left(s-\overline{m}_c^2\right)^3\left(3s-\overline{m}_c^2\right)  \nonumber\\
&&- \frac{m_s\langle \bar{s}s\rangle}{768\pi^6}\int dydz \, (y+z)(1-y-z)^2\left(s-\overline{m}_c^2\right)^2 \left(4s-\overline{m}_c^2 \right) \nonumber \\
&&- \frac{m_s\langle \bar{s}s\rangle}{512\pi^6}\int dydz \, (y+z)(1-y-z)^3\left(s-\overline{m}_c^2\right) \left(8s^2-7s\overline{m}_c^2+\overline{m}_c^4 \right)
\nonumber \\
&&+ \frac{11m_s\langle \bar{s}g_s\sigma Gs\rangle}{2048\pi^6}\int dydz  \, (y+z)(1-y-z) \left(s-\overline{m}_c^2 \right)\left(3s-\overline{m}_c^2 \right) \nonumber\\
&&+ \frac{m_s\langle \bar{s}g_s\sigma Gs\rangle}{1024\pi^6}\int dydz  \, (y+z)(1-y-z)^2 \left(15s^2-16s\overline{m}_c^2 +3\overline{m}_c^4\right) \nonumber\\
&&+\frac{ m_s m_c\langle\bar{s}s\rangle^2}{6\pi^4}\int dydz \,  \left(s-\overline{m}_c^2 \right) \nonumber\\
&&+\frac{ m_s m_c\langle\bar{s}s\rangle^2}{24\pi^4}\int dydz \,   (1-y-z)\left(3s-2\overline{m}_c^2 \right) \nonumber\\
&& -\frac{89m_s m_c\langle\bar{s}s\rangle\langle\bar{s}g_s\sigma Gs\rangle}{2304\pi^4}\int dydz  \,   \left[1+\frac{s}{2}\delta \left(s-\overline{m}_c^2 \right) \right]\nonumber\\
&&-\frac{199m_s m_c \langle\bar{s}s\rangle\langle\bar{s}g_s\sigma Gs\rangle}{2304\pi^4}\int dy   \nonumber\\
&&   +\frac{m_s\langle\bar{s}s\rangle^3}{36\pi^2}\int dy \,  \left[ 1+s\,\delta\left(s-\widetilde{m}_c^2 \right)\right]  \nonumber\\
&&+\frac{125m_s m_c\langle\bar{s}g_s\sigma Gs\rangle^2}{9216\pi^4}\int dy \left(1+\frac{s}{T^2}\right)\delta\left(s-\widetilde{m}_c^2 \right) \, ,
\end{eqnarray}

\begin{eqnarray}
\rho^{11\frac{3}{2},1}_{uss}(s)&=&\frac{1}{122880\pi^8}\int dydz \, yz(1-y-z)^4\left(s-\overline{m}_c^2\right)^4\left(7s-2\overline{m}_c^2 \right)  \nonumber\\
&&+\frac{1}{614400\pi^8}\int dydz \, yz(1-y-z)^5\left(s-\overline{m}_c^2\right)^3\left(63s^2-51s\overline{m}_c^2+ 8\overline{m}_c^4\right)  \nonumber\\
&&- \frac{m_c\left[\langle \bar{q}q\rangle+2\langle \bar{s}s\rangle\right]}{2304\pi^6}\int dydz \, (y+z)(1-y-z)^2\left(s-\overline{m}_c^2\right)^3  \nonumber \\
&&- \frac{m_c\left[\langle \bar{q}q\rangle+2\langle \bar{s}s\rangle\right]}{2304\pi^6}\int dydz \, (y+z)(1-y-z)^3\left(s-\overline{m}_c^2\right)^2\left(2s-\overline{m}_c^2\right)  \nonumber \\
&&+ \frac{11m_c\left[\langle \bar{q}g_s\sigma Gq\rangle+2\langle \bar{s}g_s\sigma Gs\rangle\right]}{12288\pi^6}\int dydz  \, (y+z)(1-y-z) \left(s-\overline{m}_c^2 \right)^2 \nonumber\\
&&+ \frac{11m_c\left[\langle \bar{q}g_s\sigma Gq\rangle+2\langle \bar{s}g_s\sigma Gs\rangle\right]}{24576\pi^6}\int dydz  \, (y+z)(1-y-z)^2 \left(s-\overline{m}_c^2 \right)\left(5s-3\overline{m}_c^2 \right) \nonumber\\
&&+\frac{\langle\bar{s}s\rangle\left[2\langle\bar{q}q\rangle+\langle\bar{s}s\rangle \right]}{72\pi^4}\int dydz \,  yz(1-y-z)\left(s-\overline{m}_c^2 \right)\left(2s-\overline{m}_c^2 \right) \nonumber
\end{eqnarray}
\begin{eqnarray}
&&-\frac{19\left[\langle\bar{s}s\rangle\langle\bar{s}g_s\sigma Gs\rangle+\langle\bar{q}q\rangle\langle\bar{s}g_s\sigma Gs\rangle+\langle\bar{s}s\rangle\langle\bar{q}g_s\sigma Gq\rangle\right]}{2304\pi^4}\int dydz \,yz \left(3s-2\overline{m}_c^2\right)\nonumber\\
&&+\frac{\langle\bar{s}s\rangle\langle\bar{s}g_s\sigma Gs\rangle+\langle\bar{q}q\rangle\langle\bar{s}g_s\sigma Gs\rangle+\langle\bar{s}s\rangle\langle\bar{q}g_s\sigma Gq\rangle}{4608\pi^4}\int dydz \,(y+z)(1-y-z) \left(5s-4\overline{m}_c^2\right)\nonumber\\
&&-\frac{m_c\langle\bar{q}q\rangle\langle\bar{s}s\rangle^2}{36\pi^2}\int dy \nonumber\\
&&-\frac{13\langle\bar{s}g_s\sigma Gs\rangle\left[2 \langle\bar{q}g_s\sigma Gq\rangle+\langle\bar{s}g_s\sigma Gs\rangle\right]}{41472\pi^4}\int dydz \, (y+z)\left[ 1+\frac{s}{4}\delta\left(s-\overline{m}_c^2 \right)\right] \nonumber\\
&&+\frac{11\langle\bar{s}g_s\sigma Gs\rangle \left[ 2\langle\bar{q}g_s\sigma Gq\rangle+\langle\bar{s}g_s\sigma Gs\rangle\right]}{4608\pi^4}\int dy  \, y(1-y)\left[ 1+\frac{s}{2}\delta\left(s-\widetilde{m}_c^2 \right)\right] \nonumber\\
&&+\frac{m_s m_c}{18432\pi^8}\int dydz \, (y+z)(1-y-z)^3\left(s-\overline{m}_c^2\right)^4  \nonumber\\
&&+\frac{m_s m_c}{73728\pi^8}\int dydz \, (y+z)(1-y-z)^4\left(s-\overline{m}_c^2\right)^3\left(7s-3\overline{m}_c^2\right)  \nonumber\\
&&- \frac{m_s\left[2\langle \bar{q}q\rangle-\langle \bar{s}s\rangle\right]}{1152\pi^6}\int dydz \, yz(1-y-z)^2\left(s-\overline{m}_c^2\right)^2 \left(5s-2\overline{m}_c^2 \right) \nonumber \\
&&+ \frac{m_s\langle \bar{s}s\rangle}{1152\pi^6}\int dydz \, yz(1-y-z)^3\left(s-\overline{m}_c^2\right) \left(35s^2-37s\overline{m}_c^2+8\overline{m}_c^4 \right) \nonumber\\
&&- \frac{m_s\langle \bar{s}g_s\sigma Gs\rangle}{384\pi^6}\int dydz  \, yz(1-y-z)^2 \left(12s^2-15s\overline{m}_c^2 +4\overline{m}_c^4\right) \nonumber\\
&&- \frac{m_s\left[\langle \bar{q}g_s\sigma Gq\rangle+\langle \bar{s}g_s\sigma Gs\rangle\right]}{12288\pi^6}\int dydz  \, (y+z)(1-y-z)^2 \left(s-\overline{m}_c^2 \right)\left(3s-2\overline{m}_c^2 \right) \nonumber\\
&&+ \frac{m_s\left[19\langle \bar{q}g_s\sigma Gq\rangle+3\langle \bar{s}g_s\sigma Gs\rangle\right]}{3072\pi^6}\int dydz  \, yz(1-y-z) \left(s-\overline{m}_c^2 \right)\left(2s-\overline{m}_c^2 \right) \nonumber\\
&&+\frac{ m_s m_c\langle\bar{s}s\rangle\left[ 5\langle\bar{q}q\rangle-\langle\bar{s}s\rangle\right]}{144\pi^4}\int dydz \,  (y+z)\left(s-\overline{m}_c^2 \right) \nonumber\\
&&-\frac{ m_s m_c\langle\bar{s}s\rangle\left[\langle\bar{q}q\rangle+\langle\bar{s}s\rangle\right] }{144\pi^4}\int dydz \,  (y+z)(1-y-z)\left(4s-3\overline{m}_c^2 \right) \nonumber\\
&& +\frac{m_s m_c\left[89\langle\bar{s}s\rangle\langle\bar{s}g_s\sigma Gs\rangle+57\langle\bar{s}s\rangle\langle\bar{q}g_s\sigma Gq\rangle+32\langle\bar{q}q\rangle\langle\bar{s}g_s\sigma Gs\rangle\right]}{9216\pi^4} \nonumber\\
&&\int dydz  \,  \left(y+z\right)\left[1+\frac{s}{3}\delta \left(s-\overline{m}_c^2 \right) \right]\nonumber\\
&&-\frac{ m_s m_c \left[231\langle\bar{s}s\rangle\langle\bar{q}g_s\sigma Gq\rangle-89\langle\bar{s}s\rangle\langle\bar{s}g_s\sigma Gs\rangle+256\langle\bar{q}q\rangle\langle\bar{s}g_s\sigma Gs\rangle\right]}{27648\pi^4}\int dy   \nonumber\\
&&+\frac{m_s m_c\left[\langle\bar{q}q\rangle\langle\bar{s}g_s\sigma Gs\rangle+\langle\bar{s}s\rangle\langle\bar{q}g_s\sigma Gq\rangle\right]}{2304\pi^4}\int dydz  \left(\frac{y}{z}+\frac{z}{y}\right)\nonumber\\
&&   +\frac{m_s\langle\bar{q}q\rangle\langle\bar{s}s\rangle^2}{27\pi^2}\int dy \,y(1-y) \left[ 1+\frac{s}{2}\delta\left(s-\widetilde{m}_c^2 \right)\right]  \nonumber\\
&&-\frac{m_s m_c\langle\bar{q}g_s\sigma Gq\rangle\langle\bar{s}g_s\sigma Gs\rangle }{4608\pi^4}\int dy   \left(\frac{1-y}{y}+\frac{y}{1-y}\right)\delta \left(s-\widetilde{m}_c^2 \right)\nonumber\\
&&-\frac{m_s m_c\langle\bar{s}g_s\sigma Gs\rangle\left[19\langle \bar{s}g_s\sigma Gs\rangle-53\langle \bar{q}g_s\sigma Gq\rangle\right]}{13824\pi^4}\int dy   \left( 1+\frac{s}{2T^2}\right) \delta\left(s-\widetilde{m}_c^2 \right)\, ,
\end{eqnarray}

\begin{eqnarray}
\widetilde{\rho}^{11\frac{3}{2},0}_{uss}(s)&=&\frac{1}{122880\pi^8}\int dydz \, (y+z)(1-y-z)^4\left(s-\overline{m}_c^2\right)^4\left(6s-\overline{m}_c^2 \right)  \nonumber\\
&&-\frac{1}{1228800\pi^8}\int dydz \, (y+z)(1-y-z)^5\left(s-\overline{m}_c^2\right)^3\left(48s^2-31s\overline{m}_c^2+ 3\overline{m}_c^4\right)  \nonumber\\
&&- \frac{m_c\left[\langle \bar{q}q\rangle+2\langle \bar{s}s\rangle\right]}{576\pi^6}\int dydz \, (1-y-z)^2\left(s-\overline{m}_c^2\right)^3  \nonumber \\
&&+ \frac{m_c\left[\langle \bar{q}q\rangle+2\langle \bar{s}s\rangle\right]}{3456\pi^6}\int dydz \,  (1-y-z)^3\left(s-\overline{m}_c^2\right)^2\left(5s-2\overline{m}_c^2\right)  \nonumber \\
&&+ \frac{11m_c\left[\langle \bar{q}g_s\sigma Gq\rangle+2\langle \bar{s}g_s\sigma Gs\rangle\right]}{3072\pi^6}\int dydz  \, (1-y-z) \left(s-\overline{m}_c^2 \right)^2 \nonumber\\
&&- \frac{11m_c\left[\langle \bar{q}g_s\sigma Gq\rangle+2\langle \bar{s}g_s\sigma Gs\rangle\right]}{6144\pi^6}\int dydz  \,  (1-y-z)^2 \left(s-\overline{m}_c^2 \right)\left(2s-\overline{m}_c^2 \right) \nonumber\\
&&+\frac{\langle\bar{s}s\rangle\left[2\langle\bar{q}q\rangle+\langle\bar{s}s\rangle \right]}{144\pi^4}\int dydz \,  (y+z)(1-y-z)\left(s-\overline{m}_c^2 \right)\left(3s-\overline{m}_c^2 \right) \nonumber\\
&&-\frac{19\left[\langle\bar{s}s\rangle\langle\bar{s}g_s\sigma Gs\rangle+\langle\bar{q}q\rangle\langle\bar{s}g_s\sigma Gs\rangle+\langle\bar{s}s\rangle\langle\bar{q}g_s\sigma Gq\rangle\right]}{2304\pi^4}\int dydz \,(y+z)  \left(2s-\overline{m}_c^2\right)\nonumber\\
&&-\frac{m_c\langle\bar{q}q\rangle\langle\bar{s}s\rangle^2}{9\pi^2}\int dy \nonumber\\
&&+\frac{11\langle\bar{s}g_s\sigma Gs\rangle\left[2\langle\bar{q}g_s\sigma Gq\rangle+\langle\bar{s}g_s\sigma Gs\rangle \right]}{9216\pi^4}\int dy  \left[ 1+s\,\delta\left(s-\widetilde{m}_c^2 \right)\right] \nonumber\\
&&+\frac{m_s m_c}{4608\pi^8}\int dydz \, (1-y-z)^3\left(s-\overline{m}_c^2\right)^4  \nonumber\\
&&-\frac{m_s m_c}{18432\pi^8}\int dydz \, (1-y-z)^4\left(s-\overline{m}_c^2\right)^3\left(3s-\overline{m}_c^2\right)  \nonumber\\
&&- \frac{m_s\left[2\langle \bar{q}q\rangle-\langle \bar{s}s\rangle\right]}{1152\pi^6}\int dydz \, (y+z)(1-y-z)^2\left(s-\overline{m}_c^2\right)^2 \left(4s-\overline{m}_c^2 \right) \nonumber \\
&&- \frac{m_s\langle \bar{s}s\rangle}{768\pi^6}\int dydz \, (y+z)(1-y-z)^3\left(s-\overline{m}_c^2\right) \left(8s^2-7s\overline{m}_c^2+\overline{m}_c^4 \right)
\nonumber \\
&&+ \frac{m_s\langle \bar{s}g_s\sigma Gs\rangle}{1536\pi^6}\int dydz  \, (y+z)(1-y-z)^2 \left(15s^2-16s\overline{m}_c^2 +3\overline{m}_c^4\right) \nonumber\\
&&+ \frac{ m_s\left[19\langle \bar{q}g_s\sigma Gq\rangle+3\langle \bar{s}g_s\sigma Gs\rangle\right]}{6144\pi^6}\int dydz  \, (y+z)(1-y-z) \left(s-\overline{m}_c^2 \right)\left(3s-\overline{m}_c^2 \right) \nonumber\\
&&+\frac{ m_s m_c\langle\bar{s}s\rangle\left[ 5\langle\bar{q}q\rangle-\langle\bar{s}s\rangle\right]}{36\pi^4}\int dydz \,  \left(s-\overline{m}_c^2 \right) \nonumber\\
&&+\frac{ m_s m_c\langle\bar{s}s\rangle\left[\langle\bar{q}q\rangle+\langle\bar{s}s\rangle \right]}{72\pi^4}\int dydz \,   (1-y-z)\left(3s-2\overline{m}_c^2 \right) \nonumber\\
&& -\frac{m_s m_c\left[89\langle\bar{s}s\rangle\langle\bar{s}g_s\sigma Gs\rangle+57\langle\bar{s}s\rangle\langle\bar{q}g_s\sigma Gq\rangle+32\langle\bar{q}q\rangle\langle\bar{s}g_s\sigma Gs\rangle\right]}{6912\pi^4}\int dydz  \,   \left[1+\frac{s}{2}\delta \left(s-\overline{m}_c^2 \right) \right]\nonumber\\
&&-\frac{m_s m_c\left[ 231\langle\bar{s}s\rangle\langle\bar{q}g_s\sigma Gq\rangle-89\langle\bar{s}s\rangle\langle\bar{s}g_s\sigma Gs\rangle+256\langle\bar{q}q\rangle\langle\bar{s}g_s\sigma Gs\rangle\right]}{6912\pi^4}\int dy   \nonumber\\
&&   +\frac{m_s\langle\bar{q}q\rangle\langle\bar{s}s\rangle^2}{54\pi^2}\int dy \,  \left[ 1+s\,\delta\left(s-\widetilde{m}_c^2 \right)\right]  \nonumber\\
&&-\frac{m_s m_c\langle\bar{s}g_s\sigma Gs\rangle\left[19\langle \bar{s}g_s\sigma Gs\rangle-269\langle \bar{q}g_s\sigma Gq\rangle\right]}{27648\pi^4} \int dy   \left( 1+\frac{s}{T^2}\right) \delta\left(s-\widetilde{m}_c^2 \right) \, ,
\end{eqnarray}

\begin{eqnarray}
\rho^{11\frac{3}{2},1}_{uus}(s)&=&\frac{1}{122880\pi^8}\int dydz \, yz(1-y-z)^4\left(s-\overline{m}_c^2\right)^4\left(7s-2\overline{m}_c^2 \right)  \nonumber\\
&&+\frac{1}{614400\pi^8}\int dydz \, yz(1-y-z)^5\left(s-\overline{m}_c^2\right)^3\left(63s^2-51s\overline{m}_c^2+ 8\overline{m}_c^4\right)  \nonumber\\
&&- \frac{m_c\left[2\langle \bar{q}q\rangle+\langle \bar{s}s\rangle\right]}{2304\pi^6}\int dydz \, (y+z)(1-y-z)^2\left(s-\overline{m}_c^2\right)^3  \nonumber \\
&&- \frac{m_c\left[2\langle \bar{q}q\rangle+\langle \bar{s}s\rangle\right]}{2304\pi^6}\int dydz \, (y+z)(1-y-z)^3\left(s-\overline{m}_c^2\right)^2\left(2s-\overline{m}_c^2\right)  \nonumber \\
&&+ \frac{11m_c\left[2\langle \bar{q}g_s\sigma Gq\rangle+\langle \bar{s}g_s\sigma Gs\rangle\right]}{12288\pi^6}\int dydz  \, (y+z)(1-y-z) \left(s-\overline{m}_c^2 \right)^2 \nonumber\\
&&+ \frac{11m_c\left[2\langle \bar{q}g_s\sigma Gq\rangle+\langle \bar{s}g_s\sigma Gs\rangle\right]}{24576\pi^6}\int dydz  \, (y+z)(1-y-z)^2 \left(s-\overline{m}_c^2 \right)\left(5s-3\overline{m}_c^2 \right) \nonumber\\
&&+\frac{\langle\bar{q}q\rangle \left[\langle\bar{q}q\rangle+2\langle\bar{s}s\rangle \right]}{72\pi^4}\int dydz \,  yz(1-y-z)\left(s-\overline{m}_c^2 \right)\left(2s-\overline{m}_c^2 \right) \nonumber\\
&&-\frac{19\left[\langle\bar{q}q\rangle\langle\bar{q}g_s\sigma Gq\rangle+\langle\bar{q}q\rangle\langle\bar{s}g_s\sigma Gs\rangle+\langle\bar{s}s\rangle\langle\bar{q}g_s\sigma Gq\rangle\right]}{2304\pi^4}\int dydz \,yz \left(3s-2\overline{m}_c^2\right)\nonumber\\
&&+\frac{\langle\bar{q}q\rangle\langle\bar{q}g_s\sigma Gq\rangle+\langle\bar{q}q\rangle\langle\bar{s}g_s\sigma Gs\rangle+\langle\bar{s}s\rangle\langle\bar{q}g_s\sigma Gq\rangle}{4608\pi^4}\int dydz \,(y+z)(1-y-z) \left(5s-4\overline{m}_c^2\right)\nonumber\\
&&-\frac{13\langle\bar{q}g_s\sigma Gq\rangle \left[\langle\bar{q}g_s\sigma Gq\rangle+2\langle\bar{s}g_s\sigma Gs\rangle \right]}{41472\pi^4}\int dydz \, (y+z)\left[ 1+\frac{s}{4}\delta\left(s-\overline{m}_c^2 \right)\right] \nonumber\\
&&+\frac{11\langle\bar{q}g_s\sigma Gq\rangle\left[\langle\bar{q}g_s\sigma Gq\rangle+2\langle\bar{s}g_s\sigma Gs\rangle \right]}{4608\pi^4}\int dy  \, y(1-y)\left[ 1+\frac{s}{2}\delta\left(s-\widetilde{m}_c^2 \right)\right] \nonumber\\
&&-\frac{m_c\langle\bar{q}q\rangle^2\langle\bar{s}s\rangle}{36\pi^2}\int dy \nonumber\\
&&+\frac{m_s m_c}{36864\pi^8}\int dydz \, (y+z)(1-y-z)^3\left(s-\overline{m}_c^2\right)^4  \nonumber\\
&&+\frac{m_s m_c}{147456\pi^8}\int dydz \, (y+z)(1-y-z)^4\left(s-\overline{m}_c^2\right)^3\left(7s-3\overline{m}_c^2\right)  \nonumber\\
&&- \frac{m_s\left[4\langle \bar{q}q\rangle-3\langle \bar{s}s\rangle\right]}{2304\pi^6}\int dydz \, yz(1-y-z)^2\left(s-\overline{m}_c^2\right)^2 \left(5s-2\overline{m}_c^2 \right) \nonumber \\
&&+ \frac{m_s\langle \bar{s}s\rangle}{2304\pi^6}\int dydz \, yz(1-y-z)^3\left(s-\overline{m}_c^2\right) \left(35s^2-37s\overline{m}_c^2+8\overline{m}_c^4 \right) \nonumber \\
&&+ \frac{m_s\left[19\langle \bar{q}g_s\sigma Gq\rangle-8\langle \bar{s}g_s\sigma Gs\rangle\right]}{3072\pi^6}\int dydz  \, yz(1-y-z) \left(s-\overline{m}_c^2 \right)\left(2s-\overline{m}_c^2 \right) \nonumber\\
&&- \frac{m_s\langle \bar{s}g_s\sigma Gs\rangle}{768\pi^6}\int dydz  \, yz(1-y-z)^2 \left(12s^2-15s\overline{m}_c^2 +4\overline{m}_c^4\right) \nonumber\\
&&- \frac{m_s\langle \bar{q}g_s\sigma Gq\rangle}{12288\pi^6}\int dydz  \, (y+z)(1-y-z)^2 \left(s-\overline{m}_c^2 \right)\left(3s-2\overline{m}_c^2 \right) \nonumber\\
&&+\frac{ m_s m_c\langle\bar{q}q\rangle\left[ 3\langle\bar{q}q\rangle-\langle\bar{s}s\rangle\right]}{144\pi^4}\int dydz \,  (y+z)\left(s-\overline{m}_c^2 \right) \nonumber\\
&&-\frac{ m_s m_c\langle\bar{q}q\rangle\langle\bar{s}s\rangle }{144\pi^4}\int dydz \,  (y+z)(1-y-z)\left(4s-3\overline{m}_c^2 \right) \nonumber
\end{eqnarray}
\begin{eqnarray}
&& +\frac{m_s m_c\left[32\langle\bar{q}q\rangle\langle\bar{s}g_s\sigma Gs\rangle+57\langle\bar{s}s\rangle\langle\bar{q}g_s\sigma Gq\rangle\right]}{9216\pi^4}\int dydz  \,  \left(y+z\right)\left[1+\frac{s}{3}\delta \left(s-\overline{m}_c^2 \right) \right]\nonumber\\
&&-\frac{m_s m_c \left[288\langle\bar{q}q\rangle\langle\bar{q}g_s\sigma Gq\rangle-57\langle\bar{s}s\rangle\langle\bar{q}g_s\sigma Gq\rangle-32\langle\bar{q}q\rangle\langle\bar{s}g_s\sigma Gs\rangle\right]}{27648\pi^4}\int dy   \nonumber\\
&&+\frac{m_s m_c\langle\bar{q}q\rangle\langle\bar{q}g_s\sigma Gq\rangle}{2304\pi^4}\int dydz  \left(\frac{y}{z}+\frac{z}{y}\right)\nonumber\\
&&   +\frac{m_s\langle\bar{q}q\rangle^2\langle\bar{s}s\rangle}{54\pi^2}\int dy \,y(1-y) \left[ 1+\frac{s}{2}\delta\left(s-\widetilde{m}_c^2 \right)\right]  \nonumber\\
&&-\frac{m_s m_c\langle\bar{q}g_s\sigma Gq\rangle^2}{9216\pi^4}\int dy   \left(\frac{1-y}{y}+\frac{y}{1-y}\right)\delta \left(s-\widetilde{m}_c^2 \right)\nonumber\\
&&+\frac{m_s m_c\langle\bar{q}g_s\sigma Gq\rangle\left[36\langle\bar{q}g_s\sigma Gq\rangle-19\langle\bar{s}g_s\sigma Gs\rangle\right]}{13824\pi^4}\int dy   \left( 1+\frac{s}{2T^2}\right) \delta\left(s-\widetilde{m}_c^2 \right) \, ,
\end{eqnarray}

\begin{eqnarray}
\widetilde{\rho}^{11\frac{3}{2},0}_{uus}(s)&=&\frac{1}{122880\pi^8}\int dydz \, (y+z)(1-y-z)^4\left(s-\overline{m}_c^2\right)^4\left(6s-\overline{m}_c^2 \right)  \nonumber\\
&&-\frac{1}{1228800\pi^8}\int dydz \, (y+z)(1-y-z)^5\left(s-\overline{m}_c^2\right)^3\left(48s^2-31s\overline{m}_c^2+ 3\overline{m}_c^4\right)  \nonumber\\
&&- \frac{m_c\left[2\langle \bar{q}q\rangle+\langle \bar{s}s\rangle\right]}{576\pi^6}\int dydz \, (1-y-z)^2\left(s-\overline{m}_c^2\right)^3  \nonumber \\
&&+ \frac{m_c\left[2\langle \bar{q}q\rangle+\langle \bar{s}s\rangle\right]}{3456\pi^6}\int dydz \,  (1-y-z)^3\left(s-\overline{m}_c^2\right)^2\left(5s-2\overline{m}_c^2\right)  \nonumber \\
&&+ \frac{11m_c\left[2\langle \bar{q}g_s\sigma Gq\rangle+\langle \bar{s}g_s\sigma Gs\rangle\right]}{3072\pi^6}\int dydz  \, (1-y-z) \left(s-\overline{m}_c^2 \right)^2 \nonumber\\
&&- \frac{11m_c\left[2\langle \bar{q}g_s\sigma Gq\rangle+\langle \bar{s}g_s\sigma Gs\rangle\right]}{6144\pi^6}\int dydz  \,  (1-y-z)^2 \left(s-\overline{m}_c^2 \right)\left(2s-\overline{m}_c^2 \right) \nonumber \\
&&+\frac{\langle\bar{q}q\rangle \left[ \langle\bar{q}q\rangle+2\langle\bar{s}s\rangle\right]}{144\pi^4}\int dydz \,  (y+z)(1-y-z)\left(s-\overline{m}_c^2 \right)\left(3s-\overline{m}_c^2 \right) \nonumber\\
&&-\frac{19\left[\langle\bar{q}q\rangle\langle\bar{q}g_s\sigma Gq\rangle+\langle\bar{q}q\rangle\langle\bar{s}g_s\sigma Gs\rangle+\langle\bar{s}s\rangle\langle\bar{q}g_s\sigma Gq\rangle\right]}{2304\pi^4}\int dydz \,(y+z)  \left(2s-\overline{m}_c^2\right)\nonumber\\
&&-\frac{m_c\langle\bar{q}q\rangle^2\langle\bar{s}s\rangle}{9\pi^2}\int dy \nonumber\\
&&+\frac{11\langle\bar{q}g_s\sigma Gq\rangle \left[ \langle\bar{q}g_s\sigma Gq\rangle+2\langle\bar{s}g_s\sigma Gs\rangle\right]}{9216\pi^4}\int dy  \left[ 1+s\,\delta\left(s-\widetilde{m}_c^2 \right)\right] \nonumber\\
&&+\frac{m_s m_c}{9216\pi^8}\int dydz \, (1-y-z)^3\left(s-\overline{m}_c^2\right)^4  \nonumber\\
&&-\frac{m_s m_c}{36864\pi^8}\int dydz \, (1-y-z)^4\left(s-\overline{m}_c^2\right)^3\left(3s-\overline{m}_c^2\right)  \nonumber\\
&&- \frac{m_s\left[4\langle \bar{q}q\rangle-3\langle \bar{s}s\rangle\right]}{2304\pi^6}\int dydz \, (y+z)(1-y-z)^2\left(s-\overline{m}_c^2\right)^2 \left(4s-\overline{m}_c^2 \right) \nonumber \\
&&- \frac{m_s\langle \bar{s}s\rangle}{1536\pi^6}\int dydz \, (y+z)(1-y-z)^3\left(s-\overline{m}_c^2\right) \left(8s^2-7s\overline{m}_c^2+\overline{m}_c^4 \right)
\nonumber \\
&&+ \frac{m_s\left[19\langle \bar{q}g_s\sigma Gq\rangle-8\langle \bar{s}g_s\sigma Gs\rangle\right]}{6144\pi^6}\int dydz  \, (y+z)(1-y-z) \left(s-\overline{m}_c^2 \right)\left(3s-\overline{m}_c^2 \right) \nonumber
\end{eqnarray}
\begin{eqnarray}
&&+ \frac{m_s\langle \bar{s}g_s\sigma Gs\rangle}{3072\pi^6}\int dydz  \, (y+z)(1-y-z)^2 \left(15s^2-16s\overline{m}_c^2 +3\overline{m}_c^4\right) \nonumber\\
&&+\frac{ m_s m_c\langle\bar{q}q\rangle \left[3\langle\bar{q}q\rangle-\langle\bar{s}s\rangle \right]}{36\pi^4}\int dydz \,  \left(s-\overline{m}_c^2 \right) \nonumber\\
&&+\frac{ m_s m_c\langle\bar{q}q\rangle\langle\bar{s}s\rangle }{72\pi^4}\int dydz \,   (1-y-z)\left(3s-2\overline{m}_c^2 \right) \nonumber\\
&& -\frac{m_s m_c\left[32\langle\bar{q}q\rangle\langle\bar{s}g_s\sigma Gs\rangle+57\langle\bar{s}s\rangle\langle\bar{q}g_s\sigma Gq\rangle\right]}{6912\pi^4}\int dydz  \,   \left[1+\frac{s}{2}\delta \left(s-\overline{m}_c^2 \right) \right]\nonumber\\
&&-\frac{m_s m_c \left[288\langle\bar{q}q\rangle\langle\bar{q}g_s\sigma Gq\rangle-57\langle\bar{s}s\rangle\langle\bar{q}g_s\sigma Gq\rangle-32\langle\bar{q}q\rangle\langle\bar{s}g_s\sigma Gs\rangle\right]}{6912\pi^4}\int dy   \nonumber\\
&&   +\frac{m_s\langle\bar{q}q\rangle^2 \langle\bar{s}s\rangle}{108\pi^2}\int dy \,  \left[ 1+s\,\delta\left(s-\widetilde{m}_c^2 \right)\right]  \nonumber\\
&&+\frac{m_s m_c\langle\bar{q}g_s\sigma Gq\rangle \left[ 144\langle\bar{q}g_s\sigma Gq\rangle-19\langle\bar{s}g_s\sigma Gs\rangle\right]}{27648\pi^4} \int dy   \left( 1+\frac{s}{T^2}\right) \delta\left(s-\widetilde{m}_c^2 \right)\, ,
\end{eqnarray}

\begin{eqnarray}
\rho^{11\frac{3}{2},1}_{uuu}(s)&=&\rho^{11\frac{3}{2},1}_{sss}(s)\mid_{m_s \to 0,\,\,\langle\bar{s}s\rangle\to\langle\bar{q}q\rangle,\,\,\langle\bar{s}g_s\sigma Gs\rangle\to\langle\bar{q}g_s\sigma Gq\rangle} \, ,  \\
\widetilde{\rho}^{11\frac{3}{2},0}_{uuu}(s)&=&\widetilde{\rho}^{11\frac{3}{2},0}_{sss}(s)\mid_{m_s \to 0,\,\,\langle\bar{s}s\rangle\to\langle\bar{q}q\rangle,\,\,\langle\bar{s}g_s\sigma Gs\rangle\to\langle\bar{q}g_s\sigma Gq\rangle} \, ,
\end{eqnarray}

\begin{eqnarray}
\rho^{11\frac{1}{2},1}_{sss}(s)&=&\frac{1}{61440\pi^8}\int dydz \, yz(1-y-z)^4\left(s-\overline{m}_c^2\right)^4\left(8s-3\overline{m}_c^2 \right)  \nonumber\\
&&-\frac{1}{307200\pi^8}\int dydz \, yz(1-y-z)^5\left(s-\overline{m}_c^2\right)^4\left(9s-4\overline{m}_c^2 \right)  \nonumber\\
&&- \frac{m_c\langle \bar{s}s\rangle}{768\pi^6}\int dydz \, (y+z)(1-y-z)^2\left(s-\overline{m}_c^2\right)^3  \nonumber \\
&&+ \frac{m_c\langle \bar{s}s\rangle}{2304\pi^6}\int dydz \, (y+z)(1-y-z)^3\left(s-\overline{m}_c^2\right)^3  \nonumber \\
&&+ \frac{11m_c\langle \bar{s}g_s\sigma Gs\rangle}{4096\pi^6}\int dydz  \, (y+z)(1-y-z) \left(s-\overline{m}_c^2 \right)^2 \nonumber\\
&&- \frac{11m_c\langle \bar{s}g_s\sigma Gs\rangle}{8192\pi^6}\int dydz  \, (y+z)(1-y-z)^2 \left(s-\overline{m}_c^2 \right)^2 \nonumber\\
&&+\frac{\langle\bar{s}s\rangle^2}{24\pi^4}\int dydz \,  yz(1-y-z)\left(s-\overline{m}_c^2 \right)\left(5s-3\overline{m}_c^2 \right) \nonumber\\
&&-\frac{19\langle\bar{s}s\rangle\langle\bar{s}g_s\sigma Gs\rangle}{384\pi^4}\int dydz \,yz \left(4s-3\overline{m}_c^2\right)\nonumber\\
&&-\frac{m_c\langle\bar{s}s\rangle^3}{36\pi^2}\int dy \nonumber\\
&&+\frac{11\langle\bar{s}g_s\sigma Gs\rangle^2}{512\pi^4}\int dy  \, y(1-y)\left[ 1+\frac{s}{3}\delta\left(s-\widetilde{m}_c^2 \right)\right] \nonumber\\
&&+\frac{m_s m_c}{12288\pi^8}\int dydz \, (y+z)(1-y-z)^3\left(s-\overline{m}_c^2\right)^4  \nonumber\\
&&-\frac{m_s m_c}{49152\pi^8}\int dydz \, (y+z)(1-y-z)^4\left(s-\overline{m}_c^2\right)^4  \nonumber\\
&&- \frac{m_s\langle \bar{s}s\rangle}{128\pi^6}\int dydz \, yz(1-y-z)^2\left(s-\overline{m}_c^2\right)^2 \left(2s-\overline{m}_c^2 \right) \nonumber
\end{eqnarray}
\begin{eqnarray}
&&- \frac{m_s\langle \bar{s}s\rangle}{384\pi^6}\int dydz \, yz(1-y-z)^3\left(s-\overline{m}_c^2\right)^2 \left(7s-4\overline{m}_c^2 \right) \nonumber \\
&&+ \frac{11m_s\langle \bar{s}g_s\sigma Gs\rangle}{1024\pi^6}\int dydz  \, yz(1-y-z) \left(s-\overline{m}_c^2 \right)\left(5s-3\overline{m}_c^2 \right) \nonumber\\
&&+ \frac{m_s\langle \bar{s}g_s\sigma Gs\rangle}{128\pi^6}\int dydz  \, yz(1-y-z)^2 \left(s-\overline{m}_c^2 \right)\left(3s-2\overline{m}_c^2 \right) \nonumber\\
&&+\frac{ m_s m_c\langle\bar{s}s\rangle^2}{24\pi^4}\int dydz \,  (y+z)\left(s-\overline{m}_c^2 \right) \nonumber\\
&&+\frac{ m_s m_c\langle\bar{s}s\rangle^2}{48\pi^4}\int dydz \,  (y+z)(1-y-z)\left(s-\overline{m}_c^2 \right) \nonumber\\
&&-\frac{89m_s m_c\langle\bar{s}s\rangle\langle\bar{s}g_s\sigma Gs\rangle}{9216\pi^4}\int dydz  \,  \left(y+z\right) \nonumber\\
&&-\frac{199m_s m_c \langle\bar{s}s\rangle\langle\bar{s}g_s\sigma Gs\rangle}{9216\pi^4}\int dy   \nonumber\\
&&+\frac{m_s m_c\langle\bar{s}s\rangle\langle\bar{s}g_s\sigma Gs\rangle}{768\pi^4}\int dydz  \left(\frac{y}{z}+\frac{z}{y}\right)\nonumber\\
&&   +\frac{m_s\langle\bar{s}s\rangle^3}{6\pi^2}\int dy \,y(1-y) \left[ 1+\frac{s}{3}\delta\left(s-\widetilde{m}_c^2 \right)\right]  \nonumber\\
&&+\frac{m_s m_c\langle\bar{s}g_s\sigma Gs\rangle^2}{2304\pi^4}\int dy \left(16+\frac{7s}{T^2}\right) \delta\left(s-\widetilde{m}_c^2 \right) \nonumber\\
&&-\frac{m_s m_c\langle\bar{s}g_s\sigma Gs\rangle^2}{3072\pi^4}\int dy   \left(\frac{1-y}{y}+\frac{y}{1-y}\right)\delta \left(s-\widetilde{m}_c^2 \right)\nonumber\\
&&-\frac{m_s m_c\langle\bar{s}g_s\sigma Gs\rangle^2}{6144\pi^4}\int dy   \left( 1+\frac{s}{T^2}\right) \delta\left(s-\widetilde{m}_c^2 \right) \, ,
\end{eqnarray}

\begin{eqnarray}
\widetilde{\rho}^{11\frac{1}{2},0}_{sss}(s)&=&\frac{1}{245760\pi^8}\int dydz \, (y+z)(1-y-z)^4\left(s-\overline{m}_c^2\right)^4\left(7s-2\overline{m}_c^2 \right)  \nonumber\\
&&-\frac{1}{1228800\pi^8}\int dydz \, (y+z)(1-y-z)^5\left(s-\overline{m}_c^2\right)^4\left(8s-3\overline{m}_c^2 \right)  \nonumber\\
&&- \frac{m_c\langle \bar{s}s\rangle}{192\pi^6}\int dydz \,  (1-y-z)^2\left(s-\overline{m}_c^2\right)^3  \nonumber \\
&&+ \frac{m_c\langle \bar{s}s\rangle}{576\pi^6}\int dydz \,  (1-y-z)^3\left(s-\overline{m}_c^2\right)^3  \nonumber \\
&&+ \frac{11m_c\langle \bar{s}g_s\sigma Gs\rangle}{1024\pi^6}\int dydz  \, (1-y-z) \left(s-\overline{m}_c^2 \right)^2 \nonumber\\
&&- \frac{11m_c\langle \bar{s}g_s\sigma Gs\rangle}{2048\pi^6}\int dydz  \, (1-y-z)^2 \left(s-\overline{m}_c^2 \right)^2 \nonumber\\
&&+\frac{\langle\bar{s}s\rangle^2}{48\pi^4}\int dydz \,  (y+z)(1-y-z)\left(s-\overline{m}_c^2 \right)\left(2s-\overline{m}_c^2 \right) \nonumber\\
&&-\frac{19\langle\bar{s}s\rangle\langle\bar{s}g_s\sigma Gs\rangle}{1536\pi^4}\int dydz \,(y+z) \left(3s-2\overline{m}_c^2\right)\nonumber\\
&&+\frac{\langle\bar{s}s\rangle\langle\bar{s}g_s\sigma Gs\rangle}{768\pi^4}\int dydz \,(1-y-z)\, s\nonumber\\
&&-\frac{m_c\langle\bar{s}s\rangle^3}{9\pi^2}\int dy \nonumber
\end{eqnarray}
\begin{eqnarray}
&&+\frac{11\langle\bar{s}g_s\sigma Gs\rangle^2}{3072\pi^4}\int dy \,  \left[ 1+\frac{s}{2}\delta\left(s-\widetilde{m}_c^2 \right)\right] \nonumber\\
&&-\frac{13\langle\bar{s}g_s\sigma Gs\rangle^2}{27648\pi^4}\int dydz  \,s\,\delta\left(s-\overline{m}_c^2 \right)  \nonumber\\
&&+\frac{m_s m_c}{3072\pi^8}\int dydz \,  (1-y-z)^3\left(s-\overline{m}_c^2\right)^4  \nonumber\\
&&-\frac{m_s m_c}{12288\pi^8}\int dydz \, (1-y-z)^4\left(s-\overline{m}_c^2\right)^4  \nonumber\\
&&- \frac{m_s\langle \bar{s}s\rangle}{1536\pi^6}\int dydz \, (y+z)(1-y-z)^2\left(s-\overline{m}_c^2\right)^2 \left(5s-2\overline{m}_c^2 \right) \nonumber \\
&&- \frac{m_s\langle \bar{s}s\rangle}{512\pi^6}\int dydz \, (y+z)(1-y-z)^3\left(s-\overline{m}_c^2\right)^2 \left(2s-\overline{m}_c^2 \right) \nonumber\\
&&+ \frac{11m_s\langle \bar{s}g_s\sigma Gs\rangle}{2048\pi^6}\int dydz  \, (y+z)(1-y-z) \left(s-\overline{m}_c^2 \right)\left(2s-\overline{m}_c^2 \right) \nonumber\\
&&+ \frac{m_s\langle \bar{s}g_s\sigma Gs\rangle}{1024\pi^6}\int dydz  \, (y+z)(1-y-z)^2 \left(s-\overline{m}_c^2 \right)\left(5s-3\overline{m}_c^2 \right) \nonumber\\
&&- \frac{m_s\langle \bar{s}g_s\sigma Gs\rangle}{2048\pi^6}\int dydz  \, (1-y-z)^2 \,s\left(s-\overline{m}_c^2 \right) \nonumber\\
&&+\frac{ m_s m_c\langle\bar{s}s\rangle^2}{6\pi^4}\int dydz \,   \left(s-\overline{m}_c^2 \right) \nonumber\\
&&+\frac{ m_s m_c\langle\bar{s}s\rangle^2}{12\pi^4}\int dydz \,  (1-y-z)\left(s-\overline{m}_c^2 \right) \nonumber\\
&&-\frac{89m_s m_c\langle\bar{s}s\rangle\langle\bar{s}g_s\sigma Gs\rangle}{2304\pi^4}\int dydz    \nonumber\\
&&-\frac{199m_s m_c \langle\bar{s}s\rangle\langle\bar{s}g_s\sigma Gs\rangle}{2304\pi^4}\int dy   \nonumber\\
&&   +\frac{m_s\langle\bar{s}s\rangle^3}{36\pi^2}\int dy \,  \left[ 1+\frac{s}{2}\delta\left(s-\widetilde{m}_c^2 \right)\right]  \nonumber\\
&&+\frac{m_s m_c\langle\bar{s}g_s\sigma Gs\rangle^2}{64\pi^4}\int dy \left(1+\frac{7s}{9T^2}\right) \delta\left(s-\widetilde{m}_c^2 \right) \nonumber\\
&&-\frac{m_s m_c\langle\bar{s}g_s\sigma Gs\rangle^2}{1536\pi^4 T^2}\int dy    \,s\,  \delta\left(s-\widetilde{m}_c^2 \right) \, ,
\end{eqnarray}

\begin{eqnarray}
\rho^{11\frac{1}{2},1}_{uss}(s)&=&\frac{1}{61440\pi^8}\int dydz \, yz(1-y-z)^4\left(s-\overline{m}_c^2\right)^4\left(8s-3\overline{m}_c^2 \right)  \nonumber\\
&&-\frac{1}{307200\pi^8}\int dydz \, yz(1-y-z)^5\left(s-\overline{m}_c^2\right)^4\left(9s-4\overline{m}_c^2 \right)  \nonumber\\
&&- \frac{m_c\left[\langle \bar{q}q\rangle+2\langle \bar{s}s\rangle\right]}{2304\pi^6}\int dydz \, (y+z)(1-y-z)^2\left(s-\overline{m}_c^2\right)^3  \nonumber \\
&&+ \frac{m_c\left[\langle \bar{q}q\rangle+2\langle \bar{s}s\rangle\right]}{6912\pi^6}\int dydz \, (y+z)(1-y-z)^3\left(s-\overline{m}_c^2\right)^3  \nonumber \\
&&+ \frac{11m_c\left[\langle \bar{q}g_s\sigma Gq\rangle+2\langle \bar{s}g_s\sigma Gs\rangle\right]}{12288\pi^6}\int dydz  \, (y+z)(1-y-z) \left(s-\overline{m}_c^2 \right)^2 \nonumber\\
&&- \frac{11m_c\left[\langle \bar{q}g_s\sigma Gq\rangle+2\langle \bar{s}g_s\sigma Gs\rangle\right]}{24576\pi^6}\int dydz  \, (y+z)(1-y-z)^2 \left(s-\overline{m}_c^2 \right)^2 \nonumber
\end{eqnarray}
\begin{eqnarray}
&&+\frac{\langle\bar{s}s\rangle \left[2\langle\bar{q}q\rangle+\langle\bar{s}s\rangle \right]}{72\pi^4}\int dydz \,  yz(1-y-z)\left(s-\overline{m}_c^2 \right)\left(5s-3\overline{m}_c^2 \right) \nonumber\\
&&-\frac{19\left[\langle\bar{s}s\rangle\langle\bar{s}g_s\sigma Gs\rangle+\langle\bar{q}q\rangle\langle\bar{s}g_s\sigma Gs\rangle+\langle\bar{s}s\rangle\langle\bar{q}g_s\sigma Gq\rangle\right]}{1152\pi^4}\int dydz \,yz \left(4s-3\overline{m}_c^2\right)\nonumber\\
&&-\frac{m_c\langle\bar{q}q\rangle\langle\bar{s}s\rangle^2}{36\pi^2}\int dy \nonumber\\
&&+\frac{11\langle\bar{s}g_s\sigma Gs\rangle \left[2\langle\bar{q}g_s\sigma Gq\rangle+\langle\bar{s}g_s\sigma Gs\rangle \right]}{1536\pi^4}\int dy \, y(1-y)\left[ 1+\frac{s}{3}\delta\left(s-\widetilde{m}_c^2 \right)\right] \nonumber\\
&&+\frac{m_s m_c}{18432\pi^8}\int dydz \, (y+z)(1-y-z)^3\left(s-\overline{m}_c^2\right)^4  \nonumber\\
&&-\frac{m_s m_c}{73728\pi^8}\int dydz \, (y+z)(1-y-z)^4\left(s-\overline{m}_c^2\right)^4  \nonumber\\
&&- \frac{m_s\left[2\langle \bar{q}q\rangle-\langle \bar{s}s\rangle\right]}{192\pi^6}\int dydz \, yz(1-y-z)^2\left(s-\overline{m}_c^2\right)^2 \left(2s-\overline{m}_c^2 \right) \nonumber \\
&&- \frac{m_s\langle \bar{s}s\rangle}{576\pi^6}\int dydz \, yz(1-y-z)^3\left(s-\overline{m}_c^2\right)^2 \left(7s-4\overline{m}_c^2 \right) \nonumber \\
&&+ \frac{m_s\langle \bar{s}g_s\sigma Gs\rangle}{192\pi^6}\int dydz  \, yz(1-y-z)^2 \left(s-\overline{m}_c^2 \right)\left(3s-2\overline{m}_c^2 \right) \nonumber\\
&&+ \frac{m_s\left[19\langle \bar{q}g_s\sigma Gq\rangle+3\langle \bar{s}g_s\sigma Gs\rangle\right]}{3072\pi^6}\int dydz  \, yz(1-y-z) \left(s-\overline{m}_c^2 \right)\left(5s-3\overline{m}_c^2 \right) \nonumber\\
&&+\frac{ m_s m_c\langle\bar{s}s\rangle\left[5\langle\bar{q}q\rangle -\langle\bar{s}s\rangle\right]}{144\pi^4}\int dydz \,  (y+z)\left(s-\overline{m}_c^2 \right) \nonumber\\
&&+\frac{ m_s m_c\langle\bar{s}s\rangle\left[\langle\bar{q}q\rangle+\langle\bar{s}s\rangle \right]}{144\pi^4}\int dydz \,  (y+z)(1-y-z)\left(s-\overline{m}_c^2 \right) \nonumber\\
&&-\frac{m_s m_c\left[89\langle\bar{s}s\rangle\langle\bar{s}g_s\sigma Gs\rangle+57\langle\bar{s}s\rangle\langle\bar{q}g_s\sigma Gq\rangle+32\langle\bar{q}q\rangle\langle\bar{s}g_s\sigma Gs\rangle\right]}{27648\pi^4}\int dydz  \,  \left(y+z\right) \nonumber\\
&&-\frac{m_s m_c \left[231\langle\bar{s}s\rangle\langle\bar{q}g_s\sigma Gq\rangle-89\langle\bar{s}s\rangle\langle\bar{s}g_s\sigma Gs\rangle+256\langle\bar{q}q\rangle\langle\bar{s}g_s\sigma Gs\rangle\right]}{27648\pi^4}\int dy   \nonumber\\
&&+\frac{m_s m_c\left[\langle\bar{q}q\rangle\langle\bar{s}g_s\sigma Gs\rangle+\langle\bar{s}s\rangle\langle\bar{q}g_s\sigma Gq\rangle\right]}{2304\pi^4}\int dydz  \left(\frac{y}{z}+\frac{z}{y}\right)\nonumber\\
&&   +\frac{m_s\langle\bar{q}q\rangle\langle\bar{s}s\rangle^2}{9\pi^2}\int dy \,y(1-y) \left[ 1+\frac{s}{3}\delta\left(s-\widetilde{m}_c^2 \right)\right]  \nonumber\\
&&-\frac{m_s m_c\langle\bar{q}g_s\sigma Gq\rangle \langle\bar{s}g_s\sigma Gs\rangle}{4608\pi^4}\int dy   \left(\frac{1-y}{y}+\frac{y}{1-y}\right)\delta \left(s-\widetilde{m}_c^2 \right)\nonumber\\
&&+\frac{17m_s m_c\langle\bar{q}g_s\sigma Gq\rangle\langle\bar{s}g_s\sigma Gs\rangle }{3456\pi^4}\int dy \left(1+\frac{8s}{17T^2}\right) \delta\left(s-\widetilde{m}_c^2 \right) \nonumber\\
&&-\frac{m_s m_c\langle\bar{s}g_s\sigma Gs\rangle\left[3\langle \bar{q}g_s\sigma Gq\rangle+19\langle \bar{s}g_s\sigma Gs\rangle\right]}{18432\pi^4}\int dy   \left( 1+\frac{s}{T^2}\right) \delta\left(s-\widetilde{m}_c^2 \right) \, ,
\end{eqnarray}

\begin{eqnarray}
\widetilde{\rho}^{11\frac{1}{2},0}_{uss}(s)&=&\frac{1}{245760\pi^8}\int dydz \, (y+z)(1-y-z)^4\left(s-\overline{m}_c^2\right)^4\left(7s-2\overline{m}_c^2 \right)  \nonumber\\
&&-\frac{1}{1228800\pi^8}\int dydz \, (y+z)(1-y-z)^5\left(s-\overline{m}_c^2\right)^4\left(8s-3\overline{m}_c^2 \right)  \nonumber\\
&&- \frac{m_c\left[\langle \bar{q}q\rangle+2\langle \bar{s}s\rangle\right]}{576\pi^6}\int dydz \,  (1-y-z)^2\left(s-\overline{m}_c^2\right)^3  \nonumber
\end{eqnarray}
\begin{eqnarray}
&&+ \frac{m_c\left[\langle \bar{q}q\rangle+2\langle \bar{s}s\rangle\right]}{1728\pi^6}\int dydz \,  (1-y-z)^3\left(s-\overline{m}_c^2\right)^3  \nonumber \\
&&+ \frac{11m_c\left[\langle \bar{q}g_s\sigma Gq\rangle+2\langle \bar{s}g_s\sigma Gs\rangle\right]}{3072\pi^6}\int dydz  \, (1-y-z) \left(s-\overline{m}_c^2 \right)^2 \nonumber\\
&&- \frac{11m_c\left[\langle \bar{q}g_s\sigma Gq\rangle+2\langle \bar{s}g_s\sigma Gs\rangle\right]}{6144\pi^6}\int dydz  \, (1-y-z)^2 \left(s-\overline{m}_c^2 \right)^2 \nonumber\\
&&+\frac{\langle\bar{s}s\rangle\left[ 2\langle\bar{q}q\rangle+\langle\bar{s}s\rangle\right]}{144\pi^4}\int dydz \,  (y+z)(1-y-z)\left(s-\overline{m}_c^2 \right)\left(2s-\overline{m}_c^2 \right) \nonumber\\
&&-\frac{19\left[\langle\bar{s}s\rangle\langle\bar{s}g_s\sigma Gs\rangle+\langle\bar{q}q\rangle\langle\bar{s}g_s\sigma Gs\rangle+\langle\bar{s}s\rangle\langle\bar{q}g_s\sigma Gq\rangle\right]}{4608\pi^4}\int dydz \,(y+z) \left(3s-2\overline{m}_c^2\right)\nonumber\\
&&+\frac{\langle\bar{s}s\rangle\langle\bar{s}g_s\sigma Gs\rangle+\langle\bar{q}q\rangle\langle\bar{s}g_s\sigma Gs\rangle+\langle\bar{s}s\rangle\langle\bar{q}g_s\sigma Gq\rangle}{2304\pi^4}\int dydz \,(1-y-z)\, s\nonumber\\
&&-\frac{m_c\langle\bar{q}q\rangle\langle\bar{s}s\rangle^2}{9\pi^2}\int dy \nonumber\\
&&+\frac{11\langle\bar{s}g_s\sigma Gs\rangle \left[2\langle\bar{q}g_s\sigma Gq\rangle+\langle\bar{s}g_s\sigma Gs\rangle \right]}{9216\pi^4}\int dy  \,  \left[ 1+\frac{s}{2}\delta\left(s-\widetilde{m}_c^2 \right)\right] \nonumber\\
&&-\frac{13\langle\bar{s}g_s\sigma Gs\rangle\left[2\langle\bar{q}g_s\sigma Gq\rangle+\langle\bar{s}g_s\sigma Gs\rangle \right]}{82944\pi^4}\int dydz  \,s\,\delta\left(s-\overline{m}_c^2 \right)  \nonumber\\
&&+\frac{m_s m_c}{4608\pi^8}\int dydz \,  (1-y-z)^3\left(s-\overline{m}_c^2\right)^4  \nonumber\\
&&-\frac{m_s m_c}{18432\pi^8}\int dydz \, (1-y-z)^4\left(s-\overline{m}_c^2\right)^4  \nonumber\\
&&- \frac{m_s\left[2\langle \bar{q}q\rangle-\langle \bar{s}s\rangle\right]}{2304\pi^6}\int dydz \, (y+z)(1-y-z)^2\left(s-\overline{m}_c^2\right)^2 \left(5s-2\overline{m}_c^2 \right) \nonumber \\
&&- \frac{m_s\langle \bar{s}s\rangle}{768\pi^6}\int dydz \, (y+z)(1-y-z)^3\left(s-\overline{m}_c^2\right)^2 \left(2s-\overline{m}_c^2 \right) \nonumber \\
&&+ \frac{m_s\langle \bar{s}g_s\sigma Gs\rangle}{1536\pi^6}\int dydz  \, (y+z)(1-y-z)^2 \left(s-\overline{m}_c^2 \right)\left(5s-3\overline{m}_c^2 \right) \nonumber\\
&&- \frac{m_s\left[\langle \bar{q}g_s\sigma Gq\rangle+\langle \bar{s}g_s\sigma Gs\rangle\right]}{6144\pi^6}\int dydz  \, (1-y-z)^2 \,s\left(s-\overline{m}_c^2 \right) \nonumber\\
&&+ \frac{m_s\left[19\langle \bar{q}g_s\sigma Gq\rangle+3\langle \bar{s}g_s\sigma Gs\rangle\right]}{6144\pi^6}\int dydz  \, (y+z)(1-y-z) \left(s-\overline{m}_c^2 \right)\left(2s-\overline{m}_c^2 \right) \nonumber\\
&&+\frac{ m_s m_c\langle\bar{s}s\rangle\left[5 \langle\bar{q}q\rangle-\langle\bar{s}s\rangle\right]}{36\pi^4}\int dydz \,   \left(s-\overline{m}_c^2 \right) \nonumber\\
&&+\frac{ m_s m_c\langle\bar{s}s\rangle\left[ \langle\bar{q}q\rangle+\langle\bar{s}s\rangle\right]}{36\pi^4}\int dydz \,  (1-y-z)\left(s-\overline{m}_c^2 \right) \nonumber\\
&&-\frac{m_s m_c\left[89\langle\bar{s}s\rangle\langle\bar{s}g_s\sigma Gs\rangle+57\langle\bar{s}s\rangle\langle\bar{q}g_s\sigma Gq\rangle+32\langle\bar{q}q\rangle\langle\bar{s}g_s\sigma Gs\rangle\right]}{6912\pi^4}\int dydz    \nonumber\\
&&-\frac{m_s m_c \left[231\langle\bar{s}s\rangle\langle\bar{q}g_s\sigma Gq\rangle-89\langle\bar{s}s\rangle\langle\bar{s}g_s\sigma Gs\rangle+256\langle\bar{q}q\rangle\langle\bar{s}g_s\sigma Gs\rangle\right]}{6912\pi^4}\int dy   \nonumber\\
&&   +\frac{m_s\langle\bar{q}q\rangle\langle\bar{s}s\rangle^2}{54\pi^2}\int dy \,  \left[ 1+\frac{s}{2}\delta\left(s-\widetilde{m}_c^2 \right)\right]  \nonumber\\
&&+\frac{9m_s m_c\langle\bar{q}g_s\sigma Gq\rangle \langle\bar{s}g_s\sigma Gs\rangle}{864\pi^4}\int dy \left(1+\frac{8s}{9T^2}\right) \delta\left(s-\widetilde{m}_c^2 \right) \nonumber\\
&&-\frac{m_s m_c\langle\bar{s}g_s\sigma Gs\rangle \left[3\langle \bar{q}g_s\sigma Gq\rangle+19\langle \bar{s}g_s\sigma Gs\rangle\right]}{13824\pi^4 T^2}\int dy    \,s\,  \delta\left(s-\widetilde{m}_c^2 \right) \, ,
\end{eqnarray}

\begin{eqnarray}
\rho^{11\frac{1}{2},1}_{uus}(s)&=&\frac{1}{61440\pi^8}\int dydz \, yz(1-y-z)^4\left(s-\overline{m}_c^2\right)^4\left(8s-3\overline{m}_c^2 \right)  \nonumber\\
&&-\frac{1}{307200\pi^8}\int dydz \, yz(1-y-z)^5\left(s-\overline{m}_c^2\right)^4\left(9s-4\overline{m}_c^2 \right)  \nonumber\\
&&- \frac{m_c\left[2\langle \bar{q}q\rangle+\langle \bar{s}s\rangle\right]}{2304\pi^6}\int dydz \, (y+z)(1-y-z)^2\left(s-\overline{m}_c^2\right)^3  \nonumber \\
&&+ \frac{m_c\left[2\langle \bar{q}q\rangle+\langle \bar{s}s\rangle\right]}{6912\pi^6}\int dydz \, (y+z)(1-y-z)^3\left(s-\overline{m}_c^2\right)^3  \nonumber \\
&&+ \frac{11m_c\left[2\langle \bar{q}g_s\sigma Gq\rangle+\langle \bar{s}g_s\sigma Gs\rangle\right]}{12288\pi^6}\int dydz  \, (y+z)(1-y-z) \left(s-\overline{m}_c^2 \right)^2 \nonumber\\
&&- \frac{11m_c\left[2\langle \bar{q}g_s\sigma Gq\rangle+\langle \bar{s}g_s\sigma Gs\rangle\right]}{24576\pi^6}\int dydz  \, (y+z)(1-y-z)^2 \left(s-\overline{m}_c^2 \right)^2 \nonumber\\
&&+\frac{\langle\bar{q}q\rangle\left[\langle\bar{q}q\rangle+2\langle\bar{s}s\rangle \right]}{72\pi^4}\int dydz \,  yz(1-y-z)\left(s-\overline{m}_c^2 \right)\left(5s-3\overline{m}_c^2 \right) \nonumber\\
&&-\frac{19\left[\langle\bar{q}q\rangle\langle\bar{q}g_s\sigma Gq\rangle+\langle\bar{q}q\rangle\langle\bar{s}g_s\sigma Gs\rangle+\langle\bar{s}s\rangle\langle\bar{q}g_s\sigma Gq\rangle\right]}{1152\pi^4}\int dydz \,yz \left(4s-3\overline{m}_c^2\right)\nonumber\\
&&-\frac{m_c\langle\bar{q}q\rangle^2\langle\bar{s}s\rangle}{36\pi^2}\int dy \nonumber\\
&&+\frac{11\langle\bar{q}g_s\sigma Gq\rangle \left[ \langle\bar{q}g_s\sigma Gq\rangle+2\langle\bar{s}g_s\sigma Gs\rangle\right]}{1536\pi^4}\int dy \, y(1-y)\left[ 1+\frac{s}{3}\delta\left(s-\widetilde{m}_c^2 \right)\right] \nonumber\\
&&+\frac{m_s m_c}{36864\pi^8}\int dydz \, (y+z)(1-y-z)^3\left(s-\overline{m}_c^2\right)^4  \nonumber\\
&&-\frac{m_s m_c}{147456\pi^8}\int dydz \, (y+z)(1-y-z)^4\left(s-\overline{m}_c^2\right)^4  \nonumber\\
&&- \frac{m_s\left[4\langle \bar{q}q\rangle-3\langle \bar{s}s\rangle\right]}{384\pi^6}\int dydz \, yz(1-y-z)^2\left(s-\overline{m}_c^2\right)^2 \left(2s-\overline{m}_c^2 \right) \nonumber \\
&&- \frac{m_s\langle \bar{s}s\rangle}{1152\pi^6}\int dydz \, yz(1-y-z)^3\left(s-\overline{m}_c^2\right)^2 \left(7s-4\overline{m}_c^2 \right) \nonumber \\
&&+ \frac{m_s\langle \bar{s}g_s\sigma Gs\rangle}{384\pi^6}\int dydz  \, yz(1-y-z)^2 \left(s-\overline{m}_c^2 \right)\left(3s-2\overline{m}_c^2 \right) \nonumber\\
&&+ \frac{m_s\left[19\langle \bar{q}g_s\sigma Gq\rangle-8\langle \bar{s}g_s\sigma Gs\rangle\right]}{3072\pi^6}\int dydz  \, yz(1-y-z) \left(s-\overline{m}_c^2 \right)\left(5s-3\overline{m}_c^2 \right) \nonumber\\
&&+\frac{ m_s m_c\langle\bar{q}q\rangle \left[ 3\langle\bar{q}q\rangle-\langle\bar{s}s\rangle\right]}{144\pi^4}\int dydz \,  (y+z)\left(s-\overline{m}_c^2 \right) \nonumber\\
&&+\frac{ m_s m_c\langle\bar{q}q\rangle\langle\bar{s}s\rangle }{144\pi^4}\int dydz \,  (y+z)(1-y-z)\left(s-\overline{m}_c^2 \right) \nonumber\\
&&-\frac{m_s m_c\left[32\langle\bar{q}q\rangle\langle\bar{s}g_s\sigma Gs\rangle+57\langle\bar{s}s\rangle\langle\bar{q}g_s\sigma Gq\rangle\right]}{27648\pi^4}\int dydz  \,  \left(y+z\right) \nonumber\\
&&-\frac{m_s m_c \left[288\langle\bar{q}q\rangle\langle\bar{q}g_s\sigma Gq\rangle-57\langle\bar{s}s\rangle\langle\bar{q}g_s\sigma Gq\rangle-32\langle\bar{q}q\rangle\langle\bar{s}g_s\sigma Gs\rangle\right]}{27648\pi^4}\int dy   \nonumber\\
&&+\frac{m_s m_c\langle\bar{q}q\rangle\langle\bar{q}g_s\sigma Gq\rangle}{2304\pi^4}\int dydz  \left(\frac{y}{z}+\frac{z}{y}\right)\nonumber
\end{eqnarray}
\begin{eqnarray}
&&   +\frac{m_s\langle\bar{q}q\rangle^2\langle\bar{s}s\rangle}{18\pi^2}\int dy \,y(1-y) \left[ 1+\frac{s}{3}\delta\left(s-\widetilde{m}_c^2 \right)\right]  \nonumber\\
&&-\frac{m_s m_c\langle\bar{q}g_s\sigma Gq\rangle^2}{9216\pi^4}\int dy   \left(\frac{1-y}{y}+\frac{y}{1-y}\right)\delta \left(s-\widetilde{m}_c^2 \right)\nonumber\\
&&+\frac{m_s m_c\langle\bar{q}g_s\sigma Gq\rangle \left[ 144\langle\bar{q}g_s\sigma Gq\rangle-19\langle\bar{s}g_s\sigma Gs\rangle\right]}{55296\pi^4}\int dy   \left( 1+\frac{s}{T^2}\right) \delta\left(s-\widetilde{m}_c^2 \right) \, ,
\end{eqnarray}

\begin{eqnarray}
\widetilde{\rho}^{11\frac{1}{2},0}_{uus}(s)&=&\frac{1}{245760\pi^8}\int dydz \, (y+z)(1-y-z)^4\left(s-\overline{m}_c^2\right)^4\left(7s-2\overline{m}_c^2 \right)  \nonumber\\
&&-\frac{1}{1228800\pi^8}\int dydz \, (y+z)(1-y-z)^5\left(s-\overline{m}_c^2\right)^4\left(8s-3\overline{m}_c^2 \right)  \nonumber\\
&&- \frac{m_c\left[2\langle \bar{q}q\rangle+\langle \bar{s}s\rangle\right]}{576\pi^6}\int dydz \,  (1-y-z)^2\left(s-\overline{m}_c^2\right)^3  \nonumber \\
&&+ \frac{m_c\left[2\langle \bar{q}q\rangle+\langle \bar{s}s\rangle\right]}{1728\pi^6}\int dydz \,  (1-y-z)^3\left(s-\overline{m}_c^2\right)^3  \nonumber \\
&&+ \frac{11m_c\left[2\langle \bar{q}g_s\sigma Gq\rangle+\langle \bar{s}g_s\sigma Gs\rangle\right]}{3072\pi^6}\int dydz  \, (1-y-z) \left(s-\overline{m}_c^2 \right)^2 \nonumber\\
&&- \frac{11m_c\left[2\langle \bar{q}g_s\sigma Gq\rangle+\langle \bar{s}g_s\sigma Gs\rangle\right]}{6144\pi^6}\int dydz  \, (1-y-z)^2 \left(s-\overline{m}_c^2 \right)^2 \nonumber\\
&&+\frac{\langle\bar{q}q\rangle \left[ 2\langle\bar{q}q\rangle+\langle\bar{s}s\rangle\right]}{144\pi^4}\int dydz \,  (y+z)(1-y-z)\left(s-\overline{m}_c^2 \right)\left(2s-\overline{m}_c^2 \right) \nonumber\\
&&-\frac{19\left[\langle\bar{q}q\rangle\langle\bar{q}g_s\sigma Gq\rangle+\langle\bar{q}q\rangle\langle\bar{s}g_s\sigma Gs\rangle+\langle\bar{s}s\rangle\langle\bar{q}g_s\sigma Gq\rangle\right]}{4608\pi^4}\int dydz \,(y+z) \left(3s-2\overline{m}_c^2\right)\nonumber\\
&&+\frac{\langle\bar{q}q\rangle\langle\bar{q}g_s\sigma Gq\rangle+\langle\bar{q}q\rangle\langle\bar{s}g_s\sigma Gs\rangle+\langle\bar{s}s\rangle\langle\bar{q}g_s\sigma Gq\rangle}{2304\pi^4}\int dydz \,(1-y-z)\, s\nonumber\\
&&-\frac{m_c\langle\bar{q}q\rangle^2\langle\bar{s}s\rangle}{9\pi^2}\int dy \nonumber\\
&&+\frac{11\langle\bar{q}g_s\sigma Gq\rangle \left[\langle\bar{q}g_s\sigma Gq\rangle+2\langle\bar{s}g_s\sigma Gs\rangle \right]}{9216\pi^4}\int dy \,  \left[ 1+\frac{s}{2}\delta\left(s-\widetilde{m}_c^2 \right)\right] \nonumber\\
&&-\frac{13\langle\bar{q}g_s\sigma Gq\rangle \left[\langle\bar{q}g_s\sigma Gq\rangle+2\langle\bar{s}g_s\sigma Gs\rangle \right]}{82944\pi^4}\int dydz  \,s\,\delta\left(s-\overline{m}_c^2 \right)  \nonumber\\
&&+\frac{m_s m_c}{9216\pi^8}\int dydz \,  (1-y-z)^3\left(s-\overline{m}_c^2\right)^4  \nonumber\\
&&-\frac{m_s m_c}{36864\pi^8}\int dydz \, (1-y-z)^4\left(s-\overline{m}_c^2\right)^4  \nonumber\\
&&- \frac{m_s\left[4\langle \bar{q}q\rangle-3\langle \bar{s}s\rangle\right]}{4608\pi^6}\int dydz \, (y+z)(1-y-z)^2\left(s-\overline{m}_c^2\right)^2 \left(5s-2\overline{m}_c^2 \right) \nonumber \\
&&- \frac{m_s\langle \bar{s}s\rangle}{1536\pi^6}\int dydz \, (y+z)(1-y-z)^3\left(s-\overline{m}_c^2\right)^2 \left(2s-\overline{m}_c^2 \right) \nonumber \\
&&+ \frac{m_s\langle \bar{s}g_s\sigma Gs\rangle}{3072\pi^6}\int dydz  \, (y+z)(1-y-z)^2 \left(s-\overline{m}_c^2 \right)\left(5s-3\overline{m}_c^2 \right) \nonumber\\
&&- \frac{m_s\langle \bar{q}g_s\sigma Gq\rangle}{6144\pi^6}\int dydz  \, (1-y-z)^2 \,s\left(s-\overline{m}_c^2 \right) \nonumber\\
&&+ \frac{m_s\left[19\langle \bar{q}g_s\sigma Gq\rangle-8\langle \bar{s}g_s\sigma Gs\rangle\right]}{6144\pi^6}\int dydz  \, (y+z)(1-y-z) \left(s-\overline{m}_c^2 \right)\left(2s-\overline{m}_c^2 \right) \nonumber
\end{eqnarray}
\begin{eqnarray}
&&+\frac{ m_s m_c\langle\bar{q}q\rangle \left[3\langle\bar{q}q\rangle-\langle\bar{s}s\rangle \right]}{36\pi^4}\int dydz \,   \left(s-\overline{m}_c^2 \right) \nonumber\\
&&+\frac{ m_s m_c\langle\bar{q}q\rangle\langle\bar{s}s\rangle }{36\pi^4}\int dydz \,  (1-y-z)\left(s-\overline{m}_c^2 \right) \nonumber\\
&&-\frac{m_s m_c\left[32\langle\bar{q}q\rangle\langle\bar{s}g_s\sigma Gs\rangle+57\langle\bar{s}s\rangle\langle\bar{q}g_s\sigma Gq\rangle\right]}{6912\pi^4}\int dydz    \nonumber\\
&&-\frac{m_s m_c \left[288\langle\bar{q}q\rangle\langle\bar{q}g_s\sigma Gq\rangle-57\langle\bar{s}s\rangle\langle\bar{q}g_s\sigma Gq\rangle-32\langle\bar{q}q\rangle\langle\bar{s}g_s\sigma Gs\rangle\right]}{6912\pi^4}\int dy   \nonumber\\
&&   +\frac{m_s\langle\bar{q}q\rangle^2\langle\bar{s}s\rangle}{108\pi^2}\int dy \,  \left[ 1+\frac{s}{2}\delta\left(s-\widetilde{m}_c^2 \right)\right]  \nonumber\\
&&+\frac{m_s m_c\langle\bar{q}g_s\sigma Gq\rangle^2}{192\pi^4}\int dy \left(1+\frac{s}{T^2}\right) \delta\left(s-\widetilde{m}_c^2 \right) \nonumber\\
&&-\frac{19m_s m_c\langle\bar{q}g_s\sigma Gq\rangle\langle\bar{s}g_s\sigma Gs\rangle }{13824\pi^4 T^2}\int dy    \,s\,  \delta\left(s-\widetilde{m}_c^2 \right) \, ,
\end{eqnarray}

\begin{eqnarray}
\rho^{11\frac{1}{2},1}_{uuu}(s)&=&\rho^{11\frac{1}{2},1}_{sss}(s)\mid_{m_s \to 0,\,\,\langle\bar{s}s\rangle\to\langle\bar{q}q\rangle,\,\,\langle\bar{s}g_s\sigma Gs\rangle\to\langle\bar{q}g_s\sigma Gq\rangle} \, ,  \\
\widetilde{\rho}^{11\frac{1}{2},0}_{uuu}(s)&=&\widetilde{\rho}^{11\frac{1}{2},0}_{sss}(s)\mid_{m_s \to 0,\,\,\langle\bar{s}s\rangle\to\langle\bar{q}q\rangle,\,\,\langle\bar{s}g_s\sigma Gs\rangle\to\langle\bar{q}g_s\sigma Gq\rangle} \, ,
\end{eqnarray}

\begin{eqnarray}
\rho^{10\frac{1}{2},1}_{sss}(s)&=&\frac{1}{245760\pi^8}\int dydz \, yz(1-y-z)^4\left(s-\overline{m}_c^2\right)^4\left(8s-3\overline{m}_c^2 \right)  \nonumber\\
&&-\frac{1}{1228800\pi^8}\int dydz \, yz(1-y-z)^5\left(s-\overline{m}_c^2\right)^4\left(9s-4\overline{m}_c^2 \right)  \nonumber\\
&&- \frac{m_c\langle \bar{s}s\rangle}{1536\pi^6}\int dydz \, (y+z)(1-y-z)^2\left(s-\overline{m}_c^2\right)^3  \nonumber \\
&&+ \frac{m_c\langle \bar{s}s\rangle}{4608\pi^6}\int dydz \, (y+z)(1-y-z)^3\left(s-\overline{m}_c^2\right)^3  \nonumber \\
&&+\frac{11m_c\langle \bar{s}g_s\sigma Gs\rangle}{8192\pi^6}\int dydz  \, (y+z)(1-y-z) \left(s-\overline{m}_c^2 \right)^2 \nonumber\\
&&-\frac{11m_c\langle \bar{s}g_s\sigma Gs\rangle}{16384\pi^6}\int dydz  \, (y+z)(1-y-z)^2 \left(s-\overline{m}_c^2 \right)^2 \nonumber\\
&&- \frac{7m_c\langle \bar{s}g_s\sigma Gs\rangle}{16384\pi^6}\int dydz  \, \left(\frac{y}{z}+\frac{z}{y}\right)(1-y-z)^2 \left(s-\overline{m}_c^2 \right)^2 \nonumber\\
&&+ \frac{m_c\langle \bar{s}g_s\sigma Gs\rangle}{8192\pi^6}\int dydz  \, \left(\frac{y}{z}+\frac{z}{y}\right)(1-y-z)^3 \left(s-\overline{m}_c^2 \right)^2 \nonumber\\
&&+\frac{\langle\bar{s}s\rangle^2}{96\pi^4}\int dydz \,  yz(1-y-z)\left(s-\overline{m}_c^2 \right)\left(5s-3\overline{m}_c^2 \right) \nonumber\\
&&-\frac{19\langle\bar{s}s\rangle\langle\bar{s}g_s\sigma Gs\rangle}{1536\pi^4}\int dydz \,yz \left(4s-3\overline{m}_c^2\right)\nonumber\\
&&-\frac{\langle\bar{s}s\rangle\langle\bar{s}g_s\sigma Gs\rangle}{3072\pi^4}\int dydz \,(y+z)(1-y-z) \left(2s-\overline{m}_c^2\right)\nonumber\\
&&-\frac{m_c\langle\bar{s}s\rangle^3}{72\pi^2}\int dy \nonumber\\
&&+\frac{13\langle\bar{s}g_s\sigma Gs\rangle^2}{110592\pi^4}\int dydz  \,(y+z) \left[1+s\,\delta \left(s-\overline{m}_c^2 \right)\right]\nonumber
\end{eqnarray}
\begin{eqnarray}
&&+\frac{11\langle\bar{s}g_s\sigma Gs\rangle^2}{2048\pi^4}\int dy   \,y(1-y) \left[1+\frac{s}{3}\delta \left(s-\widetilde{m}_c^2 \right)\right]\nonumber\\
&&+\frac{m_s m_c}{24576\pi^8}\int dydz \, (y+z)(1-y-z)^3\left(s-\overline{m}_c^2\right)^4  \nonumber\\
&&-\frac{m_s m_c}{98304\pi^8}\int dydz \, (y+z)(1-y-z)^4\left(s-\overline{m}_c^2\right)^4  \nonumber\\
&&- \frac{m_s\langle \bar{s}s\rangle}{512\pi^6}\int dydz \, yz(1-y-z)^2\left(s-\overline{m}_c^2\right)^2 \left(2s-\overline{m}_c^2 \right) \nonumber \\
&&- \frac{m_s\langle \bar{s}s\rangle}{1536\pi^6}\int dydz \, yz(1-y-z)^3\left(s-\overline{m}_c^2\right)^2 \left(7s-4\overline{m}_c^2 \right) \nonumber \\
&&+ \frac{11m_s\langle \bar{s}g_s\sigma Gs\rangle}{4096\pi^6}\int dydz  \, yz(1-y-z) \left(s-\overline{m}_c^2 \right)\left(5s-3\overline{m}_c^2 \right) \nonumber\\
&&+ \frac{m_s\langle \bar{s}g_s\sigma Gs\rangle}{512\pi^6}\int dydz  \, yz(1-y-z)^2 \left(s-\overline{m}_c^2 \right)\left(3s-2\overline{m}_c^2 \right) \nonumber\\
&&+ \frac{m_s\langle \bar{s}g_s\sigma Gs\rangle}{16384\pi^6}\int dydz  \, (y+z)(1-y-z)^2 \left(s-\overline{m}_c^2 \right)\left(3s-\overline{m}_c^2 \right) \nonumber\\
&&+\frac{ m_s m_c\langle\bar{s}s\rangle^2}{48\pi^4}\int dydz \,  (y+z)\left(s-\overline{m}_c^2 \right) \nonumber\\
&&+\frac{ m_s m_c\langle\bar{s}s\rangle^2}{96\pi^4}\int dydz \,  (y+z)(1-y-z)\left(s-\overline{m}_c^2 \right) \nonumber\\
&&-\frac{89m_s m_c\langle\bar{s}s\rangle\langle\bar{s}g_s\sigma Gs\rangle}{18432\pi^4}\int dydz  \,  \left(y+z\right) \nonumber\\
&&-\frac{199m_s m_c \langle\bar{s}s\rangle\langle\bar{s}g_s\sigma Gs\rangle}{18432\pi^4}\int dy   \nonumber\\
&&+ \frac{3m_s m_c \langle  \bar{s}s\rangle\langle \bar{s}g_s\sigma Gs\rangle}{1024\pi^4}\int dydz  \, \left(\frac{y}{z}+\frac{z}{y}\right) \nonumber\\
&&+\frac{m_s m_c\langle\bar{s}s\rangle\langle\bar{s}g_s\sigma Gs\rangle}{512\pi^4}\int dydz  \left(\frac{y}{z}+\frac{z}{y}\right)(1-y-z)\nonumber\\
&&   +\frac{m_s\langle\bar{s}s\rangle^3}{24\pi^2}\int dy \,y(1-y) \left[ 1+\frac{s}{3}\delta\left(s-\widetilde{m}_c^2 \right)\right]  \nonumber\\
&&+\frac{m_s m_c\langle\bar{s}g_s\sigma Gs\rangle^2}{4608\pi^4}\int dy \left(16+\frac{7s}{T^2}\right) \delta\left(s-\widetilde{m}_c^2 \right) \nonumber\\
&&-\frac{m_s m_c\langle\bar{s}g_s\sigma Gs\rangle^2}{3072\pi^4}\int dydz   \left(\frac{z}{y}+\frac{y}{z}\right)\delta \left(s-\overline{m}_c^2 \right)\nonumber\\
&&-\frac{17m_s m_c\langle\bar{s}g_s\sigma Gs\rangle^2}{18432\pi^4}\int dy   \left(  \frac{1-y}{y}+\frac{y}{1-y}\right) \delta\left(s-\widetilde{m}_c^2 \right)\nonumber\\
&&- \frac{m_s m_c \langle \bar{s}g_s\sigma Gs\rangle^2}{12288\pi^4}\int dy \,\left(1+\frac{s}{T^2}\right)\delta\left(s-\widetilde{m}_c^2 \right)       \, ,
\end{eqnarray}

\begin{eqnarray}
\widetilde{\rho}^{10\frac{1}{2},0}_{sss}(s)&=&\frac{1}{491520\pi^8}\int dydz \, (y+z)(1-y-z)^4\left(s-\overline{m}_c^2\right)^4\left(7s-2\overline{m}_c^2 \right)  \nonumber\\
&&-\frac{1}{2457600\pi^8}\int dydz \, (y+z)(1-y-z)^5\left(s-\overline{m}_c^2\right)^4\left(8s-3\overline{m}_c^2 \right)  \nonumber\\
&&- \frac{m_c\langle \bar{s}s\rangle}{768\pi^6}\int dydz \, (1-y-z)^2\left(s-\overline{m}_c^2\right)^3  \nonumber
\end{eqnarray}
\begin{eqnarray}
&&+ \frac{m_c\langle \bar{s}s\rangle}{2304\pi^6}\int dydz \,  (1-y-z)^3\left(s-\overline{m}_c^2\right)^3  \nonumber \\
&&+\frac{11m_c\langle \bar{s}g_s\sigma Gs\rangle}{4096\pi^6}\int dydz  \, (1-y-z) \left(s-\overline{m}_c^2 \right)^2 \nonumber\\
&&-\frac{11m_c\langle \bar{s}g_s\sigma Gs\rangle}{8192\pi^6}\int dydz  \, (1-y-z)^2 \left(s-\overline{m}_c^2 \right)^2 \nonumber\\
&&- \frac{7m_c\langle \bar{s}g_s\sigma Gs\rangle}{16384\pi^6}\int dydz  \, \left(\frac{1}{y}+\frac{1}{z} \right)(1-y-z)^2 \left(s-\overline{m}_c^2 \right)^2 \nonumber\\
&&+ \frac{m_c\langle \bar{s}g_s\sigma Gs\rangle}{8192\pi^6}\int dydz  \, \left(\frac{1}{y}+\frac{1}{z}\right)(1-y-z)^3 \left(s-\overline{m}_c^2 \right)^2 \nonumber\\
&&+\frac{\langle\bar{s}s\rangle^2}{96\pi^4}\int dydz \,  (y+z)(1-y-z)\left(s-\overline{m}_c^2 \right)\left(2s-\overline{m}_c^2 \right) \nonumber\\
&&-\frac{19\langle\bar{s}s\rangle\langle\bar{s}g_s\sigma Gs\rangle}{3072\pi^4}\int dydz \,(y+z) \left(3s-2\overline{m}_c^2\right)\nonumber\\
&&-\frac{\langle\bar{s}s\rangle\langle\bar{s}g_s\sigma Gs\rangle}{1536\pi^4}\int dydz \, (1-y-z) s\nonumber\\
&&-\frac{m_c\langle\bar{s}s\rangle^3}{36\pi^2}\int dy \nonumber\\
&&+\frac{11\langle\bar{s}g_s\sigma Gs\rangle^2}{6144\pi^4}\int dy \, \left[ 1+\frac{s}{2}\delta\left(s-\widetilde{m}_c^2 \right)\right] \nonumber\\
&&+\frac{13\langle\bar{s}g_s\sigma Gs\rangle^2}{55296\pi^4}\int dydz  \, s\,\delta \left(s-\overline{m}_c^2 \right) \nonumber\\
&&+\frac{m_s m_c}{12288\pi^8}\int dydz \, (1-y-z)^3\left(s-\overline{m}_c^2\right)^4  \nonumber\\
&&-\frac{m_s m_c}{49152\pi^8}\int dydz \, (1-y-z)^4\left(s-\overline{m}_c^2\right)^4  \nonumber\\
&&- \frac{m_s\langle \bar{s}s\rangle}{3072\pi^6}\int dydz \, (y+z)(1-y-z)^2\left(s-\overline{m}_c^2\right)^2 \left(5s-2\overline{m}_c^2 \right) \nonumber \\
&&- \frac{m_s\langle \bar{s}s\rangle}{1024\pi^6}\int dydz \, (y+z)(1-y-z)^3\left(s-\overline{m}_c^2\right)^2 \left(2s-\overline{m}_c^2 \right) \nonumber\\
&&+ \frac{11m_s\langle \bar{s}g_s\sigma Gs\rangle}{4096\pi^6}\int dydz  \, (y+z)(1-y-z) \left(s-\overline{m}_c^2 \right)\left(2s-\overline{m}_c^2 \right) \nonumber\\
&&+ \frac{m_s\langle \bar{s}g_s\sigma Gs\rangle}{2048\pi^6}\int dydz  \, (y+z)(1-y-z)^2 \left(s-\overline{m}_c^2 \right)\left(5s-3\overline{m}_c^2 \right) \nonumber\\
&&+ \frac{m_s\langle \bar{s}g_s\sigma Gs\rangle}{4096\pi^6}\int dydz  \,  (1-y-z)^2 \,s\left(s-\overline{m}_c^2 \right) \nonumber\\
&&+\frac{ m_s m_c\langle\bar{s}s\rangle^2}{24\pi^4}\int dydz \,   \left(s-\overline{m}_c^2 \right) \nonumber\\
&&+\frac{ m_s m_c\langle\bar{s}s\rangle^2}{48\pi^4}\int dydz \,   (1-y-z)\left(s-\overline{m}_c^2 \right) \nonumber\\
&&-\frac{89m_s m_c\langle\bar{s}s\rangle\langle\bar{s}g_s\sigma Gs\rangle}{9216\pi^4}\int dydz    \nonumber\\
&&-\frac{199m_s m_c \langle\bar{s}s\rangle\langle\bar{s}g_s\sigma Gs\rangle}{9216\pi^4}\int dy   \nonumber\\
&&+\frac{3m_s m_c\langle\bar{s}s\rangle\langle\bar{s}g_s\sigma Gs\rangle}{1024\pi^4}\int dydz  \left(\frac{1}{y}+\frac{1}{z}\right) \nonumber
\end{eqnarray}
\begin{eqnarray}
&&+\frac{m_s m_c\langle\bar{s}s\rangle\langle\bar{s}g_s\sigma Gs\rangle}{512\pi^4}\int dydz  \left(\frac{1}{y}+\frac{1}{z}\right)(1-y-z)\nonumber\\
&&   +\frac{m_s\langle\bar{s}s\rangle^3}{72\pi^2}\int dy \,  \left[ 1+\frac{s}{2}\delta\left(s-\widetilde{m}_c^2 \right)\right]  \nonumber\\
&&+\frac{m_s m_c\langle\bar{s}g_s\sigma Gs\rangle^2}{256\pi^4}\int dy \left(1+\frac{7s}{9T^2}\right) \delta\left(s-\widetilde{m}_c^2 \right) \nonumber\\
&&-\frac{m_s m_c\langle\bar{s}g_s\sigma Gs\rangle^2}{3072\pi^4}\int dydz   \left(\frac{1}{y}+\frac{1}{z}\right)\delta \left(s-\overline{m}_c^2 \right)\nonumber\\
&&-\frac{17m_s m_c\langle\bar{s}g_s\sigma Gs\rangle^2}{18432\pi^4}\int dy   \left(  \frac{1}{y}+\frac{1}{1-y}\right) \delta\left(s-\widetilde{m}_c^2 \right)\nonumber\\
&&- \frac{m_s m_c \langle \bar{s}g_s\sigma Gs\rangle^2}{6144\pi^4T^2}\int dy \,s\,\delta\left(s-\widetilde{m}_c^2 \right)       \, ,
\end{eqnarray}

\begin{eqnarray}
\rho^{10\frac{1}{2},1}_{uss}(s)&=&\frac{1}{245760\pi^8}\int dydz \, yz(1-y-z)^4\left(s-\overline{m}_c^2\right)^4\left(8s-3\overline{m}_c^2 \right)  \nonumber\\
&&-\frac{1}{1228800\pi^8}\int dydz \, yz(1-y-z)^5\left(s-\overline{m}_c^2\right)^4\left(9s-4\overline{m}_c^2 \right)  \nonumber\\
&&- \frac{m_c\left[\langle \bar{q}q\rangle+2\langle \bar{s}s\rangle\right]}{4608\pi^6}\int dydz \, (y+z)(1-y-z)^2\left(s-\overline{m}_c^2\right)^3  \nonumber \\
&&+ \frac{m_c\left[\langle \bar{q}q\rangle+2\langle \bar{s}s\rangle\right]}{13824\pi^6}\int dydz \, (y+z)(1-y-z)^3\left(s-\overline{m}_c^2\right)^3  \nonumber \\
&&+ \frac{11m_c\left[\langle \bar{q}g_s\sigma Gq\rangle+2\langle \bar{s}g_s\sigma Gs\rangle\right]}{24576\pi^6}\int dydz  \, (y+z)(1-y-z) \left(s-\overline{m}_c^2 \right)^2 \nonumber\\
&&+\frac{11m_c\left[\langle \bar{q}g_s\sigma Gq\rangle+2\langle \bar{s}g_s\sigma Gs\rangle\right]}{49152\pi^6}\int dydz  \, (y+z)(1-y-z)^2 \left(s-\overline{m}_c^2 \right)^2 \nonumber\\
&&- \frac{7m_c\left[\langle \bar{q}g_s\sigma Gq\rangle+2\langle \bar{s}g_s\sigma Gs\rangle\right]}{49152\pi^6}\int dydz  \, \left(\frac{y}{z}+\frac{z}{y} \right)(1-y-z)^2 \left(s-\overline{m}_c^2 \right)^2 \nonumber\\
&&+ \frac{m_c\left[\langle \bar{q}g_s\sigma Gq\rangle+2\langle \bar{s}g_s\sigma Gs\rangle\right]}{24576\pi^6}\int dydz  \, \left(\frac{y}{z}+\frac{z}{y}\right)(1-y-z)^3 \left(s-\overline{m}_c^2 \right)^2 \nonumber\\
&&+\frac{\langle\bar{s}s\rangle \left[2\langle\bar{q}q\rangle+\langle\bar{s}s\rangle \right]}{288\pi^4}\int dydz \,  yz(1-y-z)\left(s-\overline{m}_c^2 \right)\left(5s-3\overline{m}_c^2 \right) \nonumber\\
&&-\frac{19\left[\langle\bar{s}s\rangle\langle\bar{s}g_s\sigma Gs\rangle+\langle\bar{q}q\rangle\langle\bar{s}g_s\sigma Gs\rangle+\langle\bar{s}s\rangle\langle\bar{q}g_s\sigma Gq\rangle\right]}{4608\pi^4}\int dydz \,yz \left(4s-3\overline{m}_c^2\right)\nonumber\\
&&-\frac{\langle\bar{s}s\rangle\langle\bar{s}g_s\sigma Gs\rangle+\langle\bar{q}q\rangle\langle\bar{s}g_s\sigma Gs\rangle+\langle\bar{s}s\rangle\langle\bar{q}g_s\sigma Gq\rangle}{9216\pi^4}\int dydz \,(y+z)(1-y-z) \left(2s-\overline{m}_c^2\right)\nonumber\\
&&-\frac{m_c\langle\bar{q}q\rangle\langle\bar{s}s\rangle^2}{72\pi^2}\int dy \nonumber\\
&&+\frac{11\langle\bar{s}g_s\sigma Gs\rangle\left[2\langle\bar{q}g_s\sigma Gq\rangle+\langle\bar{s}g_s\sigma Gs\rangle \right]}{6144\pi^4}\int dy   \,y(1-y) \left[1+\frac{s}{3}\delta \left(s-\widetilde{m}_c^2 \right)\right]\nonumber\\
&&+\frac{13\langle\bar{s}g_s\sigma Gs\rangle \left[2\langle\bar{q}g_s\sigma Gq\rangle+\langle\bar{s}g_s\sigma Gs\rangle \right]}{331776\pi^4}\int dydz  \,(y+z) \left[1+s\,\delta \left(s-\overline{m}_c^2 \right)\right]\nonumber\\
&&+\frac{m_s m_c}{36864\pi^8}\int dydz \, (y+z)(1-y-z)^3\left(s-\overline{m}_c^2\right)^4  \nonumber\\
&&-\frac{m_s m_c}{147456\pi^8}\int dydz \, (y+z)(1-y-z)^4\left(s-\overline{m}_c^2\right)^4  \nonumber
\end{eqnarray}
\begin{eqnarray}
&&- \frac{m_s\left[2\langle \bar{q}q\rangle-\langle \bar{s}s\rangle\right]}{768\pi^6}\int dydz \, yz(1-y-z)^2\left(s-\overline{m}_c^2\right)^2 \left(2s-\overline{m}_c^2 \right) \nonumber \\
&&- \frac{m_s\langle \bar{s}s\rangle}{2304\pi^6}\int dydz \, yz(1-y-z)^3\left(s-\overline{m}_c^2\right)^2 \left(7s-4\overline{m}_c^2 \right) \nonumber \\
&&+ \frac{m_s\langle \bar{s}g_s\sigma Gs\rangle}{768\pi^6}\int dydz  \, yz(1-y-z)^2 \left(s-\overline{m}_c^2 \right)\left(3s-2\overline{m}_c^2 \right) \nonumber\\
&&+ \frac{m_s\left[19\langle \bar{q}g_s\sigma Gq\rangle+3\langle \bar{s}g_s\sigma Gs\rangle\right]}{12288\pi^6}\int dydz  \, yz(1-y-z) \left(s-\overline{m}_c^2 \right)\left(5s-3\overline{m}_c^2 \right) \nonumber\\
&&+ \frac{m_s\left[\langle \bar{q}g_s\sigma Gq\rangle+\langle \bar{s}g_s\sigma Gs\rangle\right]}{49152\pi^6}\int dydz  \, (y+z)(1-y-z)^2 \left(s-\overline{m}_c^2 \right)\left(3s-\overline{m}_c^2 \right) \nonumber\\
&&+\frac{ m_s m_c\langle\bar{s}s\rangle \left[5\langle\bar{q}q\rangle-\langle\bar{s}s\rangle \right]}{288\pi^4}\int dydz \,  (y+z)\left(s-\overline{m}_c^2 \right) \nonumber\\
&&+\frac{ m_s m_c\langle\bar{s}s\rangle \left[\langle\bar{q}q\rangle+\langle\bar{s}s\rangle \right]}{288\pi^4}\int dydz \,  (y+z)(1-y-z)\left(s-\overline{m}_c^2 \right) \nonumber\\
&&-\frac{m_s m_c\left[89\langle\bar{s}s\rangle\langle\bar{s}g_s\sigma Gs\rangle+32\langle\bar{q}q\rangle\langle\bar{s}g_s\sigma Gs\rangle+57\langle\bar{s}s\rangle\langle\bar{q}g_s\sigma Gq\rangle\right]}{55296\pi^4}\int dydz  \,  \left(y+z\right) \nonumber\\
&&-\frac{m_s m_c \left[231\langle\bar{s}s\rangle\langle\bar{q}g_s\sigma Gq\rangle+256\langle\bar{q}q\rangle\langle\bar{s}g_s\sigma Gs\rangle-89\langle\bar{s}s\rangle\langle\bar{s}g_s\sigma Gs\rangle\right]}{55296\pi^4}\int dy   \nonumber\\
&&+\frac{m_s m_c\langle\bar{s}s\rangle\left[\langle\bar{q}g_s\sigma Gq\rangle+\langle\bar{s}g_s\sigma Gs\rangle\right]}{1536\pi^4}\int dydz  \left(\frac{y}{z}+\frac{z}{y}\right)(1-y-z)\nonumber\\
&&+\frac{m_s m_c\left[9\langle\bar{s}s\rangle\langle\bar{q}g_s\sigma Gq\rangle-7\langle\bar{s}s\rangle\langle\bar{s}g_s\sigma Gs\rangle+16\langle\bar{q}q\rangle\langle\bar{s}g_s\sigma Gs\rangle\right]}{9216\pi^4}\int dydz  \left(\frac{y}{z}+\frac{z}{y}\right) \nonumber\\
&&   +\frac{m_s\langle\bar{q}q\rangle\langle\bar{s}s\rangle^2}{36\pi^2}\int dy \,y(1-y) \left[ 1+\frac{s}{3}\delta\left(s-\widetilde{m}_c^2 \right)\right]  \nonumber\\
&&-\frac{m_s m_c\langle\bar{s}g_s\sigma Gs\rangle \left[\langle\bar{q}g_s\sigma Gq\rangle+\langle\bar{s}g_s\sigma Gs\rangle \right]}{9216\pi^4}\int dydz   \left(\frac{z}{y}+\frac{y}{z}\right)\delta \left(s-\overline{m}_c^2 \right)\nonumber\\
&&-\frac{m_s m_c\langle\bar{s}g_s\sigma Gs\rangle\left[41\langle \bar{q}g_s\sigma Gq\rangle-7\langle \bar{s}g_s\sigma Gs\rangle\right]}{55296\pi^4}\int dy   \left(  \frac{1-y}{y}+\frac{y}{1-y}\right) \delta\left(s-\widetilde{m}_c^2 \right)\nonumber\\
&&-\frac{m_s m_c\langle\bar{s}g_s\sigma Gs\rangle \left[19\langle\bar{s}g_s\sigma Gs\rangle-125\langle\bar{q}g_s\sigma Gq\rangle \right]}{55296\pi^4}\int dy \left(1+\frac{s}{2T^2}\right) \delta\left(s-\widetilde{m}_c^2 \right) \nonumber\\
&&+ \frac{19m_s m_c \langle \bar{s}g_s\sigma Gs\rangle\left[\langle \bar{q}g_s\sigma Gq\rangle+\langle \bar{s}g_s\sigma Gs\rangle\right]}{110592\pi^4}\int dy \,\delta\left(s-\widetilde{m}_c^2 \right)    \, ,
\end{eqnarray}

\begin{eqnarray}
\widetilde{\rho}^{10\frac{1}{2},0}_{uss}(s)&=&\frac{1}{491520\pi^8}\int dydz \, (y+z)(1-y-z)^4\left(s-\overline{m}_c^2\right)^4\left(7s-2\overline{m}_c^2 \right)  \nonumber\\
&&-\frac{1}{2457600\pi^8}\int dydz \, (y+z)(1-y-z)^5\left(s-\overline{m}_c^2\right)^4\left(8s-3\overline{m}_c^2 \right)  \nonumber\\
&&- \frac{m_c\left[\langle \bar{q}q\rangle+2\langle \bar{s}s\rangle\right]}{2304\pi^6}\int dydz \, (1-y-z)^2\left(s-\overline{m}_c^2\right)^3  \nonumber \\
&&+ \frac{m_c\left[\langle \bar{q}q\rangle+2\langle \bar{s}s\rangle\right]}{6912\pi^6}\int dydz \,  (1-y-z)^3\left(s-\overline{m}_c^2\right)^3  \nonumber \\
&&+ \frac{11m_c\left[\langle \bar{q}g_s\sigma Gq\rangle+2\langle \bar{s}g_s\sigma Gs\rangle\right]}{12288\pi^6}\int dydz  \,  (1-y-z) \left(s-\overline{m}_c^2 \right)^2 \nonumber\\
&&- \frac{11m_c\left[\langle \bar{q}g_s\sigma Gq\rangle+2\langle \bar{s}g_s\sigma Gs\rangle\right]}{24576\pi^6}\int dydz  \,  (1-y-z)^2 \left(s-\overline{m}_c^2 \right)^2 \nonumber
\end{eqnarray}
\begin{eqnarray}
&&- \frac{7m_c\left[\langle \bar{q}g_s\sigma Gq\rangle+2\langle \bar{s}g_s\sigma Gs\rangle\right]}{49152\pi^6}\int dydz  \, \left(\frac{1}{y}+\frac{1}{z} \right)(1-y-z)^2 \left(s-\overline{m}_c^2 \right)^2 \nonumber\\
&&+ \frac{m_c\left[\langle \bar{q}g_s\sigma Gq\rangle+2\langle \bar{s}g_s\sigma Gs\rangle\right]}{24576\pi^6}\int dydz  \, \left(\frac{1}{y}+\frac{1}{z}\right)(1-y-z)^3 \left(s-\overline{m}_c^2 \right)^2 \nonumber\\
&&+\frac{\langle\bar{s}s\rangle\left[2\langle\bar{q}q\rangle+\langle\bar{s}s\rangle\right]}{288\pi^4}\int dydz \,  (y+z)(1-y-z)\left(s-\overline{m}_c^2 \right)\left(2s-\overline{m}_c^2 \right) \nonumber\\
&&-\frac{19\left[\langle\bar{s}s\rangle\langle\bar{s}g_s\sigma Gs\rangle+\langle\bar{q}q\rangle\langle\bar{s}g_s\sigma Gs\rangle+\langle\bar{s}s\rangle\langle\bar{q}g_s\sigma Gq\rangle\right]}{9216\pi^4}\int dydz \,(y+z) \left(3s-2\overline{m}_c^2\right)\nonumber\\
&&-\frac{\langle\bar{s}s\rangle\langle\bar{s}g_s\sigma Gs\rangle+\langle\bar{q}q\rangle\langle\bar{s}g_s\sigma Gs\rangle+\langle\bar{s}s\rangle\langle\bar{q}g_s\sigma Gq\rangle}{4608\pi^4}\int dydz \, (1-y-z) s\nonumber\\
&&-\frac{m_c\langle\bar{q}q\rangle\langle\bar{s}s\rangle^2}{36\pi^2}\int dy \nonumber\\
&&+\frac{13\langle\bar{s}g_s\sigma Gs\rangle \left[2\langle\bar{q}g_s\sigma Gq\rangle+\langle\bar{s}g_s\sigma Gs\rangle \right]}{165888\pi^4}\int dydz  \, s\,\delta \left(s-\overline{m}_c^2 \right) \nonumber\\
&&+\frac{11\langle\bar{s}g_s\sigma Gs\rangle \left[ 2\langle\bar{q}g_s\sigma Gq\rangle+\langle\bar{s}g_s\sigma Gs\rangle\right]}{18432\pi^4}\int dy   \, \left[1+\frac{s}{2}\delta \left(s-\widetilde{m}_c^2 \right)\right]\nonumber\\
&&+\frac{m_s m_c}{18432\pi^8}\int dydz \, (1-y-z)^3\left(s-\overline{m}_c^2\right)^4  \nonumber\\
&&-\frac{m_s m_c}{73728\pi^8}\int dydz \, (1-y-z)^4\left(s-\overline{m}_c^2\right)^4  \nonumber\\
&&- \frac{m_s\left[2\langle \bar{q}q\rangle-\langle \bar{s}s\rangle\right]}{4608\pi^6}\int dydz \, (y+z)(1-y-z)^2\left(s-\overline{m}_c^2\right)^2 \left(5s-2\overline{m}_c^2 \right) \nonumber \\
&&- \frac{m_s\langle \bar{s}s\rangle}{1536\pi^6}\int dydz \, (y+z)(1-y-z)^3\left(s-\overline{m}_c^2\right)^2 \left(2s-\overline{m}_c^2 \right) \nonumber \\
&&+ \frac{m_s\langle \bar{s}g_s\sigma Gs\rangle}{3072\pi^6}\int dydz  \, (y+z)(1-y-z)^2 \left(s-\overline{m}_c^2 \right)\left(5s-3\overline{m}_c^2 \right) \nonumber\\
&&+ \frac{m_s\left[\langle \bar{q}g_s\sigma Gq\rangle+\langle \bar{s}g_s\sigma Gs\rangle\right]}{12288\pi^6}\int dydz  \,  (1-y-z)^2 \,s\left(s-\overline{m}_c^2 \right) \nonumber\\
&&+ \frac{m_s\left[19\langle \bar{q}g_s\sigma Gq\rangle+3\langle \bar{s}g_s\sigma Gs\rangle\right]}{12288\pi^6}\int dydz  \, (y+z)(1-y-z) \left(s-\overline{m}_c^2 \right)\left(2s-\overline{m}_c^2 \right) \nonumber\\
&&+\frac{ m_s m_c\langle\bar{s}s\rangle \left[5\langle\bar{q}q\rangle -\langle\bar{s}s\rangle  \right]}{144\pi^4}\int dydz \,   \left(s-\overline{m}_c^2 \right) \nonumber\\
&&+\frac{ m_s m_c\langle\bar{s}s\rangle \left[ \langle\bar{q}q\rangle+\langle\bar{s}s\rangle\right]}{144\pi^4}\int dydz \,   (1-y-z)\left(s-\overline{m}_c^2 \right) \nonumber\\
&&-\frac{m_s m_c\left[89\langle\bar{s}s\rangle\langle\bar{s}g_s\sigma Gs\rangle+32\langle\bar{q}q\rangle\langle\bar{s}g_s\sigma Gs\rangle+57\langle\bar{s}s\rangle\langle\bar{q}g_s\sigma Gq\rangle\right]}{27648\pi^4}\int dydz    \nonumber\\
&&-\frac{m_s m_c\left[ 231\langle\bar{s}s\rangle\langle\bar{q}g_s\sigma Gq\rangle+256\langle\bar{q}q\rangle\langle\bar{s}g_s\sigma Gs\rangle-89\langle\bar{s}s\rangle\langle\bar{s}g_s\sigma Gs\rangle\right]}{27648\pi^4}\int dy   \nonumber\\
&&+\frac{m_s m_c\langle\bar{s}s\rangle \left[\langle\bar{q}g_s\sigma Gq\rangle+\langle\bar{s}g_s\sigma Gs\rangle\right]}{1536\pi^4}\int dydz  \left(\frac{1}{y}+\frac{1}{z}\right)(1-y-z)\nonumber\\
&&+\frac{m_s m_c\left[9\langle\bar{s}s\rangle\langle\bar{q}g_s\sigma Gq\rangle-7\langle\bar{s}s\rangle\langle\bar{s}g_s\sigma Gs\rangle+16\langle\bar{q}q\rangle\langle\bar{s}g_s\sigma Gs\rangle\right]}{9216\pi^4}\int dydz  \left(\frac{1}{y}+\frac{1}{z}\right) \nonumber\\
&&   +\frac{m_s\langle\bar{q}q\rangle\langle\bar{s}s\rangle^2}{108\pi^2}\int dy \,  \left[ 1+\frac{s}{2}\delta\left(s-\widetilde{m}_c^2 \right)\right]  \nonumber
\end{eqnarray}
\begin{eqnarray}
&&-\frac{m_s m_c\langle\bar{s}g_s\sigma Gs\rangle \left[\langle\bar{q}g_s\sigma Gq\rangle+\langle\bar{s}g_s\sigma Gs\rangle \right]}{9216\pi^4}\int dydz   \left(\frac{1}{y}+\frac{1}{z}\right)\delta \left(s-\overline{m}_c^2 \right)\nonumber\\
&&-\frac{m_s m_c\langle\bar{s}g_s\sigma Gs\rangle \left[41\langle\bar{q}g_s\sigma Gq\rangle-7\langle\bar{s}g_s\sigma Gs\rangle \right]}{55296\pi^4}\int dy   \left(\frac{1}{y}+\frac{1}{1-y}\right)\delta \left(s-\widetilde{m}_c^2 \right)\nonumber\\
&&-\frac{m_s m_c\langle\bar{s}g_s\sigma Gs\rangle\left[19\langle\bar{s}g_s\sigma Gs\rangle-125\langle\bar{q}g_s\sigma Gq\rangle \right] }{55296\pi^4}\int dy \left(1+\frac{s}{T^2}\right) \delta\left(s-\widetilde{m}_c^2 \right) \nonumber\\
&&+ \frac{19 m_s m_c \langle \bar{s}g_s\sigma Gs\rangle \left[\langle \bar{q}g_s\sigma Gq\rangle+\langle \bar{s}g_s\sigma Gs\rangle\right]}{55296\pi^4}\int dy \,\delta\left(s-\widetilde{m}_c^2 \right)    \, ,
\end{eqnarray}

\begin{eqnarray}
\rho^{10\frac{1}{2},1}_{uus}(s)&=&\frac{1}{245760\pi^8}\int dydz \, yz(1-y-z)^4\left(s-\overline{m}_c^2\right)^4\left(8s-3\overline{m}_c^2 \right)  \nonumber\\
&&-\frac{1}{1228800\pi^8}\int dydz \, yz(1-y-z)^5\left(s-\overline{m}_c^2\right)^4\left(9s-4\overline{m}_c^2 \right)  \nonumber\\
&&- \frac{m_c\left[2\langle \bar{q}q\rangle+\langle \bar{s}s\rangle\right]}{4608\pi^6}\int dydz \, (y+z)(1-y-z)^2\left(s-\overline{m}_c^2\right)^3  \nonumber \\
&&+ \frac{m_c\left[2\langle \bar{q}q\rangle+\langle \bar{s}s\rangle\right]}{13824\pi^6}\int dydz \, (y+z)(1-y-z)^3\left(s-\overline{m}_c^2\right)^3  \nonumber \\
&&+\frac{11m_c\left[2\langle \bar{q}g_s\sigma Gq\rangle+\langle \bar{s}g_s\sigma Gs\rangle\right]}{24576\pi^6}\int dydz  \, (y+z)(1-y-z) \left(s-\overline{m}_c^2 \right)^2 \nonumber\\
&&- \frac{11m_c\left[2\langle \bar{q}g_s\sigma Gq\rangle+\langle \bar{s}g_s\sigma Gs\rangle\right]}{49152\pi^6}\int dydz  \, (y+z)(1-y-z)^2 \left(s-\overline{m}_c^2 \right)^2 \nonumber\\
&&- \frac{7m_c\left[2\langle \bar{q}g_s\sigma Gq\rangle+\langle \bar{s}g_s\sigma Gs\rangle\right]}{49152\pi^6}\int dydz  \, \left(\frac{y}{z}+\frac{z}{y}\right)(1-y-z)^2 \left(s-\overline{m}_c^2 \right)^2 \nonumber\\
&&+ \frac{m_c\left[2\langle \bar{q}g_s\sigma Gq\rangle+\langle \bar{s}g_s\sigma Gs\rangle\right]}{24576\pi^6}\int dydz  \, \left(\frac{y}{z}+\frac{z}{y}\right)(1-y-z)^3 \left(s-\overline{m}_c^2 \right)^2 \nonumber\\
&&+\frac{\langle\bar{q}q\rangle\left[\langle\bar{q}q\rangle+2\langle\bar{s}s\rangle \right]}{288\pi^4}\int dydz \,  yz(1-y-z)\left(s-\overline{m}_c^2 \right)\left(5s-3\overline{m}_c^2 \right) \nonumber\\
&&-\frac{\langle\bar{q}q\rangle\langle\bar{q}g_s\sigma Gq\rangle+\langle\bar{q}q\rangle\langle\bar{s}g_s\sigma Gs\rangle+\langle\bar{s}s\rangle\langle\bar{q}g_s\sigma Gq\rangle}{9216\pi^4}\int dydz \,(y+z)(1-y-z) \left(2s-\overline{m}_c^2\right)\nonumber\\
&&-\frac{19\left[\langle\bar{q}q\rangle\langle\bar{q}g_s\sigma Gq\rangle+\langle\bar{q}q\rangle\langle\bar{s}g_s\sigma Gs\rangle+\langle\bar{s}s\rangle\langle\bar{q}g_s\sigma Gq\rangle\right]}{4608\pi^4}\int dydz \,yz \left(4s-3\overline{m}_c^2\right)\nonumber\\
&&-\frac{m_c\langle\bar{q}q\rangle^2\langle\bar{s}s\rangle}{72\pi^2}\int dy \nonumber\\
&&+\frac{11\langle\bar{q}g_s\sigma Gq\rangle\left[\langle\bar{q}g_s\sigma Gq\rangle+2\langle\bar{s}g_s\sigma Gs\rangle \right]}{6144\pi^4}\int dy \, y(1-y)\left[ 1+\frac{s}{3}\delta\left(s-\widetilde{m}_c^2 \right)\right] \nonumber\\
&&+\frac{13\langle\bar{q}g_s\sigma Gq\rangle \left[ \langle\bar{q}g_s\sigma Gq\rangle+2\langle\bar{s}g_s\sigma Gs\rangle\right]}{331776\pi^4}\int dydz  \,(y+z) \left[1+s\,\delta \left(s-\overline{m}_c^2 \right)\right]\nonumber\\
&&+\frac{m_s m_c}{73728\pi^8}\int dydz \, (y+z)(1-y-z)^3\left(s-\overline{m}_c^2\right)^4  \nonumber\\
&&-\frac{m_s m_c}{294912\pi^8}\int dydz \, (y+z)(1-y-z)^4\left(s-\overline{m}_c^2\right)^4  \nonumber\\
&&- \frac{m_s\left[4\langle \bar{q}q\rangle-3\langle \bar{s}s\rangle\right]}{1536\pi^6}\int dydz \, yz(1-y-z)^2\left(s-\overline{m}_c^2\right)^2 \left(2s-\overline{m}_c^2 \right) \nonumber \\
&&- \frac{m_s\langle \bar{s}s\rangle}{4608\pi^6}\int dydz \, yz(1-y-z)^3\left(s-\overline{m}_c^2\right)^2 \left(7s-4\overline{m}_c^2 \right) \nonumber
\end{eqnarray}
\begin{eqnarray}
&&+ \frac{m_s\left[19\langle \bar{q}g_s\sigma Gq\rangle-8\langle \bar{s}g_s\sigma Gs\rangle\right]}{12288\pi^6}\int dydz  \, yz(1-y-z) \left(s-\overline{m}_c^2 \right)\left(5s-3\overline{m}_c^2 \right) \nonumber\\
&&+ \frac{m_s\langle \bar{s}g_s\sigma Gs\rangle}{1536\pi^6}\int dydz  \, yz(1-y-z)^2 \left(s-\overline{m}_c^2 \right)\left(3s-2\overline{m}_c^2 \right) \nonumber\\
&&+ \frac{m_s\langle \bar{q}g_s\sigma Gq\rangle}{49152\pi^6}\int dydz  \, (y+z)(1-y-z)^2 \left(s-\overline{m}_c^2 \right)\left(3s-\overline{m}_c^2 \right) \nonumber\\
&&+\frac{ m_s m_c\langle\bar{q}q\rangle \left[3\langle\bar{q}q\rangle-\langle\bar{s}s\rangle \right]}{288\pi^4}\int dydz \,  (y+z)\left(s-\overline{m}_c^2 \right) \nonumber\\
&&+\frac{ m_s m_c\langle\bar{q}q\rangle\langle\bar{s}s\rangle }{288\pi^4}\int dydz \,  (y+z)(1-y-z)\left(s-\overline{m}_c^2 \right) \nonumber\\
&&-\frac{m_s m_c\left[32\langle\bar{q}q\rangle\langle\bar{s}g_s\sigma Gs\rangle+57\langle\bar{s}s\rangle\langle\bar{q}g_s\sigma Gq\rangle\right]}{55296\pi^4}\int dydz  \,  \left(y+z\right) \nonumber\\
&&-\frac{m_s m_c \left[288\langle\bar{q}q\rangle\langle\bar{q}g_s\sigma Gq\rangle-32\langle\bar{q}q\rangle\langle\bar{s}g_s\sigma Gs\rangle-57\langle\bar{s}s\rangle\langle\bar{q}g_s\sigma Gq\rangle\right]}{55296\pi^4}\int dy   \nonumber\\
&&+\frac{m_s m_c\langle\bar{s}s\rangle\langle\bar{q}g_s\sigma Gq\rangle}{1536\pi^4}\int dydz  \left(\frac{y}{z}+\frac{z}{y}\right)(1-y-z)\nonumber\\
&&+\frac{m_s m_c\left[16\langle\bar{q}q\rangle-7\langle\bar{s}s\rangle\right]\langle\bar{q}g_s\sigma Gq\rangle}{9216\pi^4}\int dydz  \, \left(\frac{y}{z}+\frac{z}{y}\right) \nonumber\\
&&   +\frac{m_s\langle\bar{q}q\rangle^2\langle\bar{s}s\rangle}{72\pi^2}\int dy \,y(1-y) \left[ 1+\frac{s}{3}\delta\left(s-\widetilde{m}_c^2 \right)\right]  \nonumber\\
&&-\frac{m_s m_c\langle\bar{q}g_s\sigma Gq\rangle \langle\bar{s}g_s\sigma Gs\rangle}{6912\pi^4}\int dy \left(1+\frac{s}{T^2}\right)  \delta\left(s-\widetilde{m}_c^2 \right) \nonumber\\
&&-\frac{m_s m_c\langle\bar{q}g_s\sigma Gq\rangle \langle\bar{s}g_s\sigma Gs\rangle}{9216\pi^4}\int dydz   \left(\frac{z}{y}+\frac{y}{z}\right)\delta \left(s-\overline{m}_c^2 \right)\nonumber\\
&&-\frac{m_s m_c\langle\bar{q}g_s\sigma Gq\rangle\left[24\langle\bar{q}g_s\sigma Gq\rangle-7\langle\bar{s}g_s\sigma Gs\rangle \right] }{55296\pi^4}\int dy   \left(  \frac{1-y}{y}+\frac{y}{1-y}\right) \delta\left(s-\widetilde{m}_c^2 \right)\nonumber\\
&&+\frac{m_s m_c\langle\bar{q}g_s\sigma Gq\rangle \left[ 24\langle\bar{q}g_s\sigma Gq\rangle-\langle\bar{s}g_s\sigma Gs\rangle\right]}{18432\pi^4}\int dy \,\left(1+\frac{s}{2T^2}\right)\delta\left(s-\widetilde{m}_c^2 \right)    \nonumber\\
&&+ \frac{m_s m_c \langle \bar{q}g_s\sigma Gq\rangle\langle \bar{s}g_s\sigma Gs\rangle}{36864\pi^4}\int dy \,\delta\left(s-\widetilde{m}_c^2 \right)   \, ,
\end{eqnarray}

\begin{eqnarray}
\widetilde{\rho}^{10\frac{1}{2},0}_{uus}(s)&=&\frac{1}{491520\pi^8}\int dydz \, (y+z)(1-y-z)^4\left(s-\overline{m}_c^2\right)^4\left(7s-2\overline{m}_c^2 \right)  \nonumber\\
&&-\frac{1}{2457600\pi^8}\int dydz \, (y+z)(1-y-z)^5\left(s-\overline{m}_c^2\right)^4\left(8s-3\overline{m}_c^2 \right)  \nonumber\\
&&- \frac{m_c\left[2\langle \bar{q}q\rangle+\langle \bar{s}s\rangle\right]}{2304\pi^6}\int dydz \, (1-y-z)^2\left(s-\overline{m}_c^2\right)^3  \nonumber \\
&&+ \frac{m_c\left[2\langle \bar{q}q\rangle+\langle \bar{s}s\rangle\right]}{6912\pi^6}\int dydz \,  (1-y-z)^3\left(s-\overline{m}_c^2\right)^3  \nonumber \\
&&+ \frac{11m_c\left[2\langle \bar{q}g_s\sigma Gq\rangle+\langle \bar{s}g_s\sigma Gs\rangle\right]}{12288\pi^6}\int dydz  \,  (1-y-z) \left(s-\overline{m}_c^2 \right)^2 \nonumber\\
&&- \frac{11m_c\left[2\langle \bar{q}g_s\sigma Gq\rangle+\langle \bar{s}g_s\sigma Gs\rangle\right]}{24576\pi^6}\int dydz  \,  (1-y-z)^2 \left(s-\overline{m}_c^2 \right)^2 \nonumber\\
&&- \frac{7 m_c\left[2\langle \bar{q}g_s\sigma Gq\rangle+\langle \bar{s}g_s\sigma Gs\rangle\right]}{49152\pi^6}\int dydz  \, \left(\frac{1}{y}+\frac{1}{z}\right)(1-y-z)^2 \left(s-\overline{m}_c^2 \right)^2 \nonumber
\end{eqnarray}
\begin{eqnarray}
&&+ \frac{m_c\left[2\langle \bar{q}g_s\sigma Gq\rangle+\langle \bar{s}g_s\sigma Gs\rangle\right]}{24576\pi^6}\int dydz  \, \left(\frac{1}{y}+\frac{1}{z}\right)(1-y-z)^3 \left(s-\overline{m}_c^2 \right)^2 \nonumber\\
&&+\frac{\langle\bar{q}q\rangle \left[\langle\bar{q}q\rangle+2\langle\bar{s}s\rangle \right]}{288\pi^4}\int dydz \,  (y+z)(1-y-z)\left(s-\overline{m}_c^2 \right)\left(2s-\overline{m}_c^2 \right) \nonumber\\
&&-\frac{19\left[\langle\bar{q}q\rangle\langle\bar{q}g_s\sigma Gq\rangle+\langle\bar{q}q\rangle\langle\bar{s}g_s\sigma Gs\rangle+\langle\bar{s}s\rangle\langle\bar{q}g_s\sigma Gq\rangle\right]}{9216\pi^4}\int dydz \,(y+z) \left(3s-2\overline{m}_c^2\right)\nonumber\\
&&-\frac{\langle\bar{q}q\rangle\langle\bar{q}g_s\sigma Gq\rangle+\langle\bar{q}q\rangle\langle\bar{s}g_s\sigma Gs\rangle+\langle\bar{s}s\rangle\langle\bar{q}g_s\sigma Gq\rangle}{4608\pi^4}\int dydz \, (1-y-z) s\nonumber\\
&&-\frac{m_c\langle\bar{q}q\rangle^2\langle\bar{s}s\rangle}{36\pi^2}\int dy \nonumber\\
&&+\frac{11\langle\bar{q}g_s\sigma Gq\rangle \left[\langle\bar{q}g_s\sigma Gq\rangle+2\langle\bar{s}g_s\sigma Gs\rangle \right]}{18432\pi^4}\int dy \, \left[ 1+\frac{s}{2}\delta\left(s-\widetilde{m}_c^2 \right)\right] \nonumber\\
&&+\frac{13\langle\bar{q}g_s\sigma Gq\rangle \left[\langle\bar{q}g_s\sigma Gq\rangle+2\langle\bar{s}g_s\sigma Gs\rangle \right]}{165888\pi^4}\int dydz  \, s\,\delta \left(s-\overline{m}_c^2 \right) \nonumber\\
&&+\frac{m_s m_c}{36864\pi^8}\int dydz \, (1-y-z)^3\left(s-\overline{m}_c^2\right)^4  \nonumber\\
&&-\frac{m_s m_c}{147456\pi^8}\int dydz \, (1-y-z)^4\left(s-\overline{m}_c^2\right)^4  \nonumber\\
&&- \frac{m_s\left[4\langle \bar{q}q\rangle-3\langle \bar{s}s\rangle\right]}{9216\pi^6}\int dydz \, (y+z)(1-y-z)^2\left(s-\overline{m}_c^2\right)^2 \left(5s-2\overline{m}_c^2 \right) \nonumber \\
&&- \frac{m_s\langle \bar{s}s\rangle}{3072\pi^6}\int dydz \, (y+z)(1-y-z)^3\left(s-\overline{m}_c^2\right)^2 \left(2s-\overline{m}_c^2 \right) \nonumber \\
&&+ \frac{m_s\left[19\langle \bar{q}g_s\sigma Gq\rangle-8\langle \bar{s}g_s\sigma Gs\rangle\right]}{12288\pi^6}\int dydz  \, (y+z)(1-y-z) \left(s-\overline{m}_c^2 \right)\left(2s-\overline{m}_c^2 \right) \nonumber\\
&&+ \frac{m_s\langle \bar{s}g_s\sigma Gs\rangle}{6144\pi^6}\int dydz  \, (y+z)(1-y-z)^2 \left(s-\overline{m}_c^2 \right)\left(5s-3\overline{m}_c^2 \right) \nonumber\\
&&+ \frac{m_s\langle \bar{q}g_s\sigma Gq\rangle}{12288\pi^6}\int dydz  \,  (1-y-z)^2 \,s\left(s-\overline{m}_c^2 \right) \nonumber\\
&&+\frac{ m_s m_c\langle\bar{q}q\rangle\left[3\langle\bar{q}q\rangle-\langle\bar{s}s\rangle \right]}{144\pi^4}\int dydz \,   \left(s-\overline{m}_c^2 \right) \nonumber\\
&&+\frac{ m_s m_c\langle\bar{q}q\rangle\langle\bar{s}s\rangle }{144\pi^4}\int dydz \,   (1-y-z)\left(s-\overline{m}_c^2 \right) \nonumber\\
&&-\frac{m_s m_c\left[32\langle\bar{q}q\rangle\langle\bar{s}g_s\sigma Gs\rangle+57\langle\bar{s}s\rangle\langle\bar{q}g_s\sigma Gq\rangle\right]}{27648\pi^4}\int dydz    \nonumber\\
&&-\frac{m_s m_c\left[288 \langle\bar{q}q\rangle\langle\bar{q}g_s\sigma Gq\rangle-32\langle\bar{q}q\rangle\langle\bar{s}g_s\sigma
Gs\rangle-57\langle\bar{s}s\rangle\langle\bar{q}g_s\sigma Gq\rangle\right]}{27648\pi^4}\int dy   \nonumber\\
&&+\frac{m_s m_c\langle\bar{s}s\rangle\langle\bar{q}g_s\sigma Gq\rangle}{1536\pi^4}\int dydz  \left(\frac{1}{y}+\frac{1}{z}\right)(1-y-z)\nonumber\\
&&+\frac{m_s m_c\left[16\langle\bar{q}q\rangle-7\langle\bar{s}s\rangle\right]\langle\bar{q}g_s\sigma Gq\rangle}{9216\pi^4}\int dydz  \, \left(\frac{1}{y}+\frac{1}{z}\right) \nonumber\\
&&   +\frac{m_s\langle\bar{q}q\rangle^2\langle\bar{s}s\rangle}{216\pi^2}\int dy \,  \left[ 1+\frac{s}{2}\delta\left(s-\widetilde{m}_c^2 \right)\right]  \nonumber\\
&&-\frac{m_s m_c\langle\bar{q}g_s\sigma Gq\rangle \langle\bar{s}g_s\sigma Gs\rangle}{3456\pi^4 T^2}\int dy \, s\, \delta\left(s-\widetilde{m}_c^2 \right) \nonumber
\end{eqnarray}
\begin{eqnarray}
&&-\frac{m_s m_c\langle\bar{q}g_s\sigma Gq\rangle\langle\bar{s}g_s\sigma Gs\rangle }{9216\pi^4}\int dydz   \left(\frac{1}{y}+\frac{1}{z}\right)\delta \left(s-\overline{m}_c^2 \right)\nonumber\\
&&-\frac{m_s m_c\langle\bar{q}g_s\sigma Gq\rangle \left[24 \langle\bar{q}g_s\sigma Gq\rangle-7\langle\bar{s}g_s\sigma Gs\rangle\right]}{55296\pi^4}\int dy   \left(  \frac{1}{y}+\frac{1}{1-y}\right) \delta\left(s-\widetilde{m}_c^2 \right)\nonumber\\
&&+\frac{m_s m_c\langle\bar{q}g_s\sigma Gq\rangle\left[24\langle\bar{q}g_s\sigma Gq\rangle-\langle\bar{s}g_s\sigma Gs\rangle\right]}{18432\pi^4}\int dy \,\left(1+\frac{s}{T^2}\right)\delta\left(s-\widetilde{m}_c^2 \right)    \nonumber\\
&&+ \frac{m_s m_c \langle \bar{q}g_s\sigma Gq\rangle \langle \bar{s}g_s\sigma Gs\rangle}{18432\pi^4}\int dy \,\delta\left(s-\widetilde{m}_c^2 \right)    \, ,
\end{eqnarray}

\begin{eqnarray}
\rho^{10\frac{1}{2},1}_{uuu}(s)&=&\rho^{10\frac{1}{2},1}_{sss}(s)\mid_{m_s \to 0,\,\,\langle\bar{s}s\rangle\to\langle\bar{q}q\rangle,\,\,\langle\bar{s}g_s\sigma Gs\rangle\to\langle\bar{q}g_s\sigma Gq\rangle} \, ,  \\
\widetilde{\rho}^{10\frac{1}{2},0}_{uuu}(s)&=&\widetilde{\rho}^{10\frac{1}{2},0}_{sss}(s)\mid_{m_s \to 0,\,\,\langle\bar{s}s\rangle\to\langle\bar{q}q\rangle,\,\,\langle\bar{s}g_s\sigma Gs\rangle\to\langle\bar{q}g_s\sigma Gq\rangle} \, ,
\end{eqnarray}
where $\int dydz=\int_{y_i}^{y_f}dy \int_{z_i}^{1-y}dz$, $\int dy=\int_{y_i}^{y_f}dy$, $y_{f}=\frac{1+\sqrt{1-4m_c^2/s}}{2}$,
$y_{i}=\frac{1-\sqrt{1-4m_c^2/s}}{2}$, $z_{i}=\frac{y
m_c^2}{y s -m_c^2}$, $\overline{m}_c^2=\frac{(y+z)m_c^2}{yz}$,
$ \widetilde{m}_c^2=\frac{m_c^2}{y(1-y)}$, $\int_{y_i}^{y_f}dy \to \int_{0}^{1}dy$, $\int_{z_i}^{1-y}dz \to \int_{0}^{1-y}dz$,  when the $\delta$ functions $\delta\left(s-\overline{m}_c^2\right)$ and $\delta\left(s-\widetilde{m}_c^2\right)$ appear.

\section*{Acknowledgements}
This  work is supported by National Natural Science Foundation,
Grant Number 11375063, and Natural Science Foundation of Hebei province, Grant Number A2014502017.

\end{document}